# MANAGERIAL INSIGHTS ON INVESTMENT STRATEGY IN CYBERSECURITY: FINDINGS FROM MULTI-COUNTRY RESEARCH

Silvia Tedeschi, Giacomo Marzi, Marco Balzano, Gabriele Costa

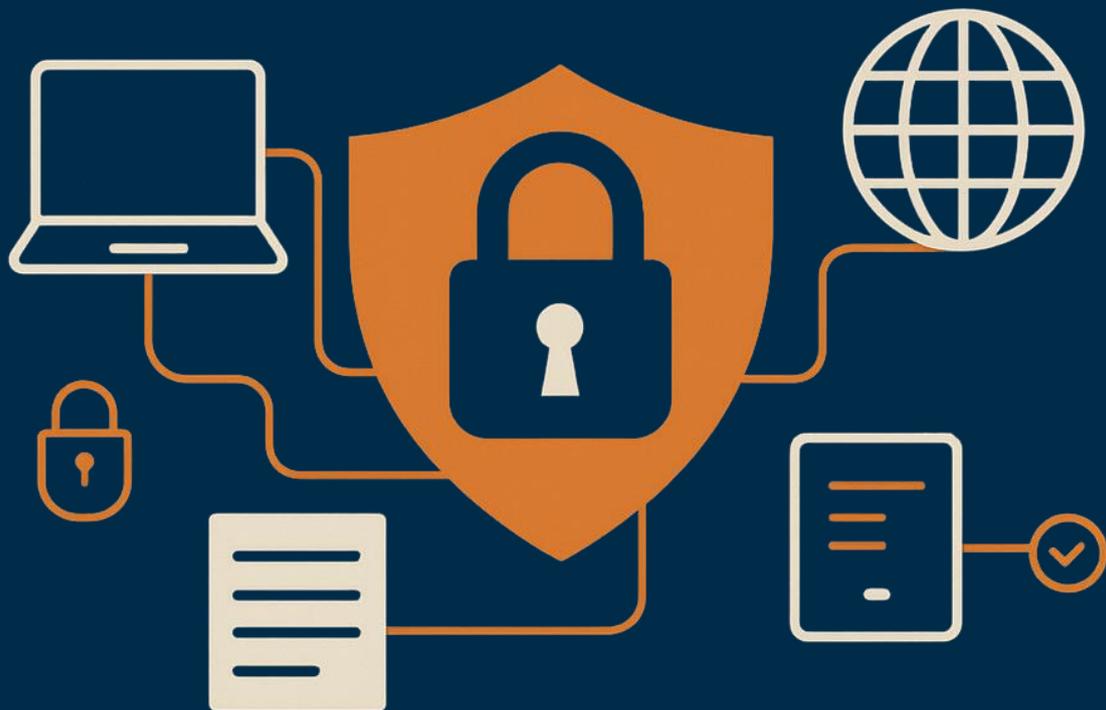

**Industry Research Report**

# TABLE OF CONTENTS

# 1. EXECUTIVE SUMMARY

Cybersecurity has shifted from a technical safeguard to a strategic priority, influencing risk management and the competitive positioning in an increasingly interconnected economy. Based on a large-scale survey of 1,083 executives and managers across Europe and North America, this report captures the contemporary realities of cybersecurity practices, perceptions, and priorities within businesses of varying size and technological maturity.

Findings show that cybersecurity is increasingly recognized as a source of competitive advantage, though resource constraints, talent shortages, and cultural resistance persistently challenge its integration into organizational strategy. Large enterprises and high-tech companies generally adopt a more proactive, investment-oriented approach, while SMEs and low-tech sectors display greater variability, often driven by budgetary limitations and competing strategic demands.

The analysis highlights critical tensions managers navigate daily: balancing innovation with security, maintaining agility without sacrificing resilience, and integrating cybersecurity seamlessly into project planning and supply chain operations. Cyber Insurance, Zero Trust Architecture, Vulnerability Assessment and Penetration Testing (VAPT), and automation emerge as areas of both opportunity and uncertainty, reflecting broader ambivalence around newer security paradigms.

Country-level comparisons show distinct patterns: the United States tends to exhibit higher investment and incident reporting, while Italy exhibits a unique combination of strong cybersecurity advocacy with lower resource allocation. Across all geographies, however, the human dimension, through employee training and leadership engagement, remains central to any cybersecurity strategy.

The results point to a maturing cybersecurity landscape, where organizations increasingly recognize security as a foundation for sustainable growth and digital trust. Yet, a gap remains between strategic intent and operational execution, particularly among smaller and less technologically advanced companies. Bridging this gap will require moving beyond compliance-driven models toward a more integrated, innovation-aligned cybersecurity culture.

# 2. INTRODUCTION: NAVIGATING THE CYBERSECURITY LANDSCAPE

## 2.1. THE STRATEGIC IMPERATIVE OF CYBERSECURITY

Cybersecurity has become a fundamental pillar of modern business strategy, extending beyond traditional IT concerns. In an era where digital assets and online operations are critical to competitiveness, companies face an increasing number of cyber threats that can lead to operational disruptions, financial losses, and reputational damage. As digital infrastructures evolve, organizations must proactively safeguard their systems, ensuring resilience and adaptability in a rapidly changing threat landscape.

## 2.2. UNDERSTANDING THE THREAT LANDSCAPE

The modern cybersecurity environment is shaped by a diverse range of threats, from data breaches and ransomware attacks to sophisticated cyber espionage. Organizations of all sizes are vulnerable, with cyber risks affecting operational continuity, regulatory compliance, and customer trust. Businesses must approach cybersecurity as a strategic risk management issue, embedding security considerations into broader corporate decision-making rather than treating it purely as a technological concern.

## 2.3. THE ROLE OF ORGANIZATIONAL STRATEGY IN CYBERSECURITY

A company's approach to cybersecurity should be aligned with its overall business strategy. This includes the adoption of a robust technological defense while also fostering a security-conscious culture among employees. Organizations that integrate cybersecurity into their strategic vision are better equipped to handle threats proactively, ensuring that security measures evolve alongside digital transformation initiatives. Strong leadership commitment and continuous staff training are essential in building resilience against cyber threats.

## 2.4. CYBERSECURITY CHALLENGES FOR BUSINESSES

While awareness of cybersecurity has increased, businesses still encounter several key challenges in implementing effective security strategies, such as:

**Resource Constraints.** Many organizations struggle with limited budgets and expertise dedicated to cybersecurity.

**Regulatory Compliance.** Adapting to complex and evolving cybersecurity regulations require continuous monitoring and investment.

**Human Factors.** Employees remain a primary vulnerability, whether through phishing attacks, weak passwords, or lack of security awareness.

**Risk Integration.** Cybersecurity is often treated as a standalone issue rather than an integral component of business risk management.

## 2.5. CYBERSECURITY AS A COMPETITIVE ADVANTAGE

Companies that prioritize cybersecurity can increasingly leverage it as a differentiator, building stronger customer trust and ensuring compliance with industry standards. A proactive cybersecurity stance mitigates risks and enhances brand reputation. Businesses that embed security into their digital transformation strategies are better positioned to navigate the complexities of an increasingly interconnected world.

## 2.6. ABOUT THE RESEARCH

This report presents insights derived from a survey conducted among 1,083 business executives and managers in both SMEs and large enterprises, offering a practical analysis of current cybersecurity practices and challenges. The goal is to capture insights on cybersecurity practices, challenges, and strategic priorities, equipping business leaders with actionable strategies to strengthen their cybersecurity posture and drive long-term resilience in an evolving digital landscape. The survey was conducted in the early months of 2025, focusing on regions including Europe, United States and United Kingdom.

# 3. MANAGERIAL PERSPECTIVES ON CYBERSECURITY: SURVEY INSIGHTS AND TRENDS

## 3.1. CYBERSECURITY: RISK OR OPPORTUNITY?

### 3.1.1. IS CYBERSECURITY A COMPETITIVE ADVANTAGE OR A NECESSARY BURDEN?

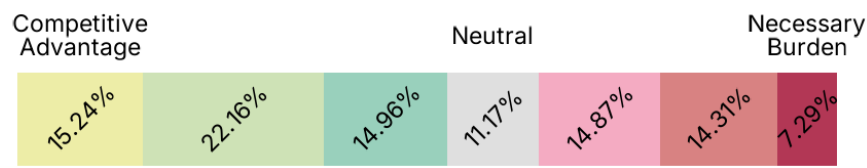

Figure 1 • Do you see cybersecurity more as a competitive advantage or a necessary burden? - 7 points Likert scale

We asked respondents whether they perceive cybersecurity as a competitive advantage or a necessary burden.

The results show a balanced distribution along the Likert scale, with one key exception: very few see it purely as a necessary burden. This suggests that cybersecurity is now regarded as a fundamental element of business strategy.

*This is also evidenced by what managers told us when we asked them openly about Cybersecurity:*

*"Cybersecurity isn't a barrier but an edge to competitive advantage that enables innovation while safeguarding our games, players, and reputation"*

*"The biggest challenge is balancing security with agility, which, when done right, becomes a competitive advantage"*

*"Cybersecurity is key to protecting data, earning customer trust, and keeping things running smoothly. We see it as a core part of innovation. While security can add complexity, it also makes our products stronger and more competitive"*

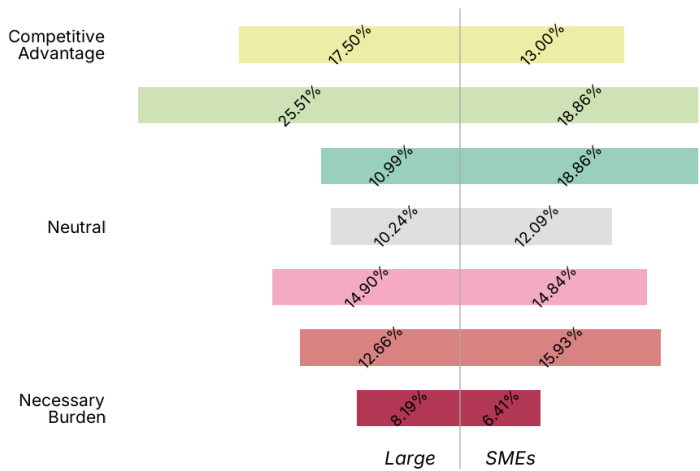

Figure 2 • Cybersecurity perception: Large Enterprises vs. SMEs

*How might company size affect the managerial perceptions about cybersecurity?*

Large enterprises tend to see cybersecurity as a clear competitive advantage, reinforcing their position in the market. SMEs, while still recognizing their strategic importance, display a more nuanced stance, with responses more evenly distributed and less concentrated at the extreme "Competitive Advantage" end of the scale.

*High tech vs Low tech: does the perception about cybersecurity change?*

High-tech companies are more likely to embrace cybersecurity as a strategic enabler, while low-tech enterprises appear more divided on their role in creating business value.

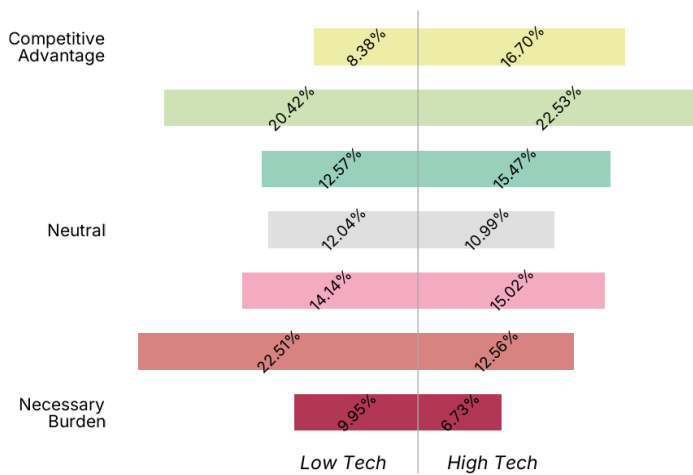

Figure 3 • Cybersecurity perception: High Tech vs. Low Tech Enterprises

### 3.1.2. CYBERSECURITY SPENDING: NECESSARY EXPENSE OR WORTHY INVESTMENT?

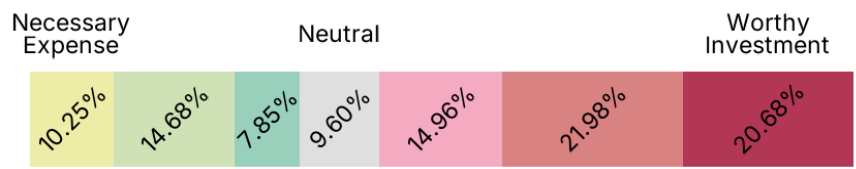

Figure 4 • Do you see cyber defense primarily as a necessary expense or a worthy investment? - 7 points Likert Scale

We posed this question to our respondents and results show a strong inclination towards seeing cyber defense as a worthy investment rather than just a necessary expense. While responses are spread across the scale, the highest concentrations are at the right end, with 21.90% and 20.60% firmly considering it a worthy investment. On the other hand, only a small minority (10.30%) see it purely as a necessary expense.

*This is further supported by the insights shared by managers when we engaged them in open discussions about Cybersecurity:*

*"Cybersecurity is very important, but budget constraints limit how much we can allocate to dedicated security measures. It's a necessary cost, but not always prioritized"*

*"The cost of implementing this will be off-putting to the board, however when you weigh it up against the ability to protect us against downtime and build more trust with customers, then it is most certainly a valuable investment that I would hope they would think about in the future"*

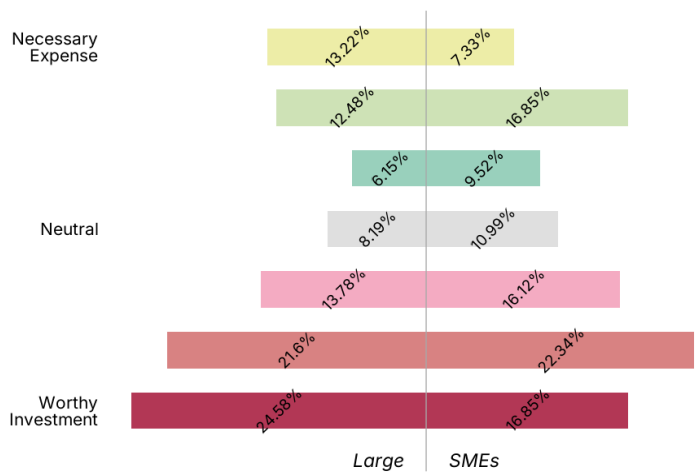

Figure 5 • Cybersecurity spending: Large Enterprises vs. SMEs

*Large vs SMEs: does the perception about cybersecurity change?*

Large enterprises perceive cybersecurity more as a "Worthy Investment", allocating a significantly larger proportion of their cybersecurity budget to this category compared to SMEs. This suggests a more proactive and strategic stance, perhaps recognizing the long-term benefits and the potential for greater returns from robust security measures.

On the other hand, SMEs seem to prioritize "Necessary Expenses", dedicating a larger share (16.85%) compared to large enterprises (13.22%). This implies a more reactive approach, focusing on essential security measures that are deemed absolutely required. Their higher spending in the "Neutral" category (10.99%) compared to their "Worthy Investment" allocation further reinforces this focus on immediate needs.

The difference in perspective likely stems from several factors. Large enterprises, with their greater public visibility and potentially larger attack surfaces, probably feel a more acute need to proactively invest in comprehensive cybersecurity.

The potential financial and reputational damage from a cyber incident is often much higher for larger organizations, justifying a more investment-oriented approach.

SMEs, often operating with tighter budgets and potentially facing different scales of threats, might perceive cybersecurity more as a cost to be managed. While no less important, their immediate financial constraints might lead them to prioritize essential expenses over more strategic investments.

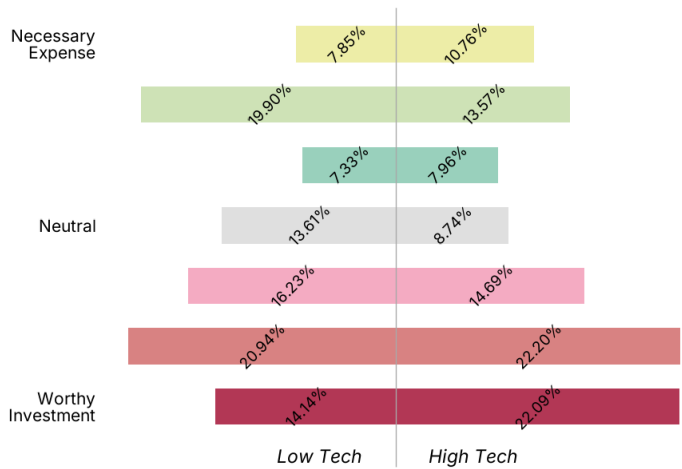

Figure 6 • Cybersecurity spending: Low Tech vs. High Tech Enterprises

*How does technological intensity shape the perception of cybersecurity?*

High-tech companies overwhelmingly recognize cybersecurity as a key investment, clustering around the highest values on the scale. In contrast, low-tech enterprises show a more divided stance, with responses concentrated around both 2 and 6 values, signaling that while some see the value in cybersecurity, others still perceive it as a cost-driven necessity rather than a growth enabler.

### 3.1.3. CYBER THREATS: GREATER CONCERN INSIDE OR OUTSIDE EUROPE?

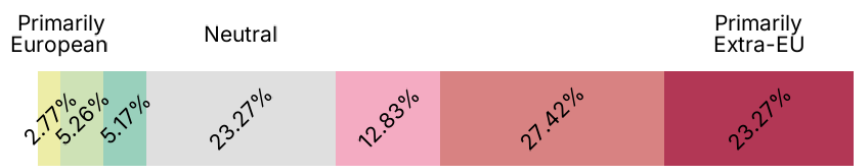

Figure 7 • Are you more concerned about cyber threats from within Europe or from outside Europe? - 7 points Likert Scale

When asked about the perceived origin of cyber threats, responses show a clear inclination towards "Primarily Extra-EU", indicating that a significant portion of organizations see external, non-European threats as the predominant risk. However, a secondary segment of respondents remains neutral, suggesting that they perceive cyber threats as equally originating from both within and outside the EU.

*While this perspective is not directly emphasized by managers' responses during our open dialogue on Cybersecurity, several of them alluded to the global and cross-border nature*

*of cybersecurity risks, highlighting concerns that span beyond national or continental boundaries.*

*"The number and severity of cyber threats continues to grow, both from other companies, as well as from foreign governments and other hostile parties"*

*"We operate on a global scale, in different markets and countries […] it is vital that our assets are protected at all times"*

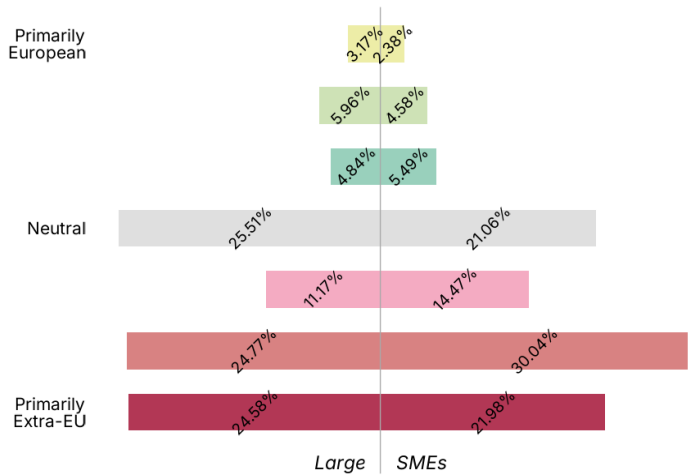

Figure 8 • Cyber Threats: Large Enterprises vs. SMEs

*How does company size influence the perception of Cybersecurity?*

Empirical evidence shows that company size does not play a major role in shaping this perception, as SMEs and large enterprises exhibit broadly similar views.

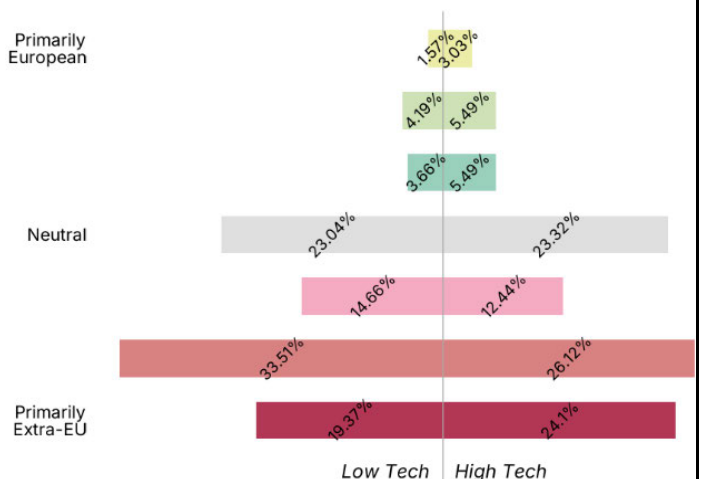

Figure 9 • Cyber Threats: Low Tech vs. High Tech Enterprises

*How does technological intensity shape the perception of cybersecurity?*

Similarly, the distinction between high-tech and low-tech sectors do not show significant differences; in fact, the results indicate a clear tendency to perceive cybersecurity threats as more pronounced from outside the EU.

### 3.1.4. SECURITY MEASURES: PRODUCTIVITY BARRIER OR ESSENTIAL FOR PROTECTION?

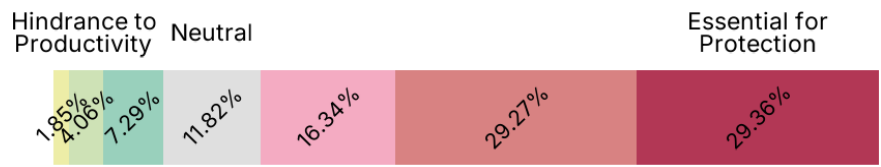

Figure 10 • Do you feel that security measures in your organization are more of a hindrance to productivity or essential for protection? - 7 points Likert Scale

We engaged respondents in a discussion about their views on security protocols which can be perceived as walking a fine line between protection and operational efficiency.

There is a clear consensus: the vast majority of managers view security measures as essential safeguards rather than obstacles to productivity.

*These managers' perceptions provide additional evidence, as shown in our open inquiries about Cybersecurity:*

*"Due to Society's inclination to the Internet, there is the need to protect your space and environment against internet fraudsters. [...] It's key to integrate security measures to help curb cybercrimes"*

*"Cybersecurity is a main component of any new idea plan or innovation plan. [...] With no strong security measures, even the most incredible ideas can become risky"*

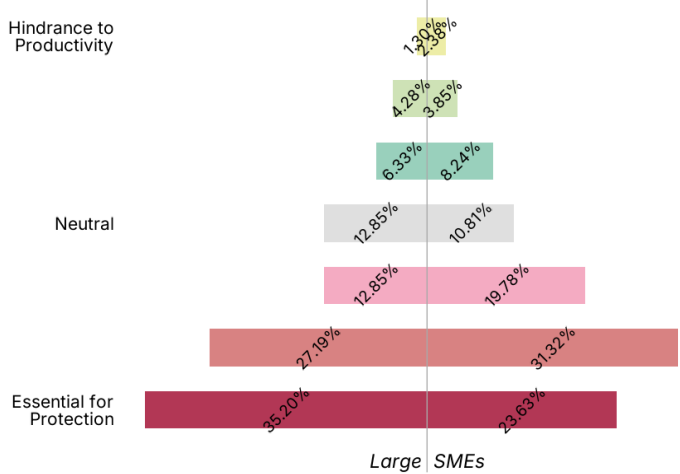

Figure 11 • Security Measures: Large Enterprises vs. SMEs

*How does company size influence the perception of Cybersecurity?*

Company size does not play a major role in shaping this perception, as SMEs and large enterprises exhibit broadly similar views.

However, there are slight differences at the extreme end of the spectrum; large enterprises are more likely to lean toward the value of "Essential for Protection", while SMEs tend to remain at the preceding value.

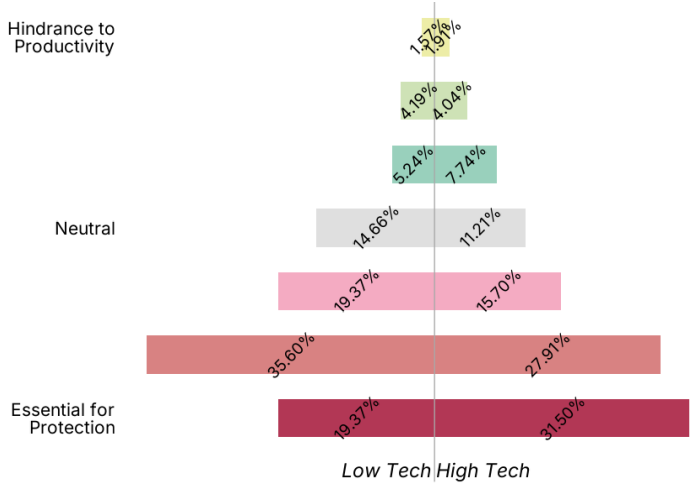

*How does technological intensity shape the perception of cybersecurity?*

A similar result can be observed: where low-tech organizations tend to remain at the preceding value, while high-tech organizations move toward the extreme with more significant responses.

Figure 12 • Security Measures: Low Tech vs. High Tech Enterprises

### 3.1.5. CYBER REGULATIONS: FRIEND OR FOE?

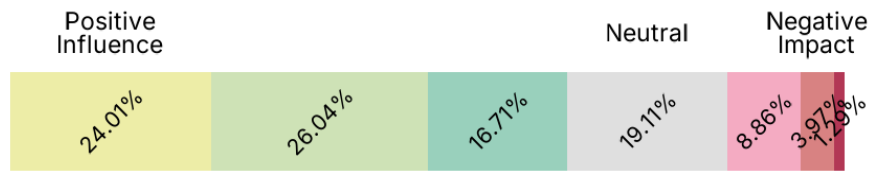

Figure 13 • How do you perceive the impact of cyber regulations on your business: as a positive influence or a negative impact? - 7 points Likert Scale

We explored how respondents view cyber regulations, while compliance can introduce complexity, around 50% of managers view regulatory frameworks as having a positive influence, ensuring better security standards and trust.

*This perspective is reinforced by what managers directly expressed when we asked them about Cybersecurity:*

*"Staying ahead of security requirements helps organizations comply with regulations and avoid potential penalties. Overall, integrating cybersecurity into technological developments is at the same time about protection and enabling innovation in a secure and trusted environment"*

*"Cybersecurity is essential for any business today. Incorporating it into your innovation strategy can offer numerous benefits. It safeguards critical information, upholds customer trust, and ensures compliance with necessary regulations"*

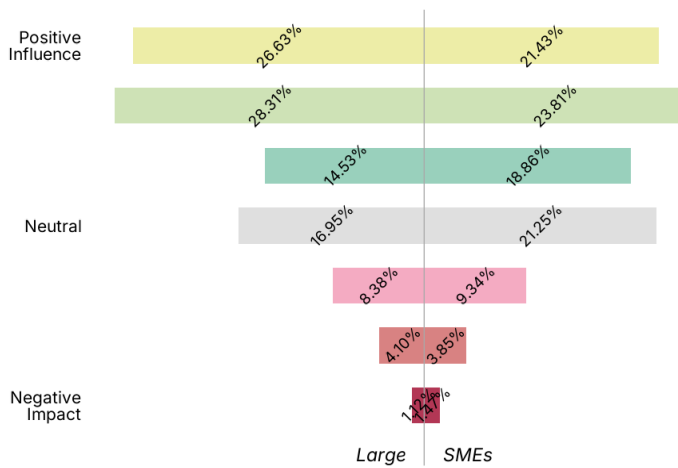

Figure 14 • Cyber Regulations perception: Large Enterprises vs. SMEs

*How does company size influence the perception of Cybersecurity?*

Both large enterprises and SMEs lean towards the positive side of the scale. SMEs are more inclined towards the neutral values, while large enterprises shift slightly more towards the extreme value of "Positive Influence".

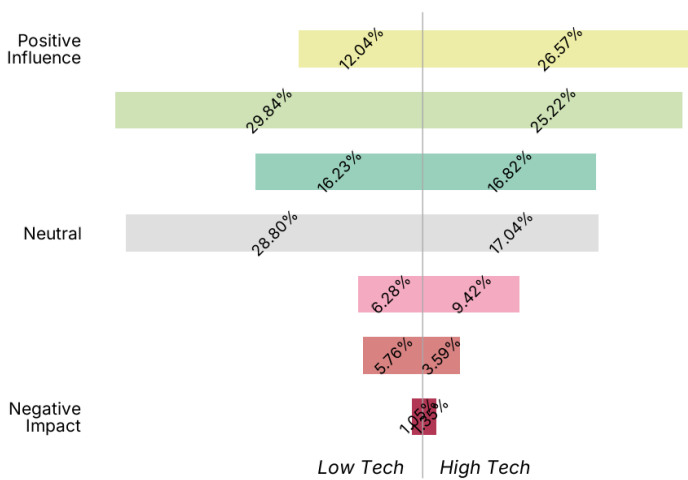

Figure 15 • Cyber Regulations perception: Low Tech vs. High Tech Enterprises

*How does technological intensity shape the perception of cybersecurity?*

For low-tech enterprises, the distribution of responses is notably unbalanced, with a significant concentration in the neutral and medium positive range, and only a minimal representation at the extreme values.

In contrast, high-tech enterprises exhibit a more balanced distribution, with responses gradually increasing from "Negative Impact" to "Positive Influence".

### 3.1.6. SUPPLY CHAIN SECURITY: PRIORITY OR AFTERTHOUGHT?

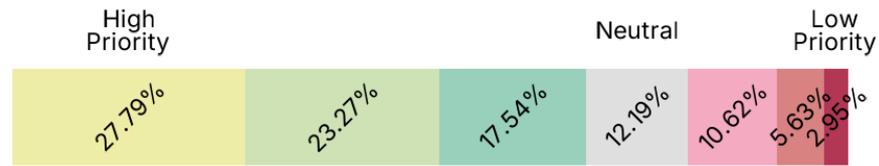

Figure 16 • Is supply chain security a high priority or a low priority in your organization? - 7 points Likert Scale

Our inquiry examined how respondents address security challenges within their supply chains, reflecting the growing external dimensions of cybersecurity. For most organizations, securing third-party vendors and suppliers is a top priority. This underscores the growing recognition that a company's cybersecurity posture is only as strong as its weakest link.

*Managers' responses during our open dialogue on Cybersecurity also highlight this viewpoint.*

*"I see cybersecurity as a critical component of our operations, especially as we increasingly rely on digital systems for production, supply chain management, and customer interactions"*

*"Cybersecurity is crucial in manufacturing, protecting technical products, IoT systems, and supply chain data from threats"*

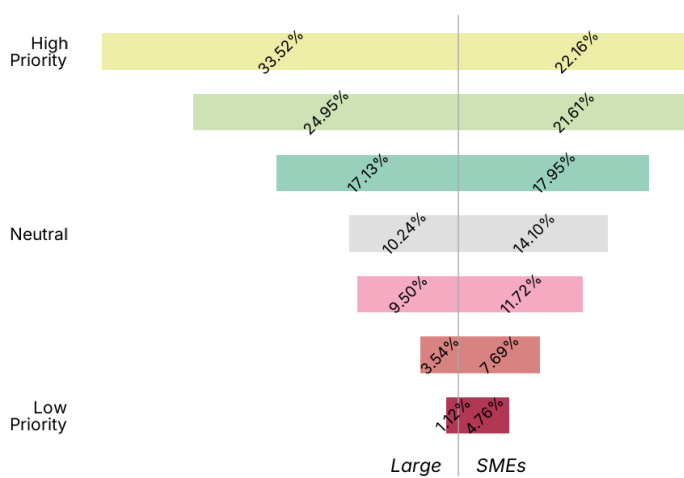

Figure 17 • Supply Chain Security: Large Enterprises vs. SMEs

*How does company size influence the perception of Cybersecurity?*

There is a gradual increase in prioritization of supply chain security for both large enterprises and SMEs. However, SMEs show a higher concentration of responses at the low priority end, while large enterprises exhibit a stronger inclination toward the high priority extreme, suggesting that large enterprises consider supply chain security to be a higher priority compared to SMEs.

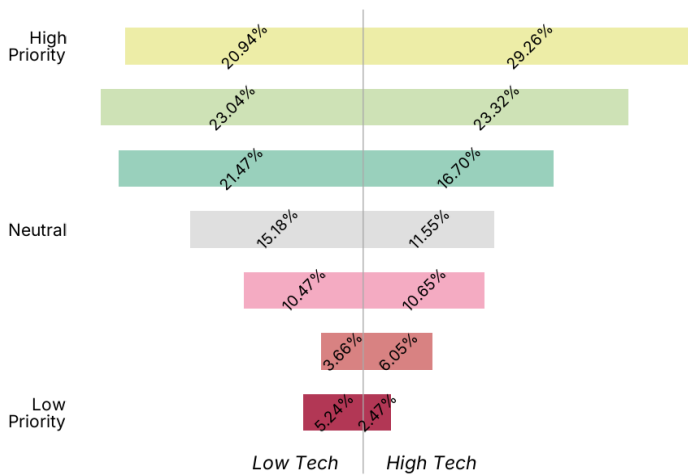

Figure 18 • Supply Chain Security: Low Tech vs. High Tech Enterprises

*How does technological intensity shape the perception of cybersecurity?*

No significant differences emerge. So, the perceived importance of supply chain security is not necessarily driven by the technological sophistication of a company but rather by other factors, such as industry-specific risks.

### 3.1.7. CYBERSECURITY TALENT: SCARCE OR ACCESSIBLE?

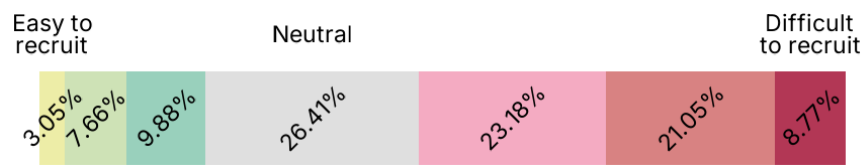

Figure 19 • How would you describe the ease of recruiting Cybersecurity talent: easy or difficult? - 7 points Likert Scale

The ability to attract and retain cybersecurity professionals is a pressing challenge. Delving into respondents' perspectives we see that most organizations struggle with recruitment, highlighting a widening skills gap. This reinforces the need for companies to invest in talent development and strategic partnerships to close the expertise deficit.

*The opinions of managers illustrate this point:*

*"[...] The main challenge we face is recruiting skilled talent to fill relevant roles. We hope new technological developments will reduce our headcount needs"*

*"The company can't afford to be compromised for even a short spell of time it is difficult for us to recruit quality cybersecurity officers and even to train new recruits in our current measures"*

*How does company size influence the perception of Cybersecurity?*

Recruiting cybersecurity talent is a challenge that seems to present the same level of difficulty for large enterprises and SMEs.

The highest percentage of responses is positioned around the neutral value of the scale, followed by significant percentages indicating "Difficult to recruit", with lower values at the extreme end.

Overall, this suggests that both large enterprises and SMEs face considerable challenges in attracting cybersecurity talent.

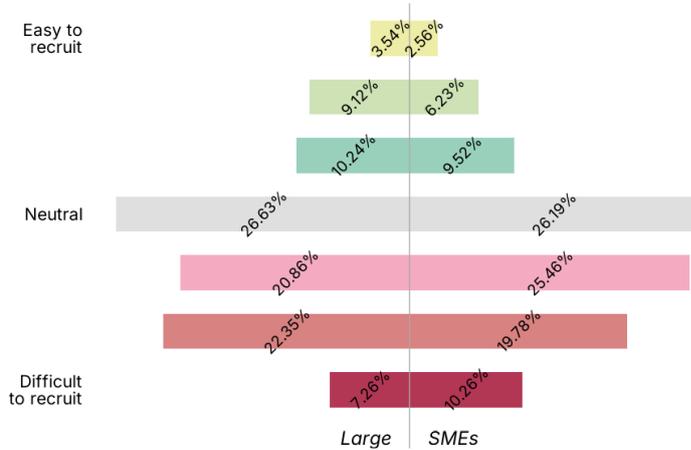

Figure 20 • Recruiting Cybersecurity talent: Large Enterprises vs. SMEs

*How does technological intensity shape the perception of cybersecurity?*

Differences in technology level do not appear to influence the difficulty in recruiting cybersecurity talent. This further emphasizes that neither company size nor technological intensity significantly impact the challenge, highlighting talent acquisition as a widespread and persistent issue.

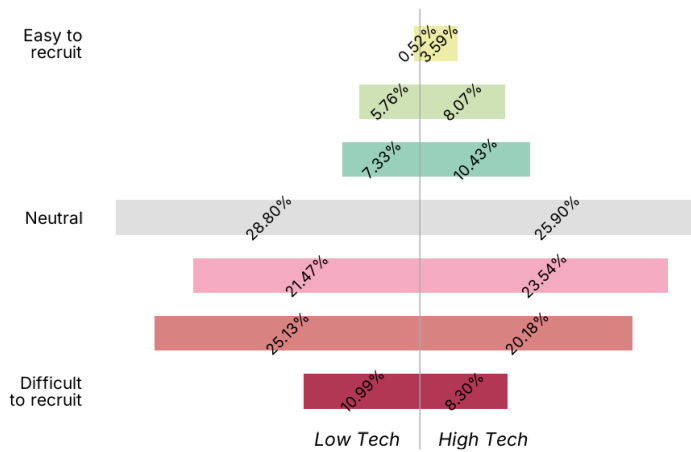

Figure 21 • Recruiting Cybersecurity talent: Low Tech vs. High Tech Enterprises

## 3.2. EMERGING TECH AND CYBERSECURITY EFFECTIVENESS

### 3.2.1. EMERGING TECHNOLOGIES: THREAT OR OPPORTUNITY?

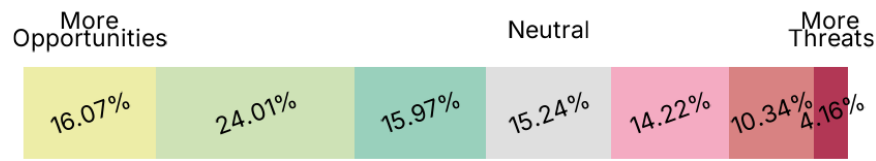

Figure 22 • Do You See Emerging Technologies as offering more opportunities or more threats to your Cybersecurity posture? - 7 points Likert Scale

With rapid technological advancements, companies are divided on whether emerging tech strengthens or complicates cybersecurity efforts. While some see AI, automation, and cloud solutions as powerful enablers, the distribution of responses is nuanced.

The majority of respondents indicate that emerging technologies provide more opportunities in the realm of cybersecurity, although they do not strongly align with the extreme ends of the Likert scale. Most responses fall within the intermediate values, reflecting a cautious optimism.

*This is reflected in the feedback we received from managers:*

*"There are many challenges such as complexity and cost, balancing security and usability, and rapid technological changes while integrating security measures into new technological developments. In spite of challenges regarding new technological developments, we gain opportunities such as enhanced protection, innovative and competitive advantages, and improved customer confidence"*

*"Secure development and deployment of new technology and initiatives through this method protects the organization from threats and vulnerabilities"*

*"Cybersecurity is a crucial pillar of any innovation strategy, ensuring that new technologies remain resilient against evolving threats. It plays a dual role: enabling trust and compliance while fostering a secure environment for innovation to thrive"*

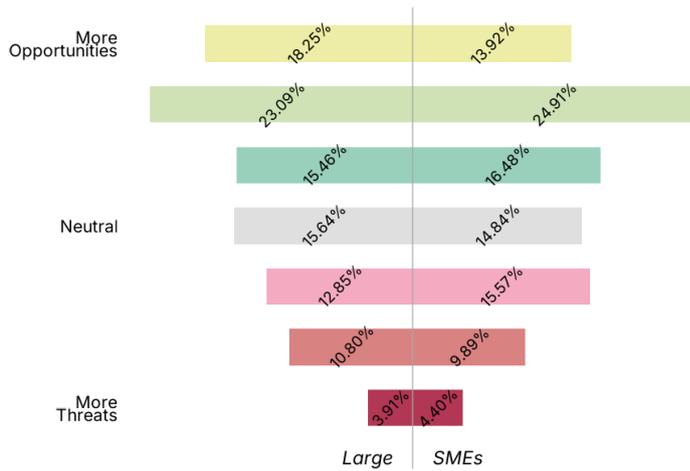

Figure 23 • Emerging Technologies: Large Enterprises vs. SMEs

*How does company size influence the perception of Cybersecurity?*

Asking managers from SMEs and large enterprises, the responses show no significant differences, aside from a slight variation of a few percentage points at the extremes of the scale.

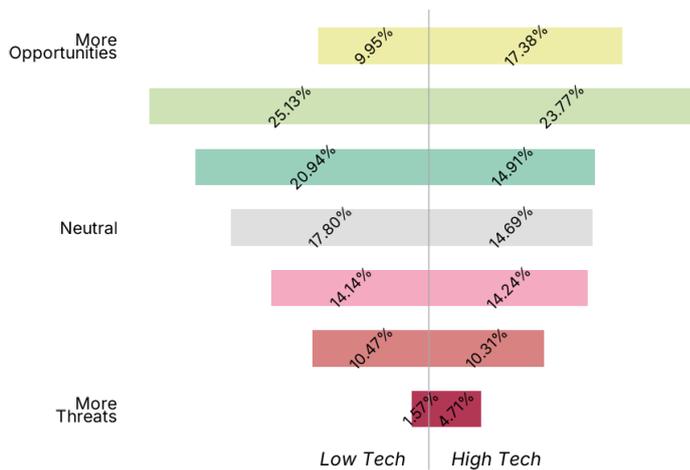

Figure 24 • Emerging Technologies: Low Tech vs. High Tech Enterprises

*How does technological intensity shape the perception of cybersecurity?*

Here the contrast becomes more pronounced. Low-tech companies are noticeably less inclined to perceive emerging technologies as a source of cybersecurity opportunities compared to their high-tech counterparts.

This suggests that while advanced industries are more confident in leveraging new technologies for security enhancements, traditional sectors may still struggle with adoption or perceive higher risks in implementation.

### 3.2.2. INCIDENT READINESS: PREPARED OR EXPOSED?

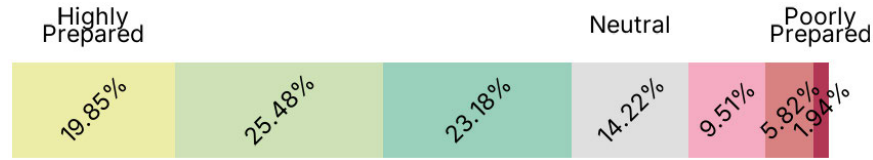

Figure 25 • How prepared is your organization to handle a Cybersecurity incident? - 7 points Likert Scale

The reality of cyber threats is no longer a question of "if" but "when". Inviting respondents to share their opinion, the majority of them feel either highly prepared or prepared to face these threats, while only a small percentage recognize significant gaps in their readiness.

*This perspective is reflected in what managers express about the measures they take to prepare for incidents.*

*"It's necessary to do and a big impact factor is the company is struck by an incident. Basically, we're doing risk management and trying to avoid the worst-case scenarios"*

*"Incident response planning is crucial, ensuring rapid mitigation strategies"*

*"Planning for emergency scenarios is critically important. We rate it as one of our highest priorities, as being prepared for potential cybersecurity incidents is essential to mitigate risks and ensure swift recovery in the event of a breach"*

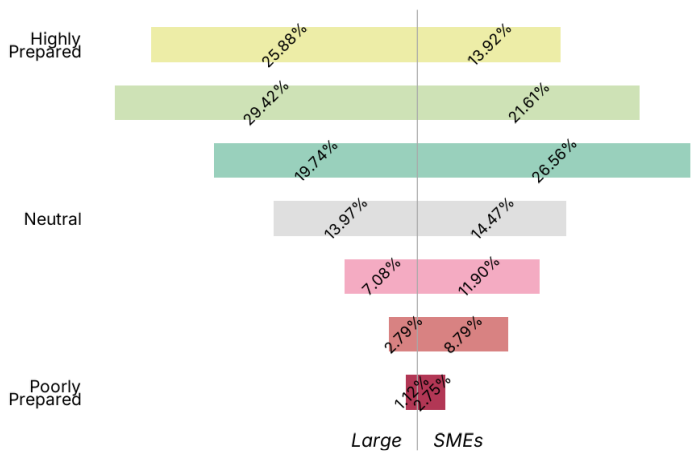

Figure 26 • Incident Readiness: Large Enterprises vs. SMEs

*How does company size influence the perception of Cybersecurity?*

Analyzing the responses from large enterprises and SMEs, we observe that large enterprises are more inclined toward "Highly Prepared" compared to SMEs.

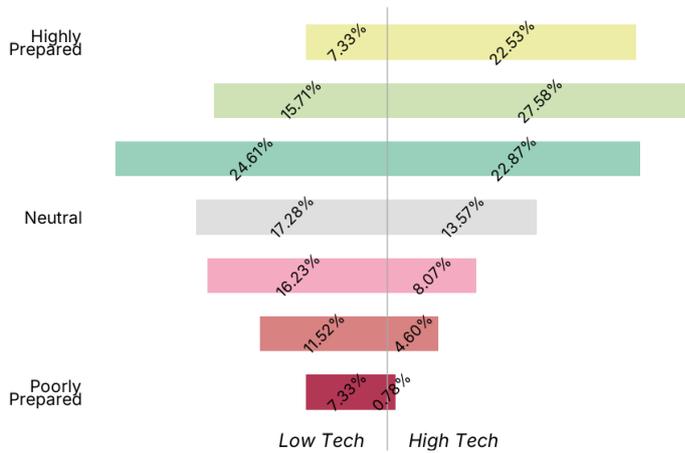

Figure 27 • Incident Readiness: Low Tech vs. High Tech Enterprises

*How does technological intensity shape the perception of cybersecurity?*

According to our sample, high-tech companies exhibit a stronger confidence in their cybersecurity preparedness differently to Low-tech companies. This suggests that larger and more technologically advanced organizations may have greater resources, expertise, or strategic focus on cybersecurity readiness compared to their smaller or less tech-driven counterparts.

### 3.2.3. CYBER INSURANCE: ESSENTIAL OR OPTIONAL?

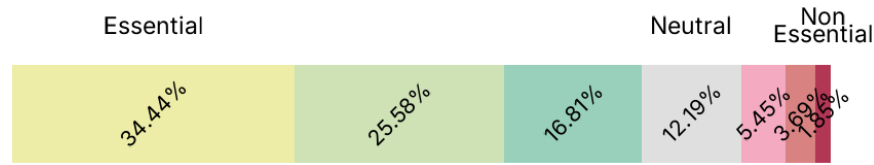

Figure 28 • Is Cyber Insurance essential or Non-Essential for your organizations? - 7 Points Likert scale

Cyber insurance is increasingly being recognized as a necessary safety net. When engaging respondents in a discussion on the topic, a significant portion considered it essential for risk mitigation, while only a very small percentage remained neutral or skeptical about its value.

*The insights shared by managers suggest that cyber insurance is viewed as a means to ensure operational continuity, while also being seen either as a compliance obligation or as an unavoidable expense to mitigate major incidents:*

*"Cyber Insurance is a necessary evil that we must plan for and allocate funds for in order to maintain a smooth operating business"*

*"As a PE backed firm, we are required to have certain levels of security and insurance"*

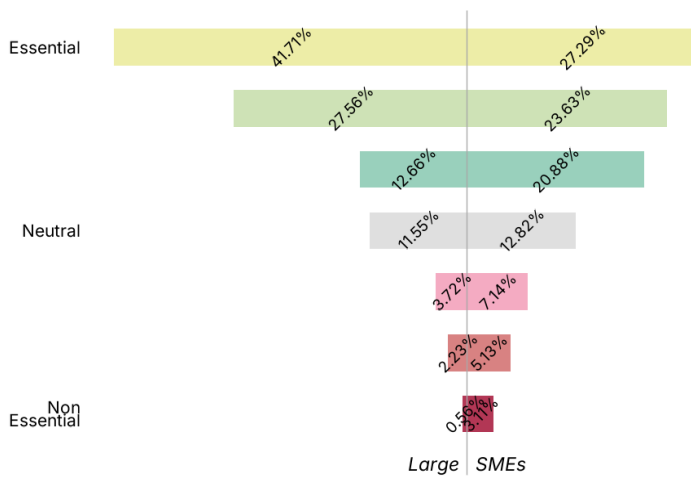

Figure 29 • Cyber Insurance: Large Enterprises vs. SMEs

*How does company size influence the perception of Cybersecurity?*

We can observe a clear distinction: at the extreme end of the scale, large enterprises exceed SMEs by more than 10 percentage points in defining cyber insurance as essential.

Meanwhile, the responses from SMEs are more evenly distributed across the upper end of the scale, spanning from the neutral value to the highest levels of perceived necessity.

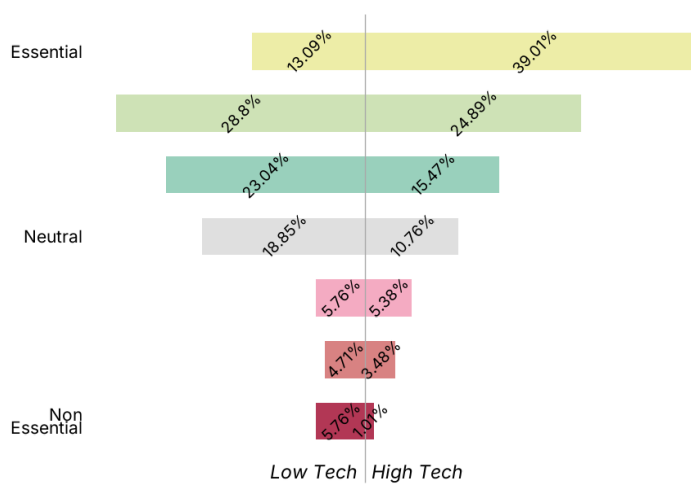

Figure 30 • Cyber Insurance: Low Tech vs. High Tech Enterprises

*How does technological intensity shape the perception of cybersecurity?*

High-tech companies show a remarkable 20 percentage point increase at the "Essential" extreme compared to low-tech enterprises.

This suggests that high-tech organizations, likely more aware of evolving cyber risks, are significantly more inclined to view cyber insurance as a crucial component of their risk management strategy. In contrast, low-tech companies may still be weighing its necessity, potentially due to differing threat exposure, financial constraints, or lower digitalization levels.

### 3.2.4. ZERO TRUST ARCHITECTURE: EFFECTIVE OR OVERHYPED?

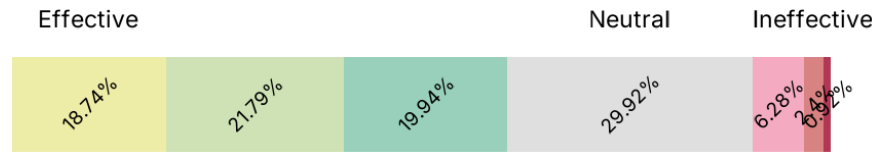

Figure 31 • How effective do you find Zero Trust Architecture for your Cybersecurity needs? - 7 Points Likert Scale

To better understand attitudes towards Zero Trust architecture we posed this key question to our respondents. A significant portion of managers hesitate to take a definitive stance on its effectiveness, suggesting that many may still be unfamiliar with its principles or unsure of its practical impact. Despite this, the overall trend leans towards recognizing Zero Trust as an effective approach, reflecting a growing interest in its potential to enhance cybersecurity.

*This perspective is reflected in the candid feedback we received from managers:*

*"Zero-Trust Architecture – Implementing zero-trust principles (never assume trust, always verify) can improve overall security in digital transformation efforts"*

*"The key is to integrate security seamlessly, leveraging advanced threat intelligence, zero-trust architecture, and automation to ensure protection without compromising efficiency"*

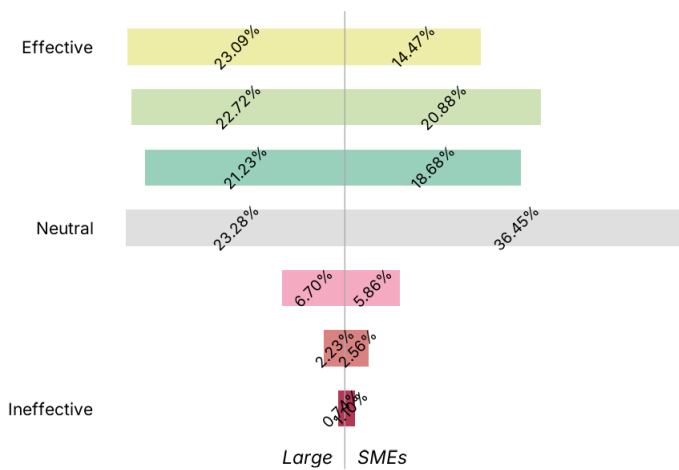

Figure 32 • Zero Trust Architecture: Large Enterprises vs. SMEs

*How does company size influence the perception of Cybersecurity?*

The neutral response emerges as the most selected option, particularly among SMEs. This indicates a greater degree of uncertainty or limited exposure to Zero Trust among smaller organizations.

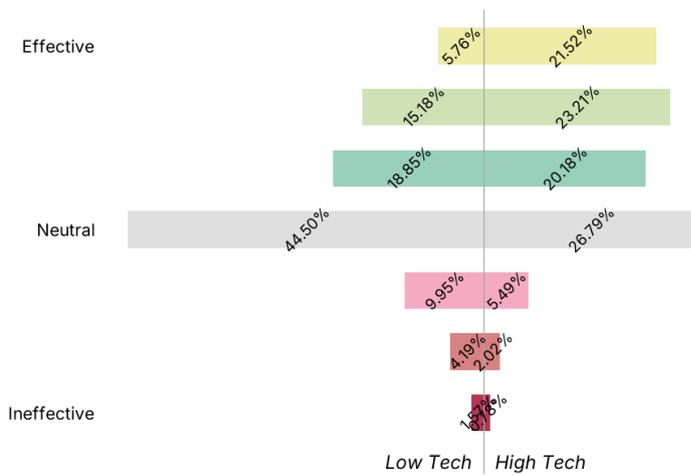

Figure 33 • Zero Trust Architecture: Low Tech vs. High Tech Enterprises

*How does technological intensity shape the perception of cybersecurity?*

With nearly half of low-tech respondents choosing the neutral position. Additionally, only a small fraction of low-tech companies considers Zero Trust to be highly effective, suggesting that adoption and confidence in this framework remain much stronger within high-tech industries.

This highlights the need for increased awareness and education on Zero Trust, especially in sectors that may not yet fully grasp its strategic advantages.

### 3.2.5. BUDGET FOCUS: INNOVATION OR SECURITY?

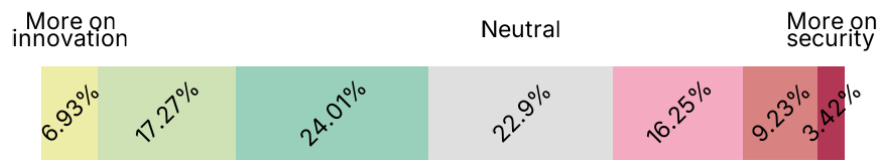

Figure 34 • In budget allocation, do you prioritize more on innovation or more on security? - 7 Points Likert Scale

Resource allocation is always a strategic decision. Considering that innovation drives growth and is an important key factor, the survey shows that many organizations are not prioritizing cybersecurity investments alongside digital transformation, but they remain in a neutral position, indicating a possible balance between the two variables in budget allocation.

*An exploration of managers' opinions shows a clear alignment with these findings:*

*"It is a constant battle between growth and trying to squeeze in more time for cybersecurity. Cybersecurity can often be seen as unnecessary, and a burden which delays development and growth, but we are slowly changing this view"*

*"In a mature organization, cybersecurity should strictly react to investments and implementation of innovation strategy. However, it sometimes can slow down development of new ideas so approach to innovation and cybersecurity should be balanced"*

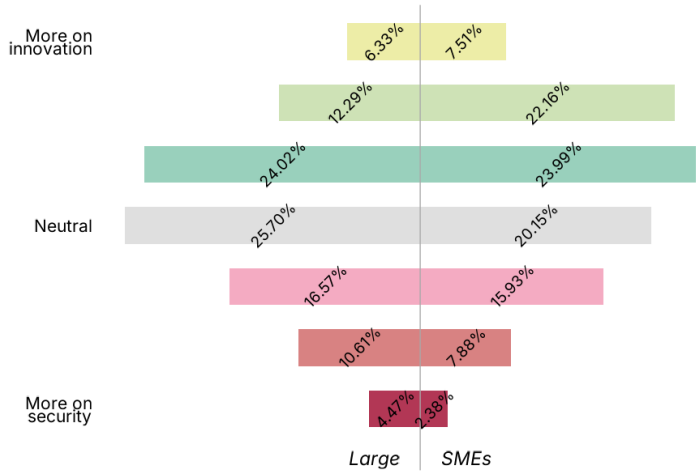

Figure 35 • Budget allocation: Large Enterprises vs. SMEs

*How does company size influence the perception of Cybersecurity?*

Both large enterprises and SMEs have a similar distribution across the scale. However, SMEs show a slightly stronger tendency to prioritize innovation, with a few percentage points more at the end of the spectrum.

This slight tilt may stem from SMEs' strategic focus on growth and scaling, where innovation is often perceived as the primary lever for expanding market reach and achieving competitiveness.

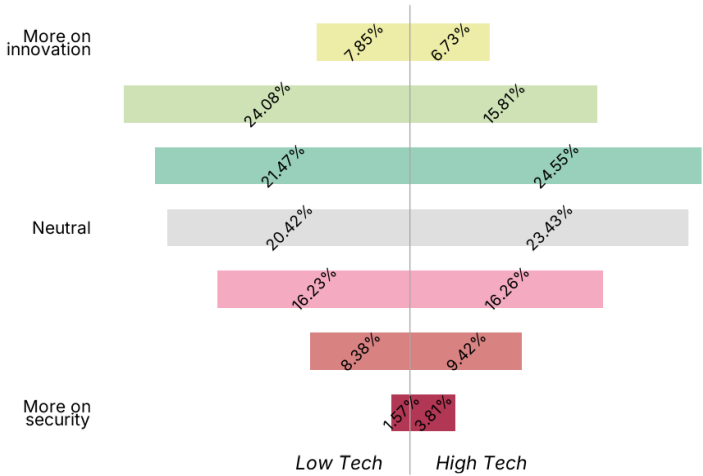

Figure 36 • Budget allocation: Low Tech vs. High Tech Enterprises

*How does technological intensity shape the perception of cybersecurity?*

Low-tech companies are a little more inclined to invest in innovation over security.

This behavior is likely driven by a desire to increase their technological maturity and modernize their operations, especially in response to digital transformation trends.

### 3.2.6. EMPLOYEE TRAINING: CORE OR SECONDARY PRIORITY?

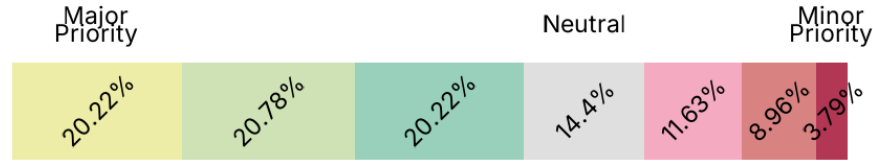

Figure 37 • Is employee training in Cybersecurity a major priority or a minor priority for your organization? - 7 points Likert Scale

Human error remains one of the leading causes of cyber incidents, prompting us to explore how respondents prioritize employee training. Results show that most organizations recognize employee training as a major priority, reinforcing the idea that a well-informed workforce is the first line of defense against cyber threats.

*This perspective is reinforced by what managers shared when we invited them to express their thoughts openly.*

*"Remaining vigilant with cybersecurity is always a challenge, because all security systems rely on the people that use them, so cybersecurity education and training is a must"*

*"I think the costs will be a big challenge and more so training the senior staff on new security measures. There is a clear need for it going forward"*

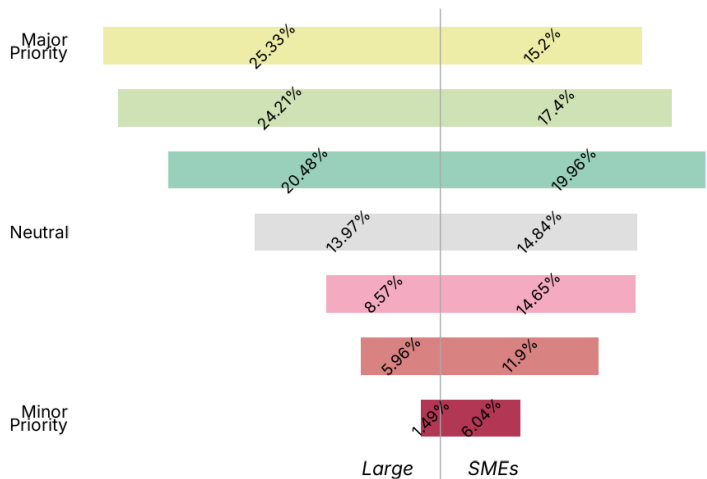

Figure 38 • Employee training in Cybersecurity: Large Enterprises vs. SMEs

*Large vs SMEs: does the perception about cybersecurity change?*

Large enterprises are more concentrated in the upper half of the scale, suggesting they view cybersecurity training as a major priority. SMEs, instead, cluster in the lower range, indicating it's often a less urgent concern. This gap likely reflects differences in resources and compliance obligations, with larger companies better equipped to invest in structured training, while SMEs face tighter budgets and competing priorities.

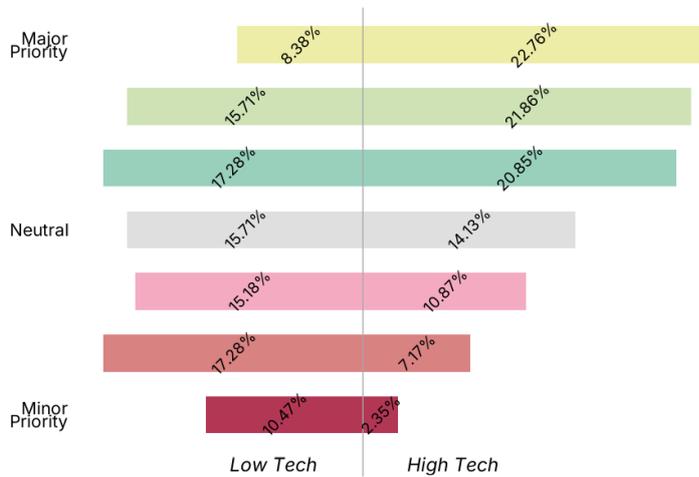

Figure 39 • Employee training in Cybersecurity: Low Tech vs. High Tech Enterprises

*High tech vs Low tech: does the perception about cybersecurity change?*

High-tech companies lean strongly toward the "major priority" end of the scale, reinforcing their more proactive posture toward cybersecurity preparedness. In contrast, low-tech companies are more likely to consider training a minor priority, with their responses skewed toward the lower end of the scale.

This gap may stem from differences in digital maturity and exposure to cyber risk.

### 3.2.7. OUTSOURCING CYBERSECURITY: SAFE OR RISKY?

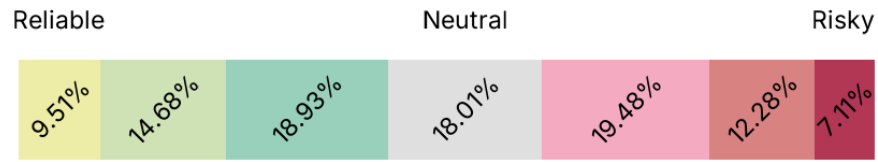

Figure 40 • How do you view outsourcing Cybersecurity? - 7 points Likert Scale

Delegating cybersecurity to external partners is an increasingly common strategy, especially for organizations lacking in-house expertise. However, trust and control remain central concerns. Our inquiry focused on this topic reflects this tension: the majority of respondents do not see outsourcing as either entirely reliable or entirely risky. Most answers are concentrated around the neutral midpoints of the scale, suggesting that while outsourcing is being considered or adopted, many companies are still hesitant to fully endorse it due to perceived risks, particularly the loss of control over sensitive security operations.

*The views of managers provide further evidence of their inclination to outsource for various reasons:*

*"Our main challenge is that our primary personnel are not trained to identify attacks. Therefore, we outsource"*

*"I think the company's we outsource our software to take care of it. It's important but we don't do it on our end. Everything is done in the cloud these days"*

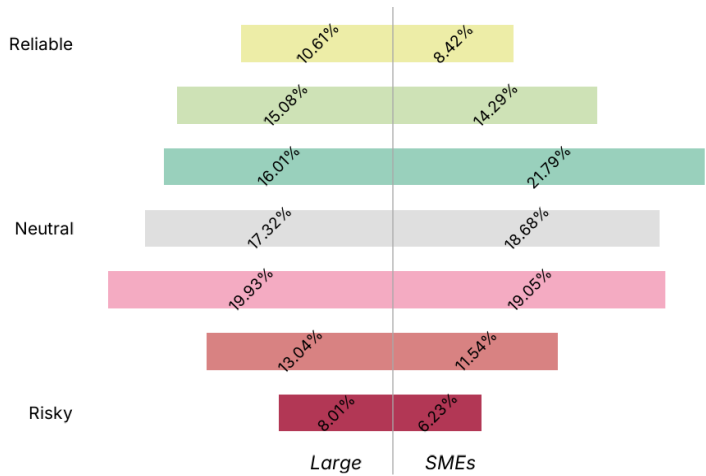

Figure 41 • Outsourcing Cybersecurity: Large Enterprises vs. SMEs

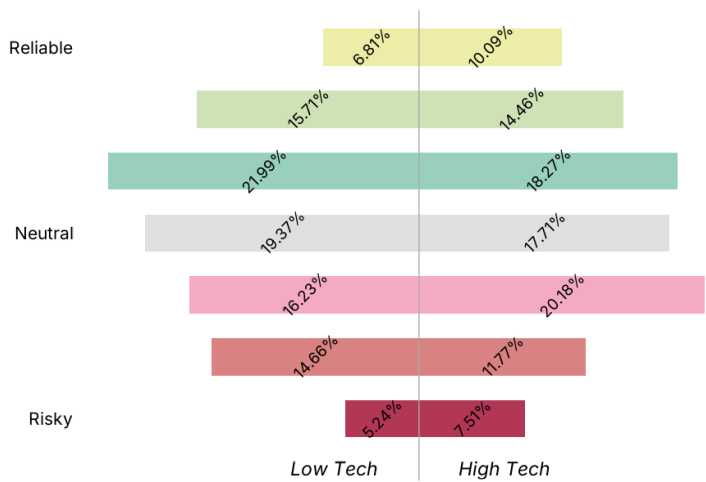

Figure 42 • Outsourcing Cybersecurity: Low Tech vs. High Tech Enterprises

*Large vs SMEs: does the perception about cybersecurity change?*

Both SMEs and large enterprises show a similar ambivalence toward outsourcing, likely due to concerns about effectiveness and accountability. SMEs may rely on it out of necessity but fear losing control, while larger companies view it as a complement to internal teams, yet share similar doubts.

*High tech vs Low tech: does the perception about cybersecurity change?*

Once again, the majority of answers cluster around the neutral middle of the scale, indicating a widespread cautious approach toward outsourcing cybersecurity.

### 3.2.8. CYBERSECURITY AUTOMATION: BENEFIT OR RISK?

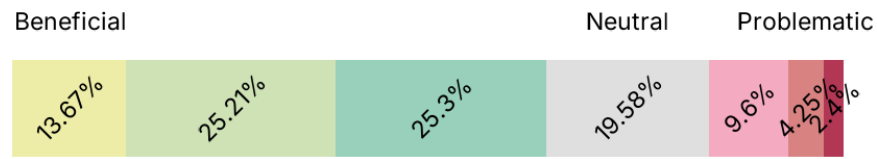

Figure 43 • Do you find automation in Cybersecurity to be beneficial or problematic? - 7 points Likert Scale

Automation is revolutionizing cybersecurity, but not everyone sees it as a universal solution. While many respondents highlight its benefits in threat detection and response, others remain cautious about over-reliance on automated systems.

*The opinions of managers illustrate this point:*

*"By integrating robust security measures early in the development process, we can build trustworthy and scalable solutions. The main opportunity lies in leveraging AI and automation to enhance security while reducing manual oversight"*

*"The challenge lies in balancing security with speed and continuous integration and delivery of products, but automation and proactive risk management help maintain protection without slowing progress"*

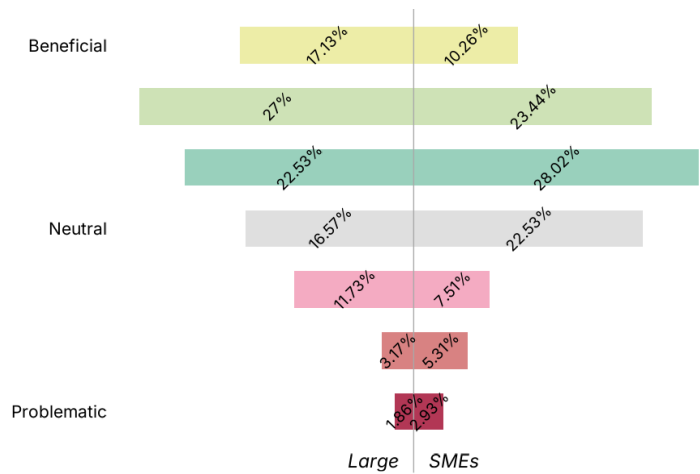

Figure 44 • Automation in Cybersecurity: Large Enterprises vs. SMEs

*Large vs SMEs: does the perception about cybersecurity change?*

Large enterprises tend to lean more towards the extreme end of the "Beneficial" scale, while SMEs are more neutral, leaning slightly towards the "Problematic" side.

This divergence may arise from the resources and infrastructure of large organizations, enabling them to leverage automated solutions, while SMEs face budget and expertise limitations, causing hesitancy in embracing automation.

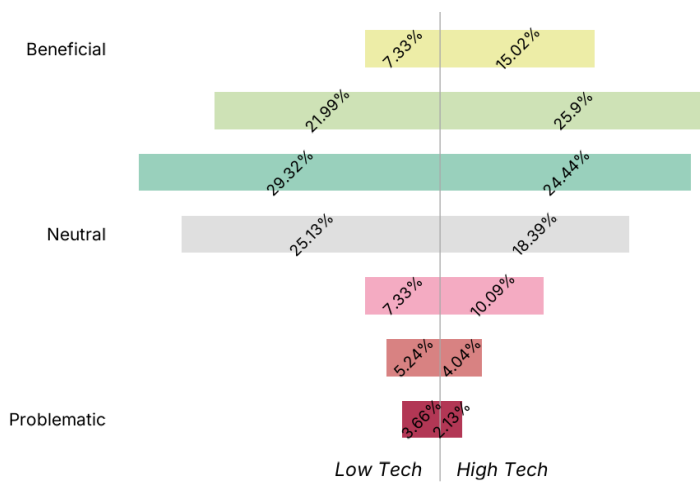

Figure 45 • Automation in Cybersecurity: Low Tech vs. High Tech Enterprises

*High tech vs Low tech: does the perception about cybersecurity change?*

Low-tech enterprises are positioned in the mid-neutral range, slightly inclined towards the "Beneficial" side, whereas high-tech enterprises are more distinctly aligned with the extreme "Beneficial" end of the scale. This pattern underscores the potential advantages that high-tech companies experience from automation, likely due to their inherent adaptability and access to advanced technologies.

## 3.3. FROM INVESTMENT TO IMPACT: CYBERSECURITY BUDGETS AND INCIDENT HISTORY

### 3.3.1. CYBERSECURITY BUDGETS

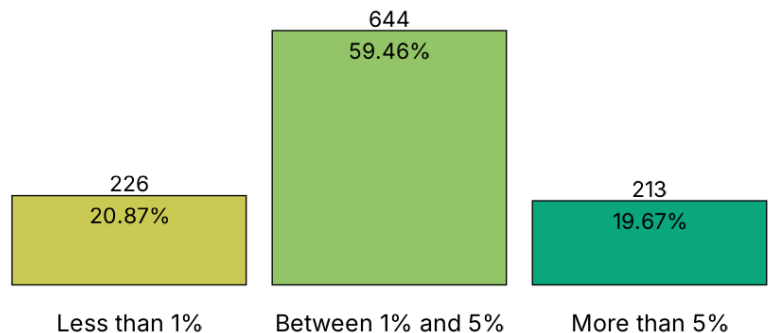

Figure 46 • Percentage of Revenue invested in Cybersecurity

Survey results show varying levels of revenue invested in cybersecurity. While there is a balance between the number of companies investing less than 1% and those investing more than 5%, the majority allocate between 1% and 5% of their revenue to cybersecurity measures. This trend underscores the critical need for organizations to evaluate their investment strategies, ensuring they allocate sufficient resources to protect against evolving threats and secure their digital environments.

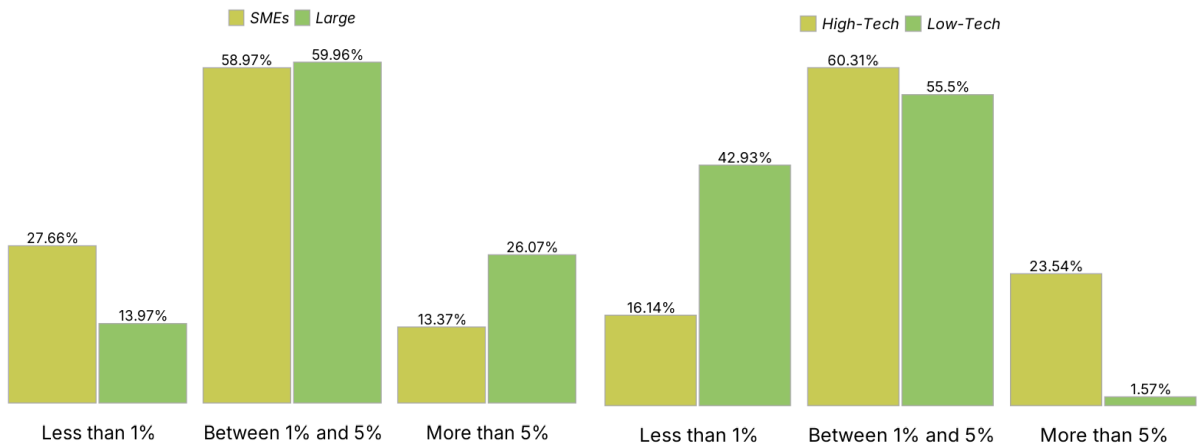

Figure 47 • Revenue invested in Cybersecurity: Large Enterprises vs. SMEs & Low Tech vs. High Tech Enterprises

Comparing the results between SMEs and large enterprises, we observe that while their median investment levels in cybersecurity are similar, differences emerge at the extremes.

Large enterprises are more inclined to allocate a higher percentage of their revenue to cybersecurity.

This contrast becomes even more pronounced when shifting the focus to high-tech versus low-tech enterprises: while larger and more technologically advanced companies recognize cybersecurity as a strategic priority, many low-tech enterprises remain hesitant or constrained in their spending. This reluctance could expose them to heightened vulnerabilities, highlighting the need for greater awareness and industry-wide efforts to bridge this investment gap.

### 3.3.2. CYBERSECURITY INCIDENT HISTORY

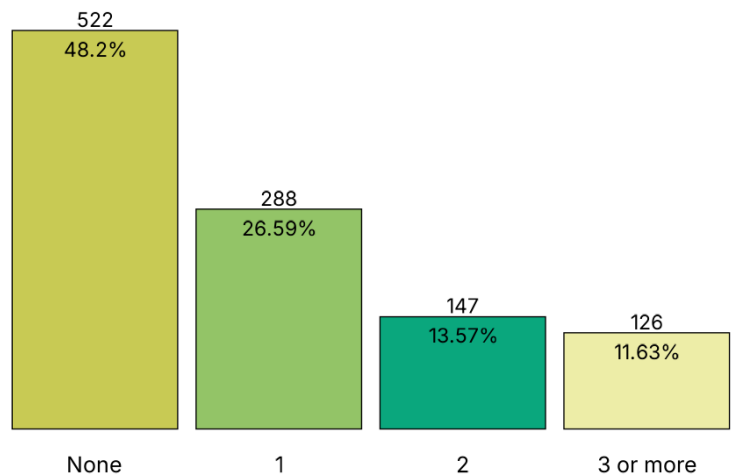

Figure 48 • Cybersecurity Incidents in the last 5 years

Regarding cybersecurity incidents, a notable percentage of respondents have not encountered any cyber incidents in the past five years, reinforcing the importance of proactive investments in prevention and mitigation.

The results show a decreasing trend in the number of incidents reported, which is a positive sign indicating that these preventive measures may be effective.

This highlights the value of a proactive approach in protecting assets and promoting a culture of security awareness within organizations.

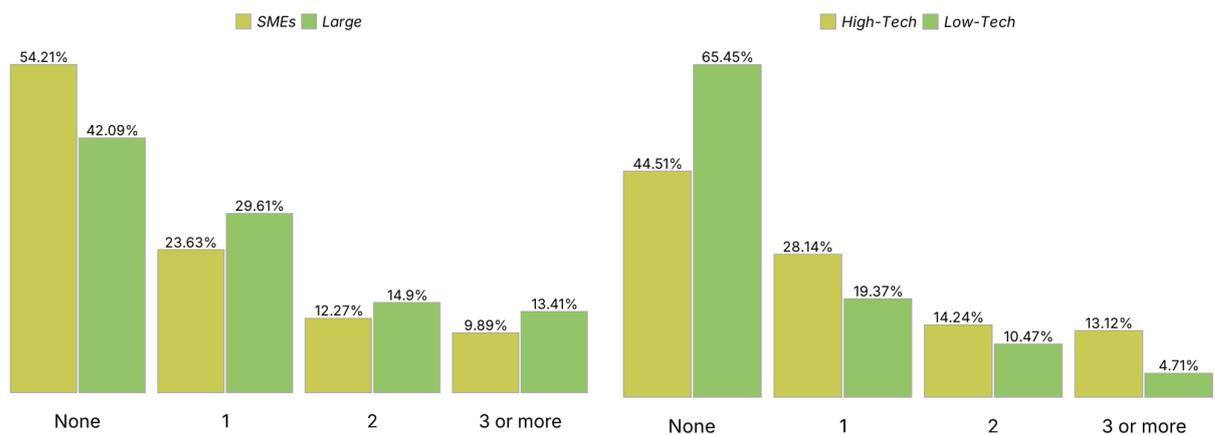

Figure 49 • Cybersecurity Incidents in the last 5 years: Large Enterprises vs. SMEs and Low Tech vs. High Tech Enterprises

Comparing the results between SMEs and large enterprises, we see that SMEs report fewer cyber incidents in the last five years than large organizations. This may be due to SMEs being less exposed to high-profile attacks and having fewer resources that attract cybercriminals.

Exploring the differences by sector, we report that high-tech enterprises report more incidents than low-tech ones. This could be attributed to the fact that high-tech companies often handle more sensitive data and are at the forefront of digital innovation, making them prime targets for cyberattacks. Additionally, the rapid pace of technological change in the high-tech sector can lead to vulnerabilities that are exploited by cybercriminals.

## 3.4. ACTIONS AGAINST CYBERATTACKS

### 3.4.1. CURRENT ACTIONS TO MITIGATE CYBERATTACK RISKS

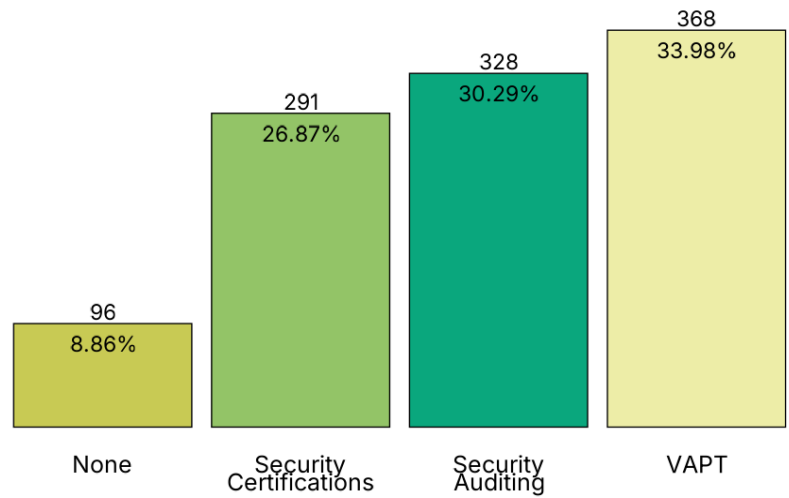

Figure 50 • Current Actions to Mitigate Cyberattack Risks

When asked about current actions to mitigate cyberattack risks, a diversified approach emerges. While a small portion of respondents report taking no specific action, the majority indicate concrete measures are in place. Security certifications are adopted by 27%, followed closely by security auditing and Vulnerability Assessment and Penetration Testing (VAPT) activities, suggesting a growing emphasis on proactive and technical assessments of vulnerability.

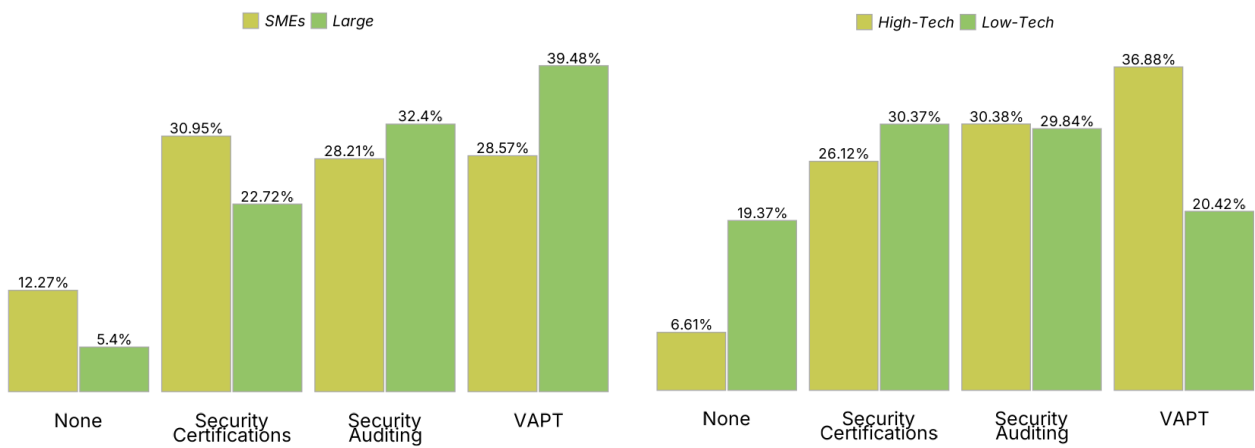

Figure 51 • Comparative results about Current Actions to Mitigate Cyberattack Risks- Large Enterprises vs. SMEs and Low Tech vs. High Tech Enterprises

When comparing SMEs and large enterprises, a clear pattern emerges: larger organizations are more likely to adopt advanced practices such as security auditing and VAPT, while SMEs tend to report either no action or rely primarily on security certifications. This may reflect differences in available resources, internal expertise, and regulatory pressure, larger companies often have the capacity to invest in in-depth assessments, whereas SMEs may opt for less resource-intensive or more externally driven measures.

A similar distribution appears when comparing high-tech and low-tech companies. However, these segments show greater divergence at the extremes: low-tech companies are more likely to report taking no action, while high-tech companies lead in the adoption of VAPT. In contrast, both groups remain closely aligned when it comes to the adoption of security certifications and security auditing.

### 3.4.2. PLANNED ACTIONS TO MITIGATE CYBER ATTACKS IN THE NEXT 3 YEARS

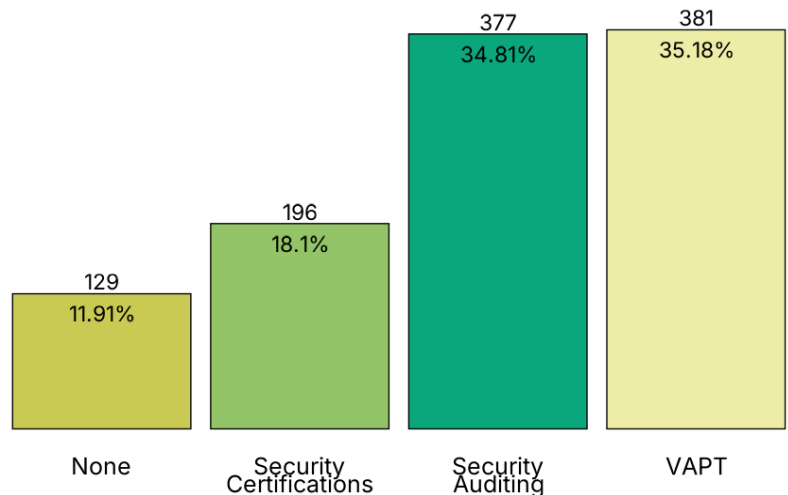

Figure 52 • Planned Actions to Mitigate Cyber Attacks in the Next 3 Years

The distribution of actions planned for the next three years closely mirrors the actions already taken, indicating that organizations are likely to sustain their current approach to cybersecurity measures. This suggests a continuity in strategy, where companies plan to build on existing initiatives and maintain a similar focus on mitigating cyber threats moving forward, rather than introducing major shifts in their cybersecurity priorities.

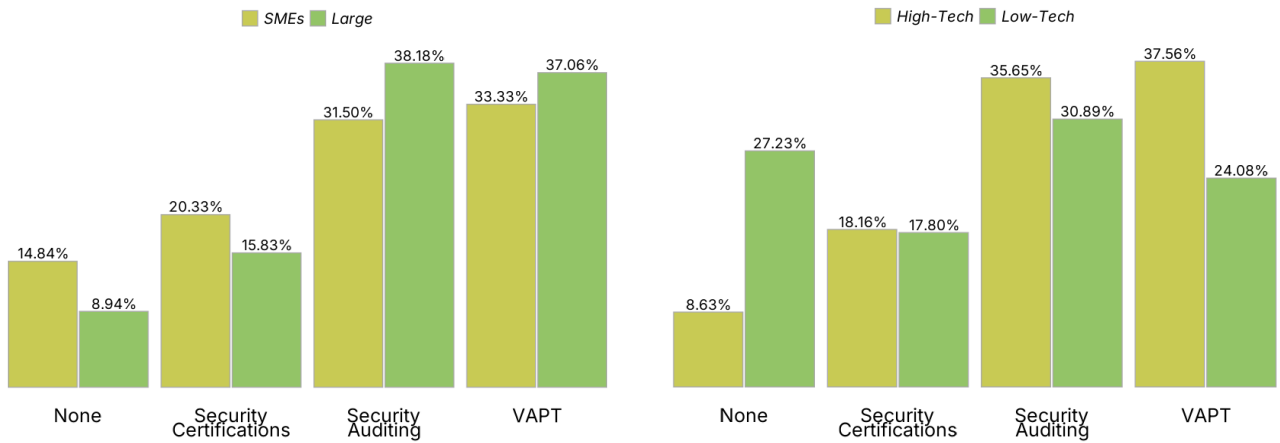

Figure 53 • Comparative results about Planned Actions to Mitigate Cyber Attacks in the Next 3 Years: Large Enterprises vs. SMEs and Low Tech vs. High Tech Enterprises

When comparing organizations by size, whether large or SMEs, the trend from no action to VAPT remains consistent, though the values differ, likely reflecting their varying investment capabilities. However, what is more concerning is the comparison between high-tech and low-tech sectors: low-tech organizations stand out with a significantly higher percentage of no action responses, indicating a lower level of sensitivity and a different approach to cybersecurity.

## 3.5. CYBERSECURITY IN PROJECT MANAGEMENT: INTEGRATED OR OVERLOOKED?

### 3.5.1. CYBERSECURITY BUDGETS ALLOCATED IN PROJECT PLANNING

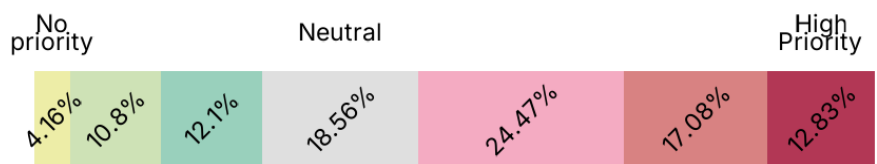

Figure 54 • How much priority do you place on including Cybersecurity budgets during project planning? - 7 points Likert Scale

We engaged respondents in a discussion about recognizing cybersecurity as a crucial component of project planning and allocating dedicated budgets for its implementation.

The responses show a notable skew towards the "High Priority" direction. However, most responses fall within the positive mid-range of the scale rather than firmly declaring it as a "Pure High Priority". The trend shows rising awareness, but also a need for stronger integration of cybersecurity into project planning.

*The opinions of managers illustrate this point:*

*"We place a high priority on including cybersecurity budgets during project planning"*

*"Cybersecurity is expensive but having a budget allocated for cybersecurity is crucial in any organization"*

*"Cybersecurity is a critical part of project planning, and it needs dedicated budgets, clear tasks, and timelines to make sure it's done right from the start"*

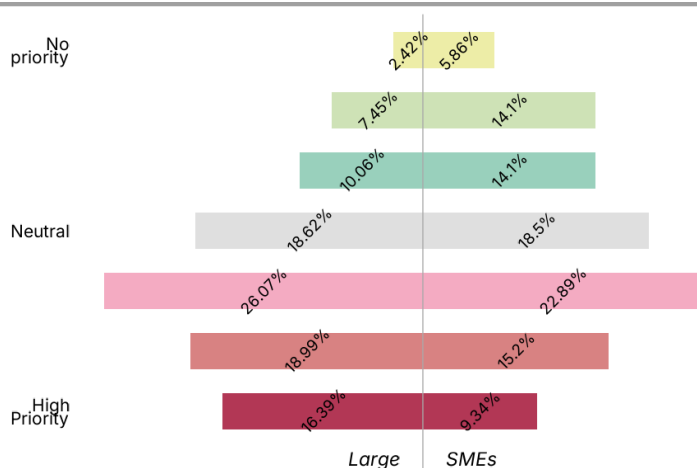

Figure 55 • Cybersecurity budgets allocated in project planning: Large Enterprises vs. SMEs

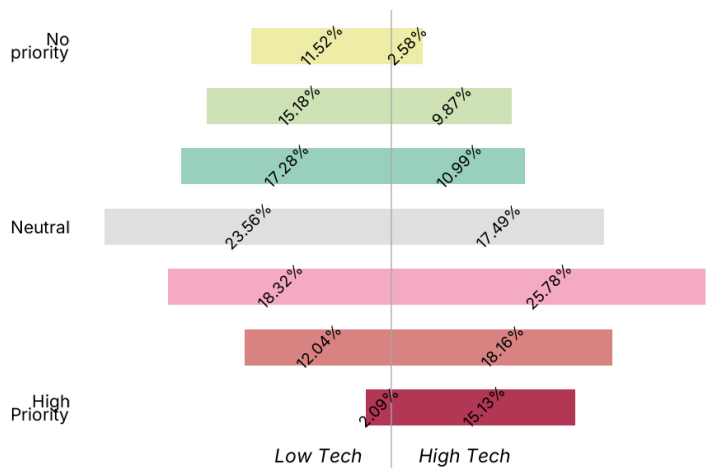

Figure 56 • Cybersecurity budgets allocated in project planning: Low Tech vs. High Tech Enterprises

*Large vs SMEs: does the perception about cybersecurity change?*

Large enterprises tend to lean toward the "High Priority" end of the scale, while SMEs remain closer to the neutral range. This suggests that larger organizations are more likely to integrate cybersecurity into their planning, whereas SMEs may still struggle to position it within their budget strategies.

*High tech vs Low tech: does the perception about cybersecurity change?*

Low-tech enterprises are predominantly positioned in the neutral range or even indicating "No Priority". In fact, they report declaring "No Priority" approximately five times more frequently than their high-tech counterparts. Conversely, high-tech enterprises exhibit a strong inclination towards the "High Priority" end of the scale, with responses significantly moving away from neutral values.

### 3.5.2. TASKS AND DEFINED TIMELINES SPECIFICALLY DEDICATED FOR IMPLEMENTING CYBERSECURITY MEASURES

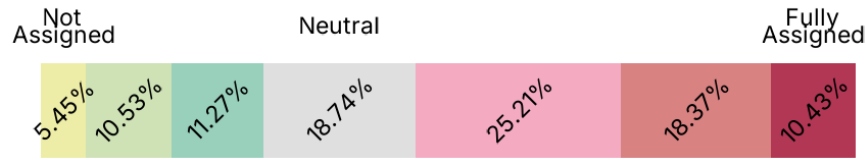

Figure 57 • Are clear tasks and defined timelines specifically allocated for implementing Cybersecurity measures within your projects? - 7 points Likert Scale

In addition to budget allocation, we invited respondents to share their opinions on whether clear tasks and defined timelines are specifically allocated for implementing cybersecurity measures within projects shows a similar trend in responses.

While many managers recognize the importance of establishing these elements, the majority of responses tend to cluster in the positive mid-range of the scale rather than fully committing to the practice.

*They also emphasize this perspective in their open comments:*

*"[...] clear definition of tasks and timelines in well-planned projects, cybersecurity measures are implemented in stages - from the early design phase, through testing, to operational monitoring and incident response. Standards are key [...]"*

*"We try to incorporate clear tasks and timelines for cybersecurity within projects, but resource allocation is often a challenge"*

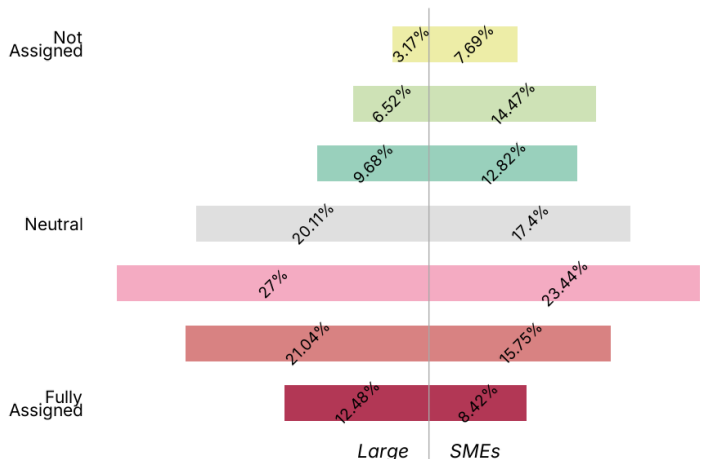

Figure 58 • Tasks and defined timelines specifically allocated for implementing Cybersecurity: Large Enterprises vs. SMEs

*Large vs SMEs: does the perception about cybersecurity change?*

It becomes evident that SMEs are more inclined to lean towards the "Not Assigned" end of the scale.

In contrast, large enterprises tend to remain closer to the neutral-positive values, indicating a greater likelihood of having tasks and timelines that are "Fully Assigned". However, overall, the percentage of responses indicating "Fully Assigned" is low for both groups.

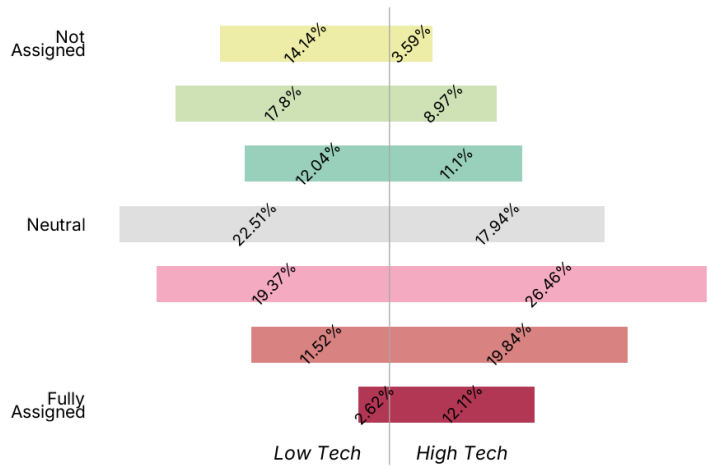

Figure 59 • Tasks and defined timelines specifically allocated for implementing Cybersecurity: Low Tech vs. High Tech Enterprises

*High tech vs Low tech: does the perception about cybersecurity change?*

Low-tech enterprises show notably higher concentrations in the lower end of the scale, particularly in the "Not Assigned" categories, and report almost no instances of "Fully Assigned" tasks. This contrast highlights the difficulties these companies face in prioritizing and structuring cybersecurity within their projects. In comparison, high-tech enterprises show a clearer tendency to define tasks and timelines, suggesting that their greater technological maturity may support a more effective integration of cybersecurity into project planning.

### 3.5.3. TEAM'S RESOURCES ALLOCATED TO CYBERSECURITY ACTIVITIES

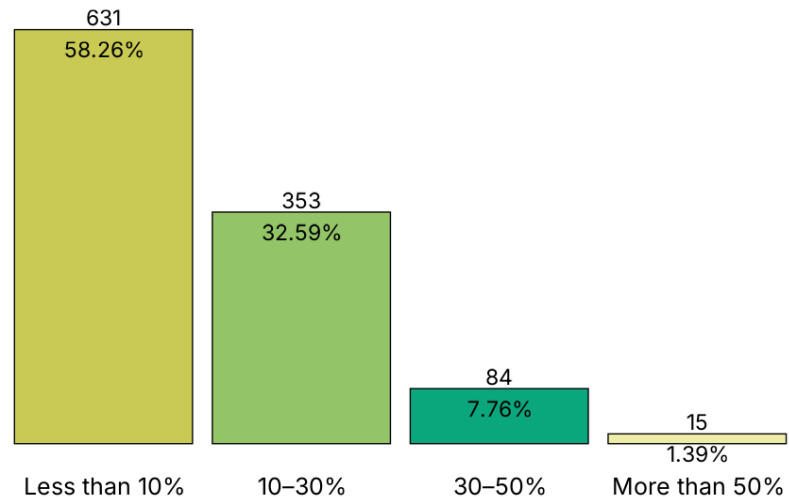

Figure 60 • Percentage of project team's resources allocated to Cybersecurity

Regarding the resources allocated to cybersecurity activities within teams, the responses show a decreasing trend, starting with the majority indicating "Less than 10%" of their resources dedicated to these activities. This trend is concerning, as it suggests that many organizations may not be prioritizing adequate investment in cybersecurity.

This underscores the urgent need for teams to reassess their resource allocation strategies to ensure that cybersecurity receives the attention and funding necessary to effectively mitigate risks and protect critical assets.

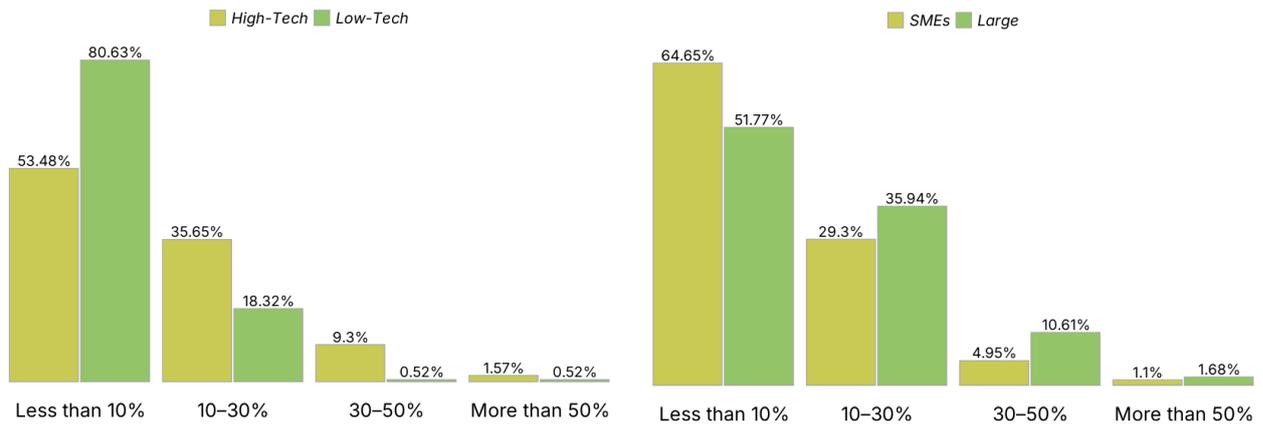

Figure 61 • Comparative results about project team's resources allocated to Cybersecurity: Large Enterprises vs. SMEs and Low Tech vs. High Tech Enterprises

SMEs tend to allocate "Less than 10%" of their resources to cybersecurity activities more frequently than large enterprises. In contrast, large organizations show higher percentages in the "10-30%" and "30-50%" ranges, indicating a greater willingness to invest more substantially in cybersecurity measures.

Focusing on the sector, low-tech enterprises allocate fewer team resources than high-tech enterprises. In fact, the majority of low-tech companies report that they allocate less than 10% of their resources.

This disparity highlights the potential resource limitations that SMEs and low-tech enterprises face in prioritizing cybersecurity.

### 3.5.4. PLANNING FOR EMERGENCY SCENARIOS

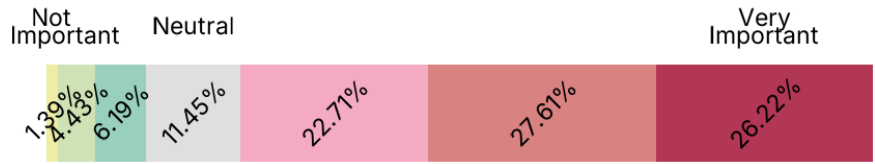

Figure 62 • How would you rate the importance of planning for emergency scenarios such as responses to potential Cybersecurity Incidents? - 7 points Likert Scale

We sought to uncover the thoughts of respondents regarding the importance of planning for emerging scenarios in response to potential cybersecurity incidents. The majority of respondents consider this type of planning to be very important, which is a positive outlook. This recognition reflects an awareness of the evolving nature of cyber threats and the need for organizations to be proactive in their response strategies, ultimately enhancing their resilience and preparedness in the face of potential incidents.

*The opinions of managers illustrate this point:*

*"Planning for emergency scenarios is critically important. We rate it as one of our highest priorities, as being prepared for potential cybersecurity incidents is essential to mitigate risks and ensure swift recovery in the event of a breach"*

*"Cybersecurity is crucial for protecting sensitive data and maintaining system integrity. It involves regular assessments, dedicated budgets and integration into overall strategy, emergency planning and post project reviews are essential for long term security"*

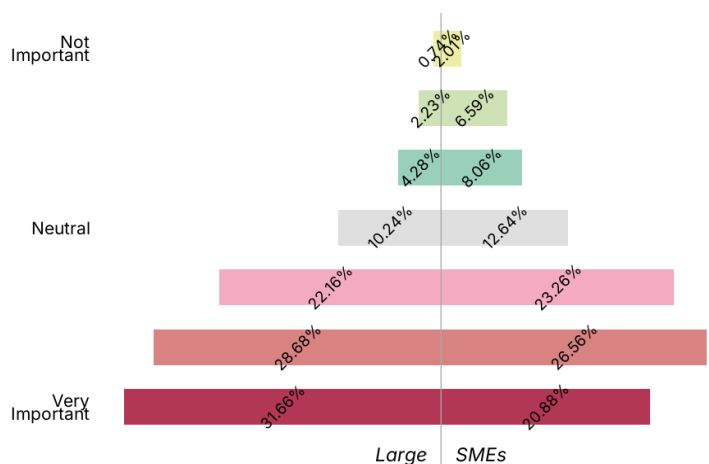

Figure 63 • Importance of planning for emergency scenarios: Large Enterprises vs. SMEs

*Large vs SMEs: does the perception about cybersecurity change?*

Regarding the importance of planning for emergency scenarios, both groups show an inclination towards the "Very Important" end of the scale. Larger organizations are more likely to prioritize comprehensive emergency planning, while SMEs show a more moderate stance.

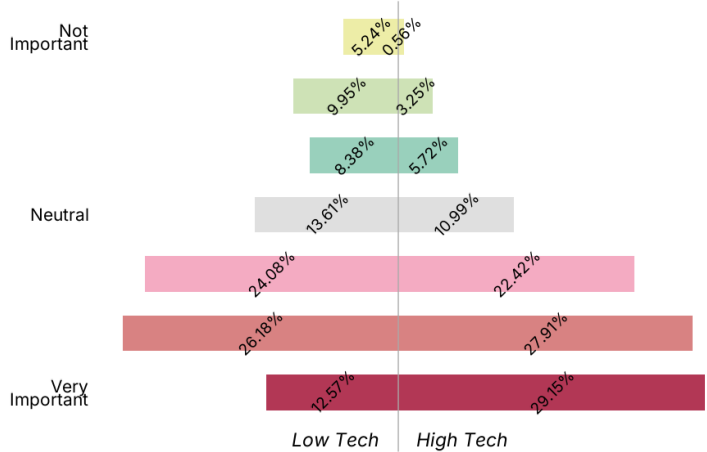

Figure 64 • Importance of planning for emergency scenarios: Low Tech vs. High Tech Enterprises

*High tech vs Low tech: does the perception about cybersecurity change?*

The differences between low-tech and high-tech enterprises are more pronounced. High-tech enterprises exhibit a markedly higher percentage of respondents rating this planning as "Very Important" compared to low-tech enterprises. This finding suggests that high-tech companies are more attuned to the necessity of preparing for potential cybersecurity incidents, likely due to their increased exposure to cyber risks and a deeper understanding of the implications of these threats.

### 3.5.5. POST-PROJECT VULNERABILITY ASSESSMENTS AND LONG-TERM CYBERSECURITY PROTECTIONS

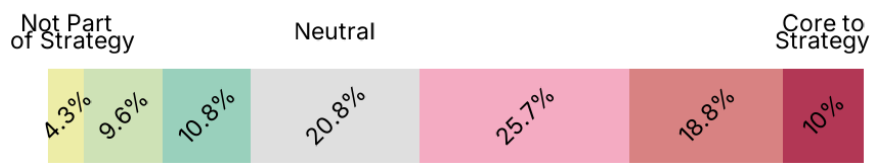

Figure 65 • To what extent do post-project vulnerability assessments and long-term Cybersecurity protection form part of your strategy? - 7 points Likert Scale

We explored how respondents perceive the post-project vulnerability assessments and long-term cybersecurity protection as part of organizational strategy. The answers have been rather neutral, showing no significant bias toward either extreme of the scale: "Not Part Of Strategy" or "Core To Strategy". This neutrality suggests that while organizations recognize the relevance of these practices, there may be uncertainty or inconsistency in their implementation.

*The opinions of managers illustrate this point:*

*"Emergency response planning is essential, as proactive strategies help mitigate potential risks and minimize disruptions. Post-project vulnerability assessments and long-term*

*protections are integral to maintaining security, ensuring continued resilience as new technologies evolve"*

*"Cybersecurity doesn't end with the completion of a project - it requires ongoing monitoring and risk assessment"*

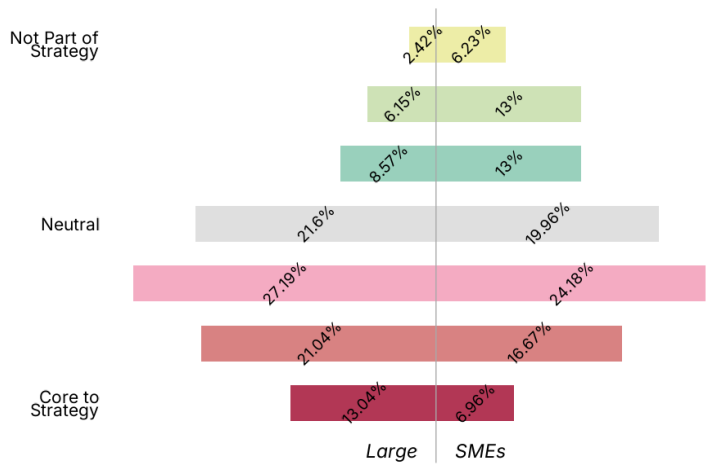

Figure 66 • Long-term Cybersecurity protection: Large Enterprises vs. SMEs

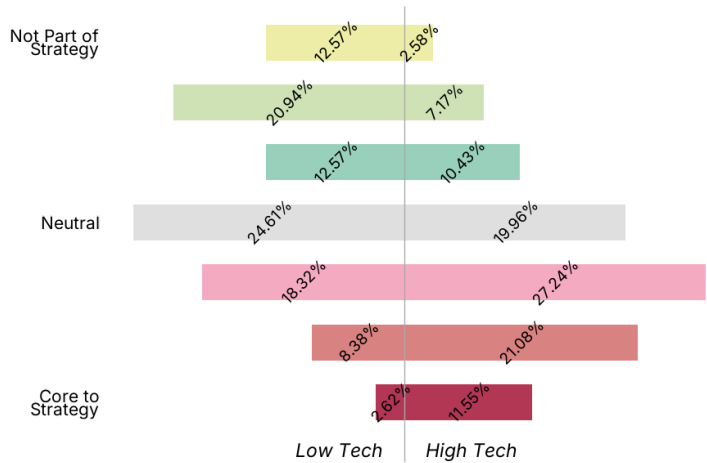

Figure 67 • Long-term Cybersecurity protection: Low Tech vs. High Tech Enterprises

*Large vs SMEs: does the perception about cybersecurity change?*

The distribution of results is quite similar. SMEs show a greater inclination towards the "Not Part of Strategy" end of the scale compared to large enterprises, which tend to lean more towards the "Core to Strategy" side. Despite this trend, neither group moves towards the extreme values of the scale.

*High tech vs Low tech: does the perception about cybersecurity change?*

Low-tech enterprises are more likely to gravitate towards the neutral range of the scale and lean towards the "Not Part of Strategy" end, indicating a lack of emphasis on long-term cybersecurity protection. In contrast, high-tech enterprises rarely indicate that such measures are "Not Part of Strategy", and they are significantly more inclined to categorize these practices as "Core to Strategy".

# 4. MAIN CHALLENGES AND OPPORTUNITIES FOR MANAGERS

*What sentiments and perceptions emerge from the open feedback shared by managers, and how do these insights shed light on the challenges they face?*

Figure 68 • Sentiment Analysis of Open Responses from Managers

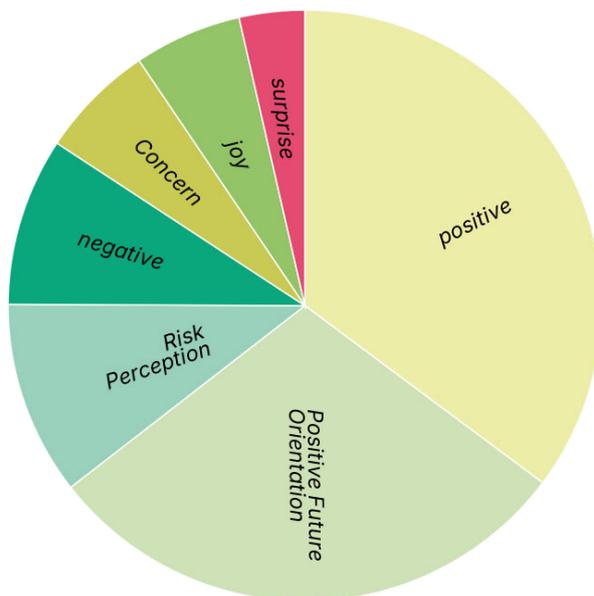

### Cybersecurity is both a necessity and an opportunity

Managers across industries increasingly view cybersecurity as a strategic enabler rather than solely a matter of defense. It supports trust, regulatory compliance, and long-term innovation. As one respondent put it: "*Cybersecurity is about building smarter, safer products that people trust*".

### Budget and internal expertise remain major barriers

Many managers cite tight budgets, competing priorities, and a lack of skilled personnel as key obstacles. This leads to reactive approaches, where cybersecurity investments are made after incidents, rather than proactively.

### Training gaps and cultural resistance are common

A recurring theme is the lack of awareness and buy-in across teams. Several managers note that cybersecurity is still seen as a "cost center" or "IT-only issue", making company-wide training and cultural change a significant challenge.

### Balancing speed and security is a daily trade-off

Especially in innovation-driven environments, security is often at odds with speed and usability. Managers report pressure to deliver fast, sometimes at the cost of robust security. The challenge is to make cybersecurity part of the innovation process: "*secure by design*", rather than a blocker.

*Automation and integration offer promising paths forward*

Tools like AI-based detection, Zero Trust architectures, and automated testing are seen as ways to scale protection without slowing progress. These are especially promising for companies with limited human resources.

*Compliance and regulation are both drivers and constraints*

While some companies are driven to invest in cybersecurity due to industry standards or customer requirements, others see regulation as a burden.

The ability to align innovation with compliance is cited as a key competitive differentiator.

*Cybersecurity still suffers from fragmentation*

In many organizations, cybersecurity is not yet fully embedded in project planning or strategic innovation processes. Few respondents report having fully assigned tasks, budgets, and timelines, especially in low-tech and smaller companies.

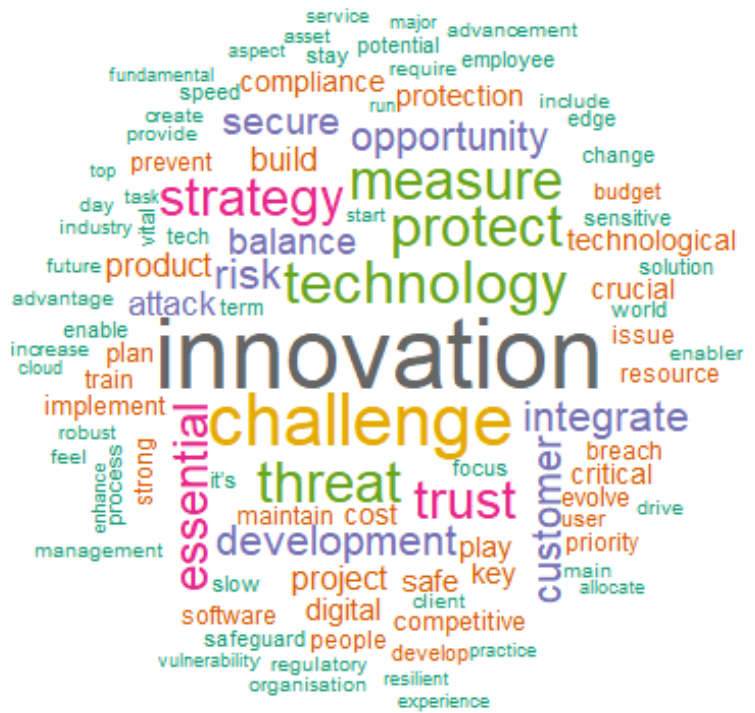

Figure 69 • Most Frequent Terms in Managers' Open Responses

*Country- and sector-level disparities persist*

Data shows that geographic differences impact both action and perception. For instance, the USA tends to invest more proactively, while Europe (including Italy) reports fewer planned actions and lower team resource allocation.

*Company size and tech maturity matter: up to a point*

Larger and high-tech companies tend to prioritize cybersecurity more, with clearer structures, dedicated budgets, and advanced practices like VAPT or Zero Trust. SMEs and low-tech companies, on the other hand, more often struggle with limited resources, less defined responsibilities, and lower perceived urgency.

# 5. CYBERSECURITY PERCEPTION BY GEOGRAPHICAL REGION

## 5.1. SURVEY RESPONSES BY COUNTRY

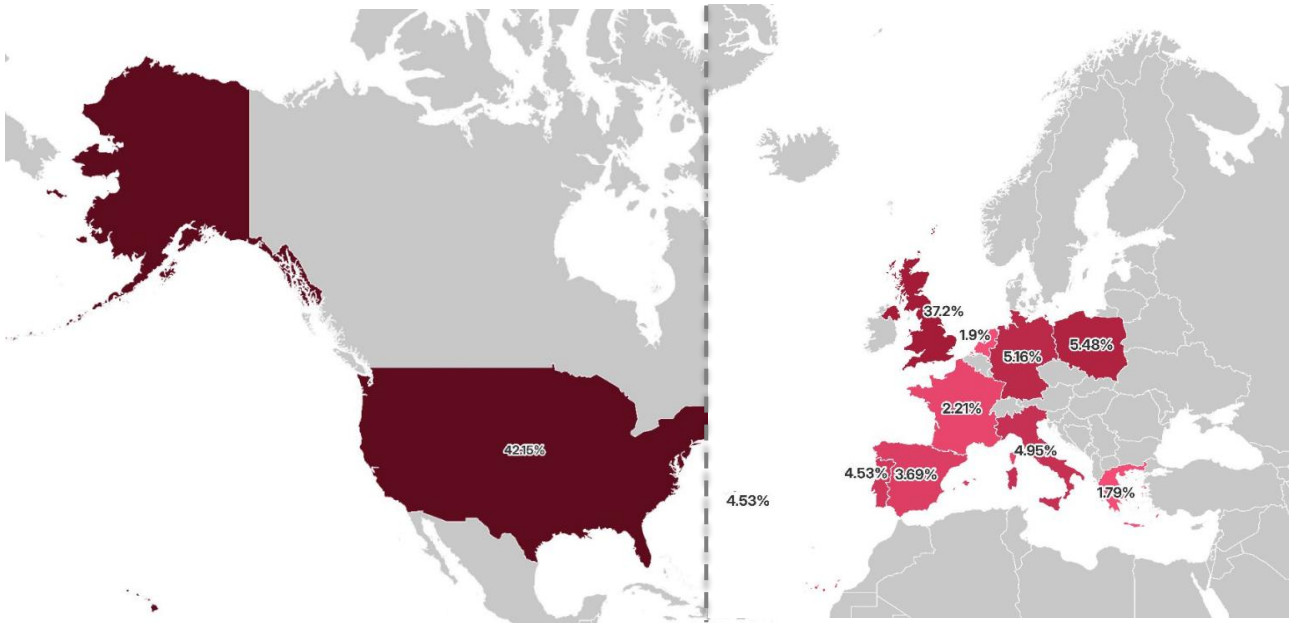

Figure 70 • Heat Map of Survey Responses by Country

Understanding how cybersecurity is perceived across different regions helps organizations benchmark their own practices and anticipate external expectations in international contexts. This section provides a comparative overview of survey responses from Europe (EU), the United Kingdom (UK), the United States (USA), and Italy.

By grouping all regional insights here, we offer a focused lens on how country-specific factors influence views on cybersecurity investment, innovation, risk, and compliance. While many patterns are consistent, notable differences emerge. The USA often shows stronger optimism or more polarized positions, the UK aligns closely on innovation topics, and Italy sometimes stands apart with unique or more cautious responses. Europe, in general, reflects a broader range of perspectives, often occupying a middle ground.

This analysis supports decision-makers in adapting cybersecurity strategies to regional dynamics and recognizing how cultural, regulatory, and market conditions can shape attitudes and actions.

*How do different geographical regions perceive cybersecurity: as a competitive advantage or a necessary burden?*

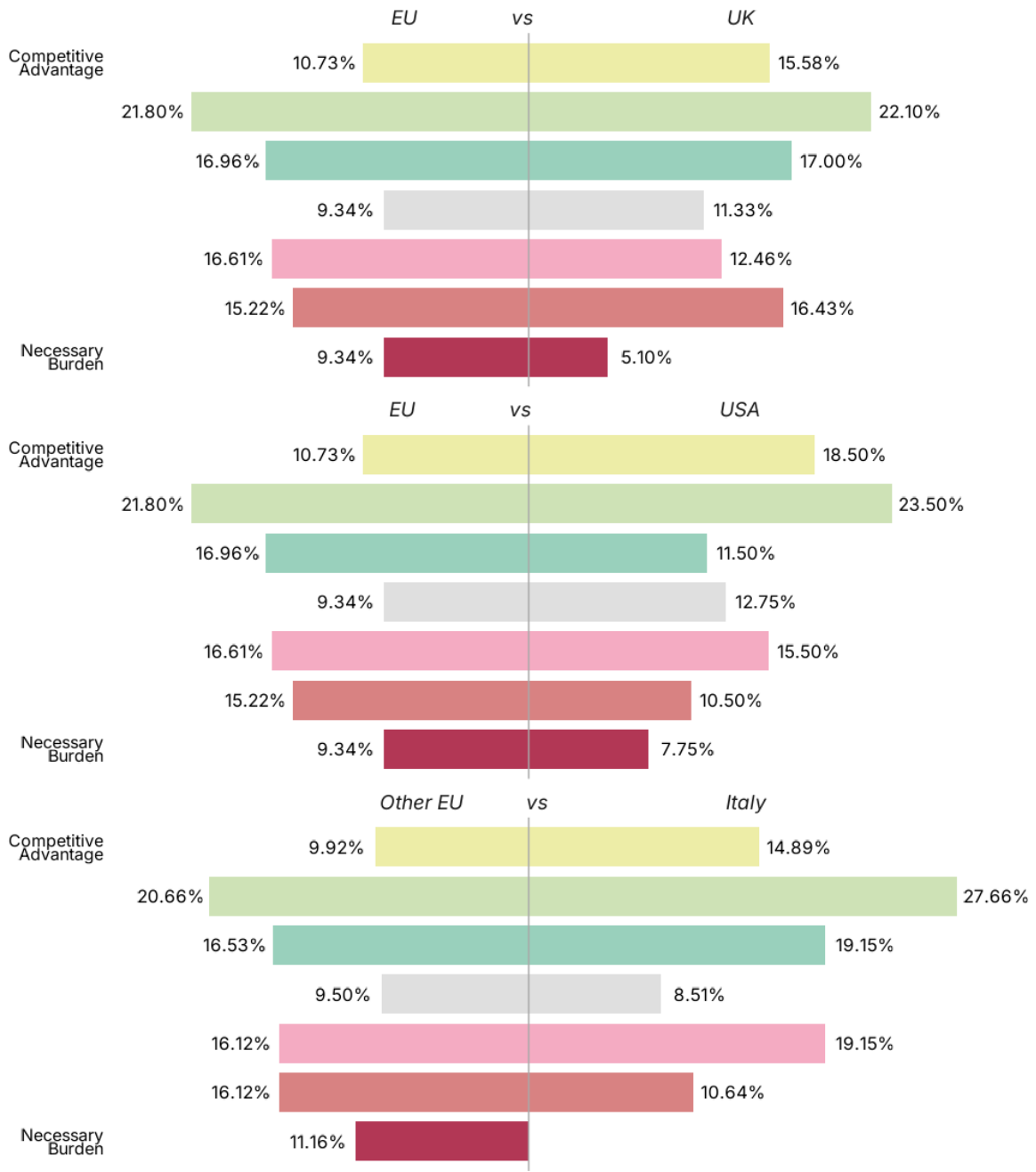

Figure 71 • Cybersecurity perception by country

The USA tends to lean more toward viewing cybersecurity as a competitive advantage, but it also records a higher percentage for those who perceive it as a necessary burden. Italy stands out as the country that most firmly believes in cybersecurity as a competitive advantage, reporting zero responses for it being perceived as a necessary burden.

*How do different geographical regions see cyber defense: as a necessary expense or a worthy investment?*

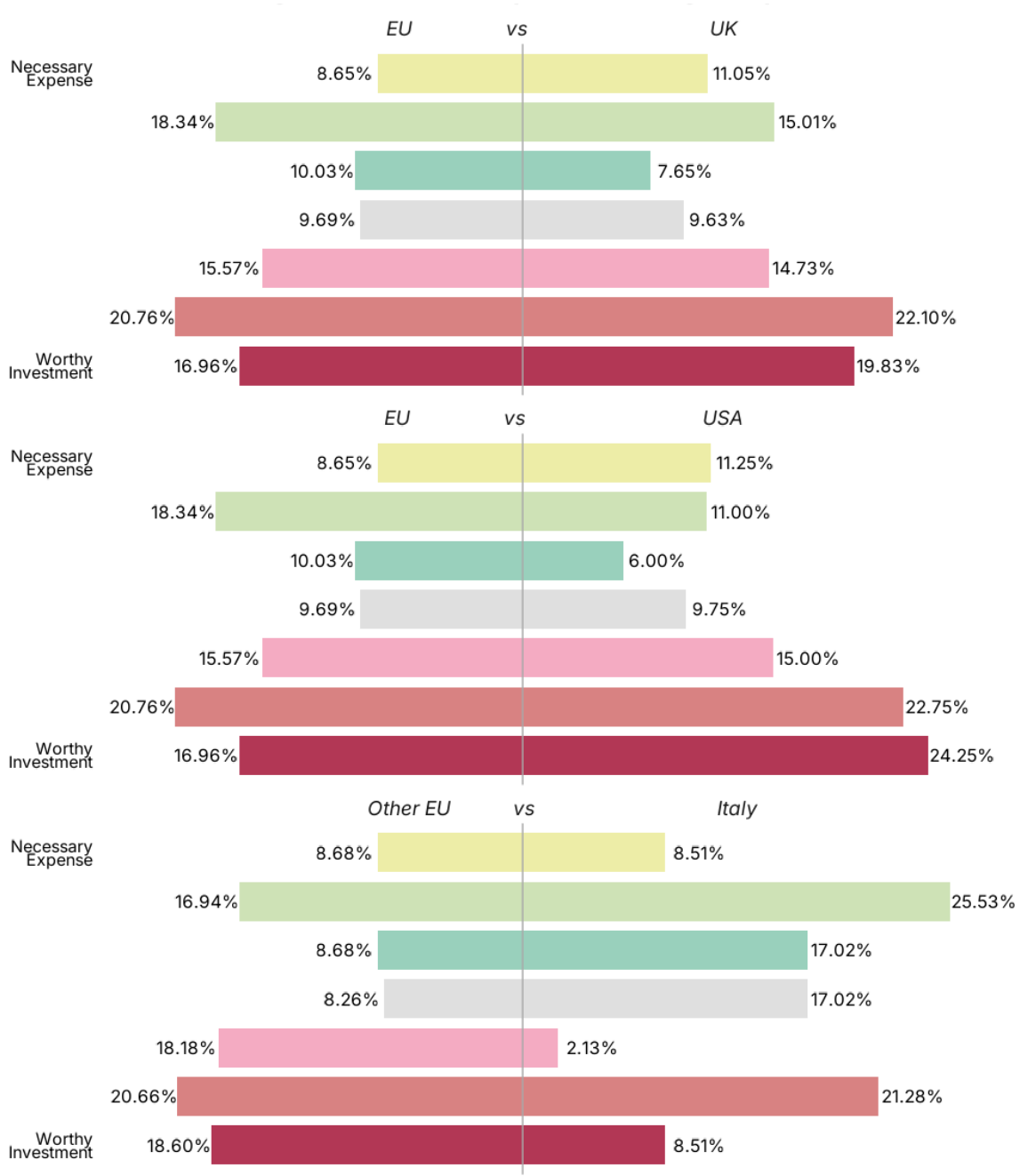

Figure 72 • Cybersecurity spending by Country

The UK and USA are more inclined than Europe to perceive cybersecurity as a worthy investment, while Italy exhibits a much stronger tendency toward intermediate values on the scale rather than extreme positions.

*How do concerns about cyber threats vary across geographical regions: are they more focused on risks from within or outside Europe?*

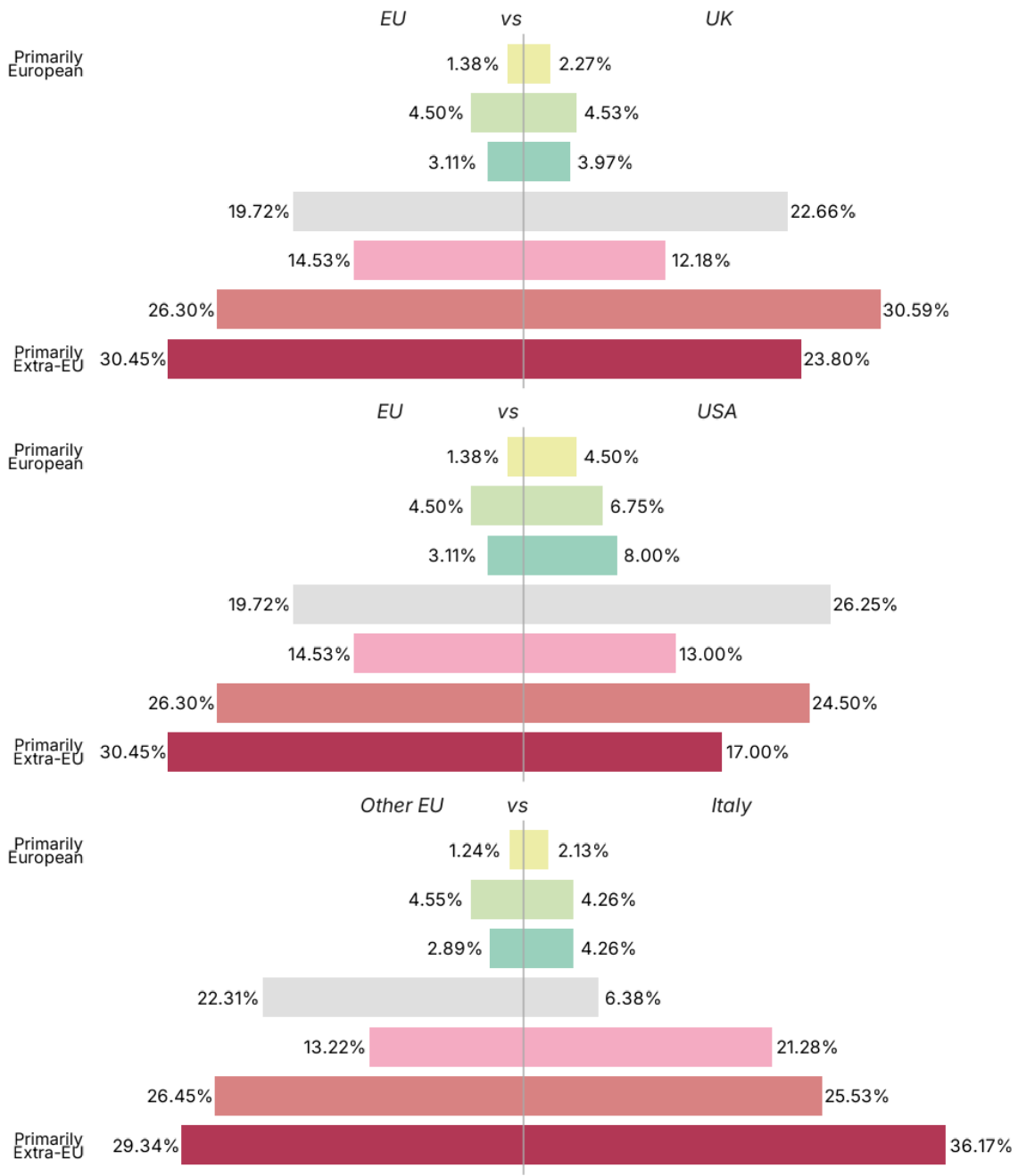

Figure 73 • Cyber Threats perception by Country

According to our sample, Italy is the most inclined to perceive cyber threats as primarily aligned with extra-EU trends, which is consistent with the views expressed by other countries.

*How do different geographical regions feel about security measures in their organizations: more of a hindrance to productivity or essential for protection?*

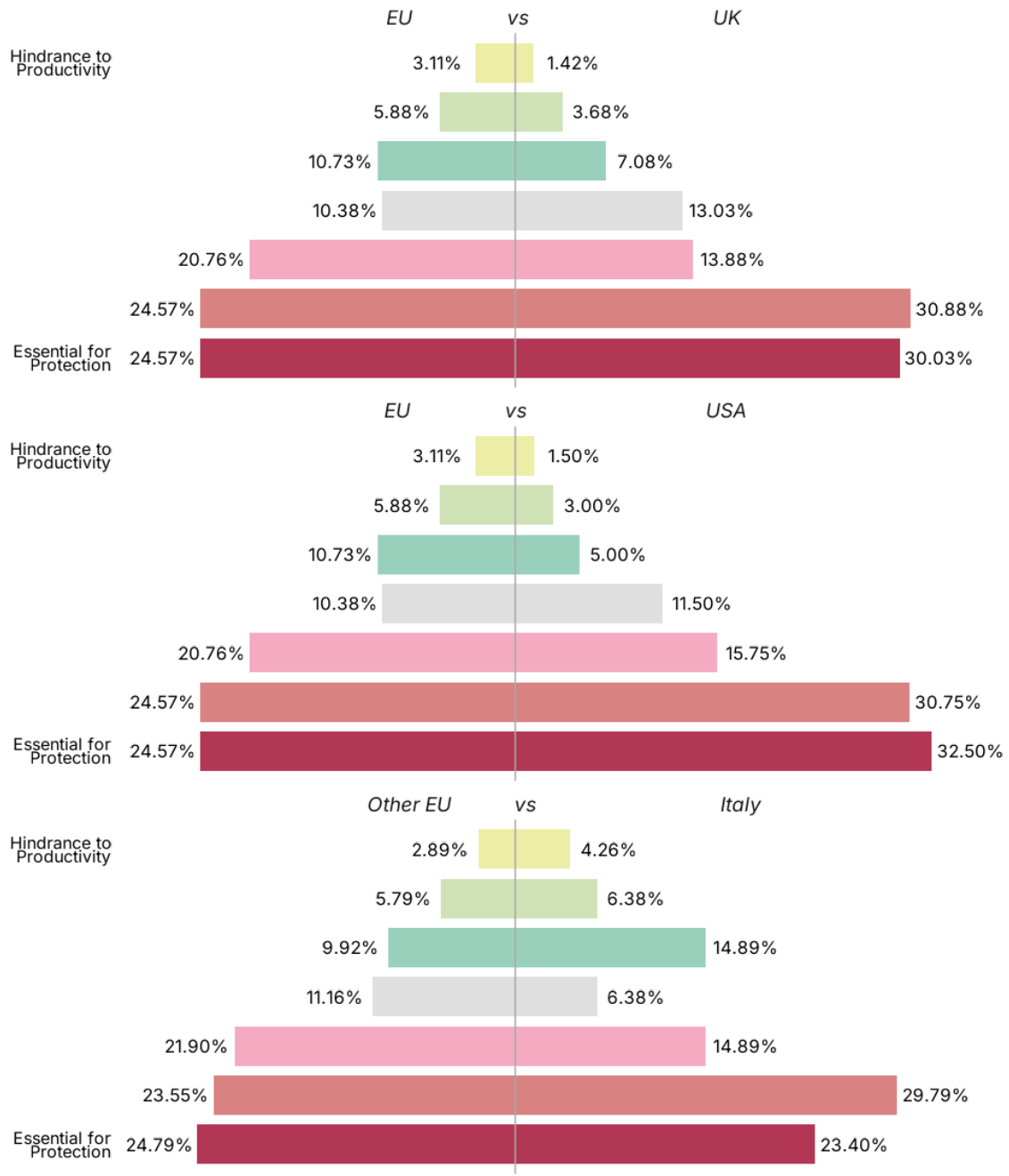

Figure 74 • Security Measures perception by Country

The results shown in the graph highlight that, for all participants without distinction, UK, USA, Europe, and Italy, security measures are considered essential for protection.

*How do different geographical regions perceive the impact of cyber regulations: as a positive influence or a negative impact?*

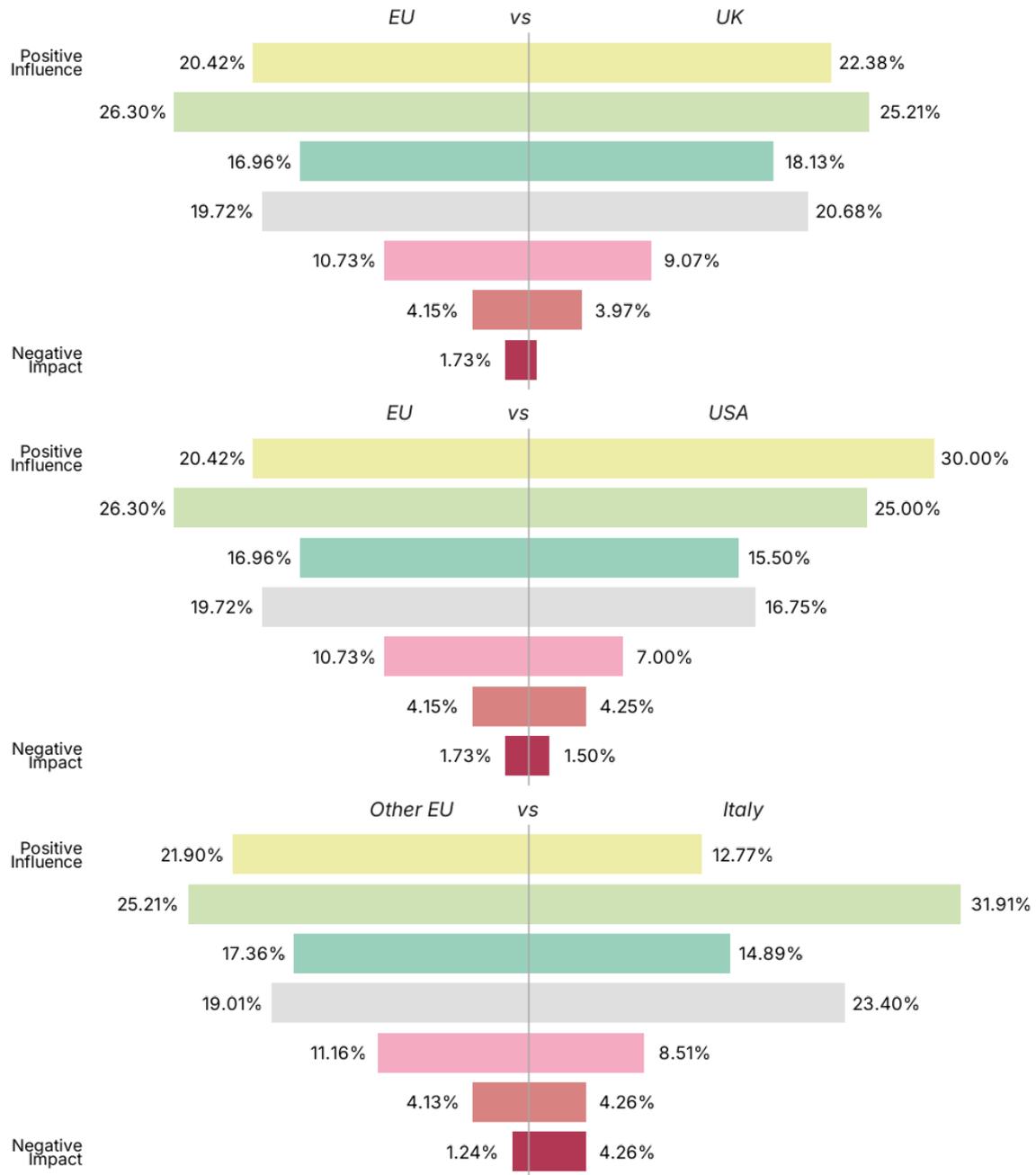

Figure 75 • Cyber Regulations perception by Country

The USA stands out for its strong belief that cybersecurity regulations have a positive impact on the entire business, while Italy tends to be less extreme, remaining at the lower end of the scale. The EU positions itself at a mid-level, similar to the UK, leaning more towards the moderately positive values on the scale.

*How do different geographical regions perceive supply chain security: a high priority or a low priority?*

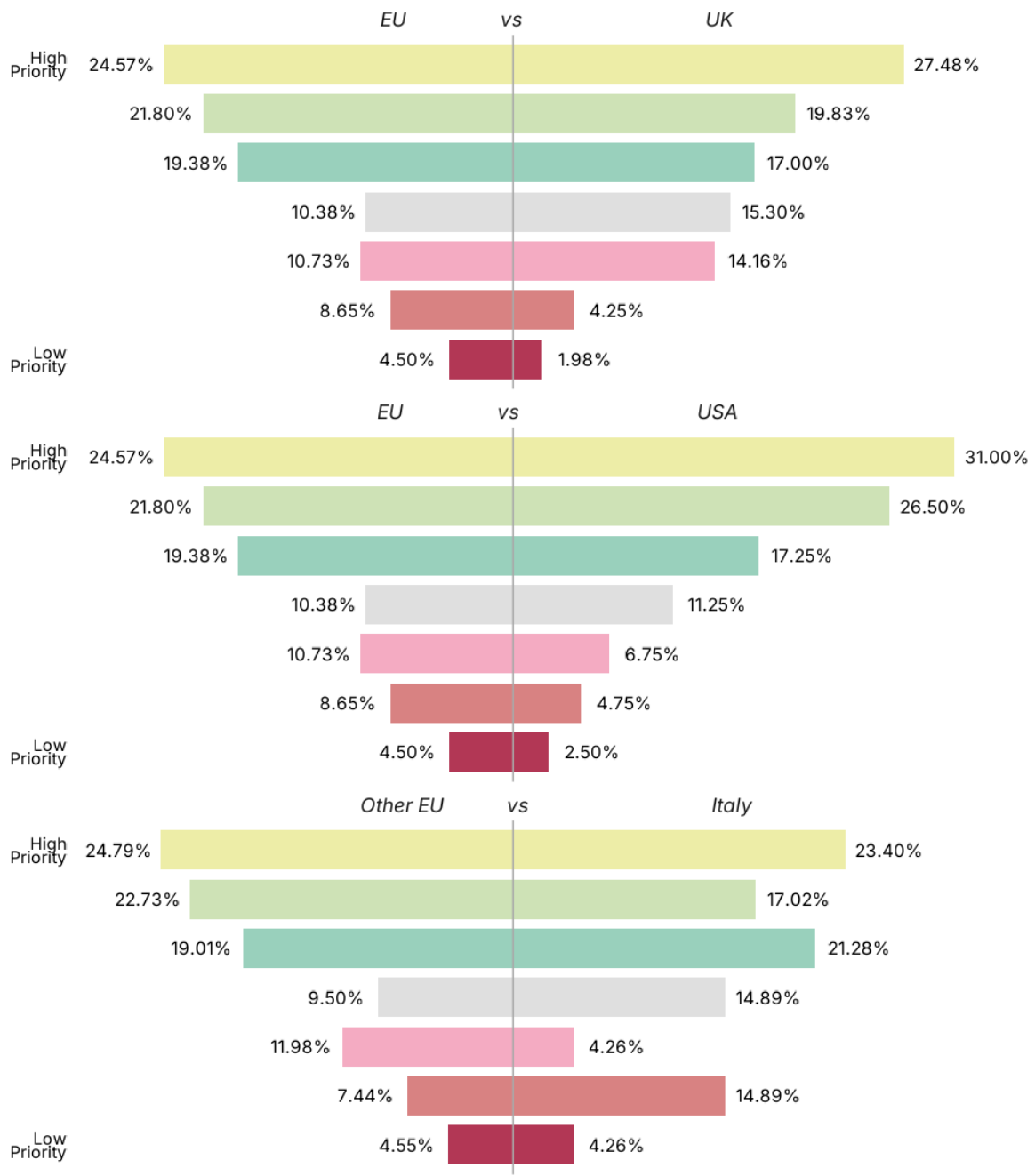

Figure 76 • Supply Chain Security perception by Country

While the USA, EU, and UK exhibit similar behavior and show the same distribution of responses, indicating that they consider supply chain security a high priority, Italy presents a different picture. Almost 15% of Italian respondents believe it has a low priority.

*How do different geographical regions describe the ease of recruiting cybersecurity talent: is it easy or difficult?*

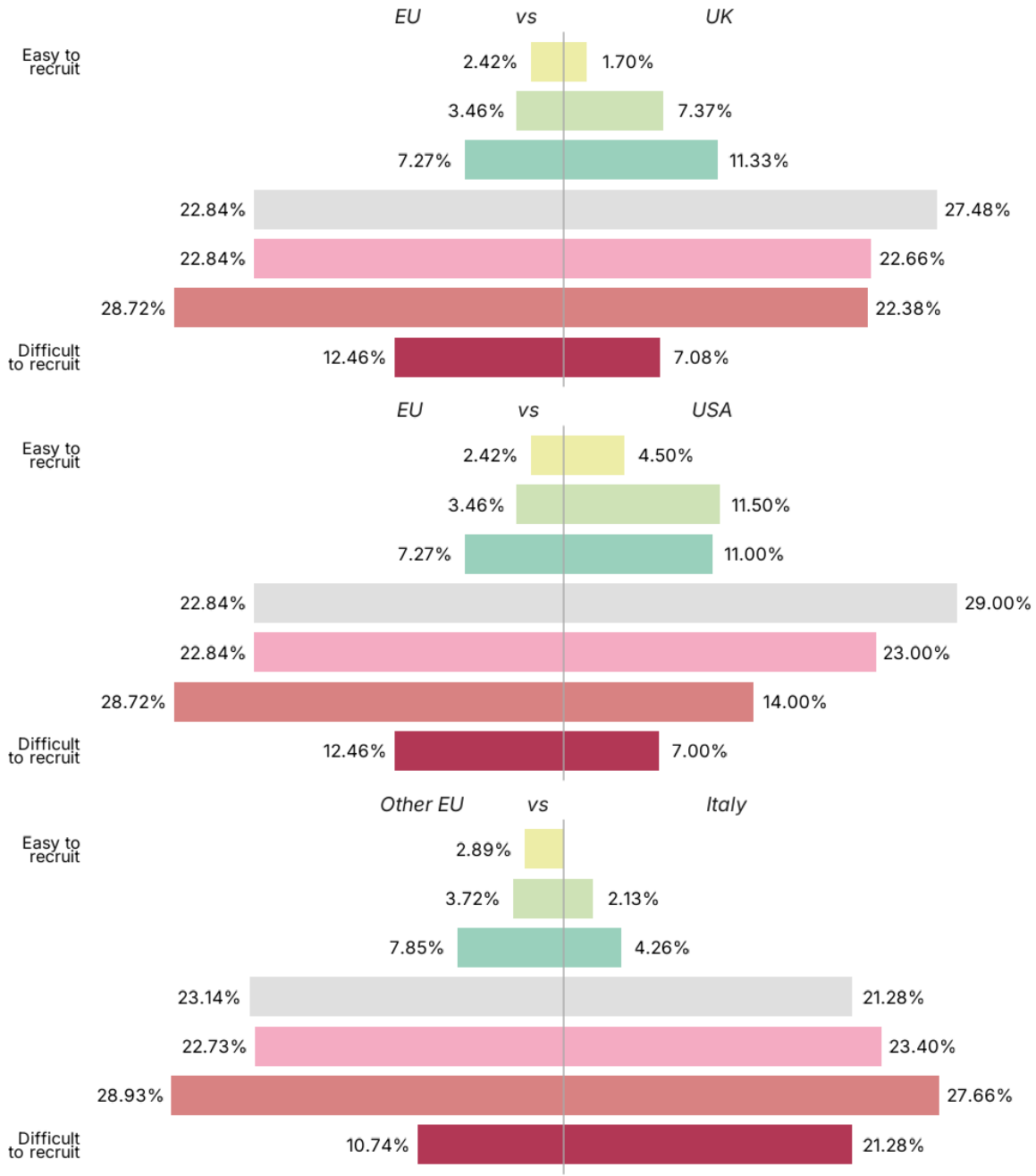

Figure 77 • Recruiting Cybersecurity talent by Country

The UK, USA, and EU appear to experience a similar moderate to high level of difficulty in recruiting cybersecurity talent. In contrast, Italy faces an even greater challenge, struggling more significantly with this issue.

*How do different geographical regions view emerging technologies: as opportunities or threats to their cybersecurity posture?*

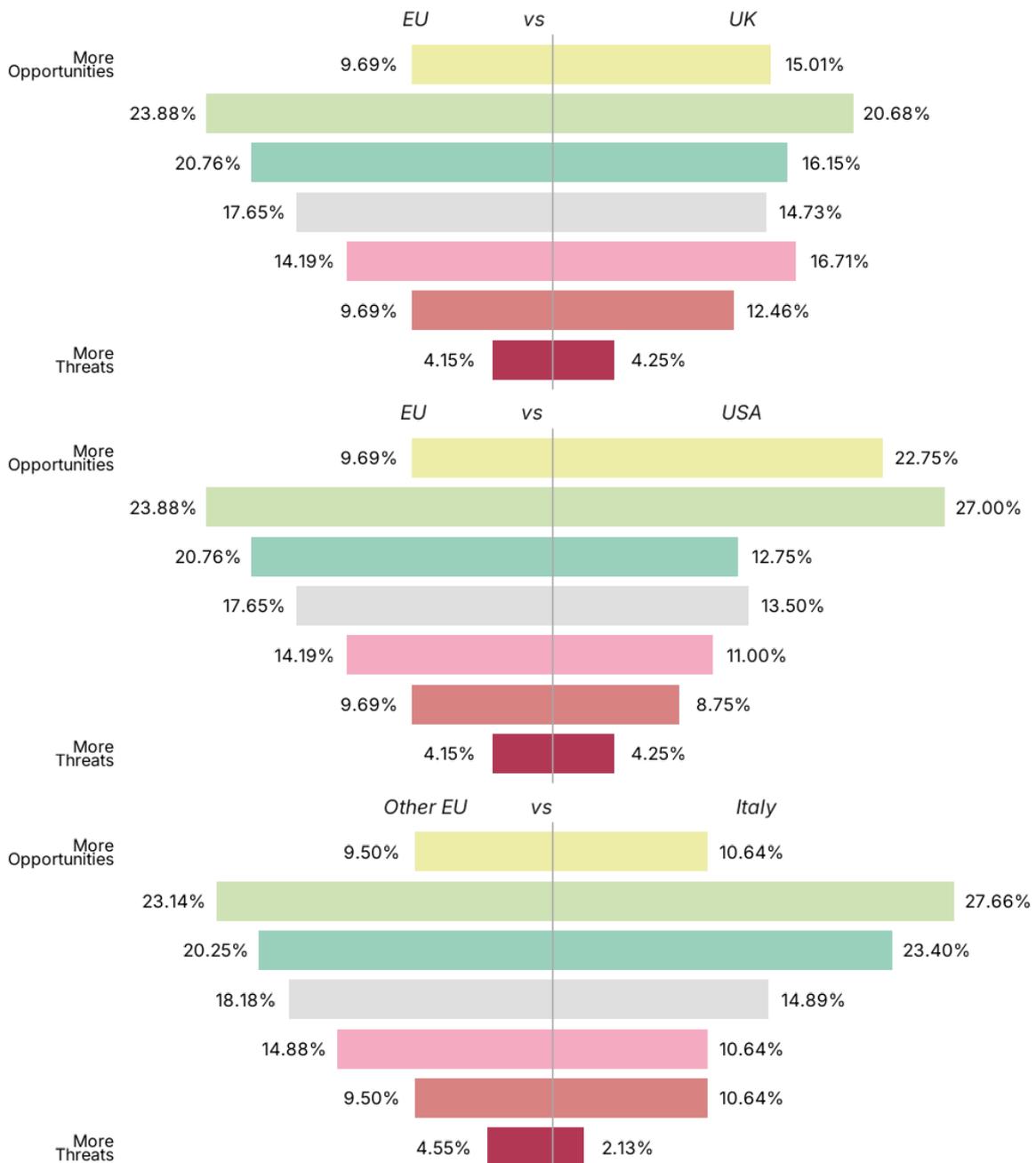

Figure 78 • Emerging Technologies perception by Country

While the UK and EU do not lean toward the extreme end of the scale, the USA appears more convinced that new technologies provide greater opportunities for cybersecurity.

*How prepared are different geographical regions to handle a cybersecurity incident?*

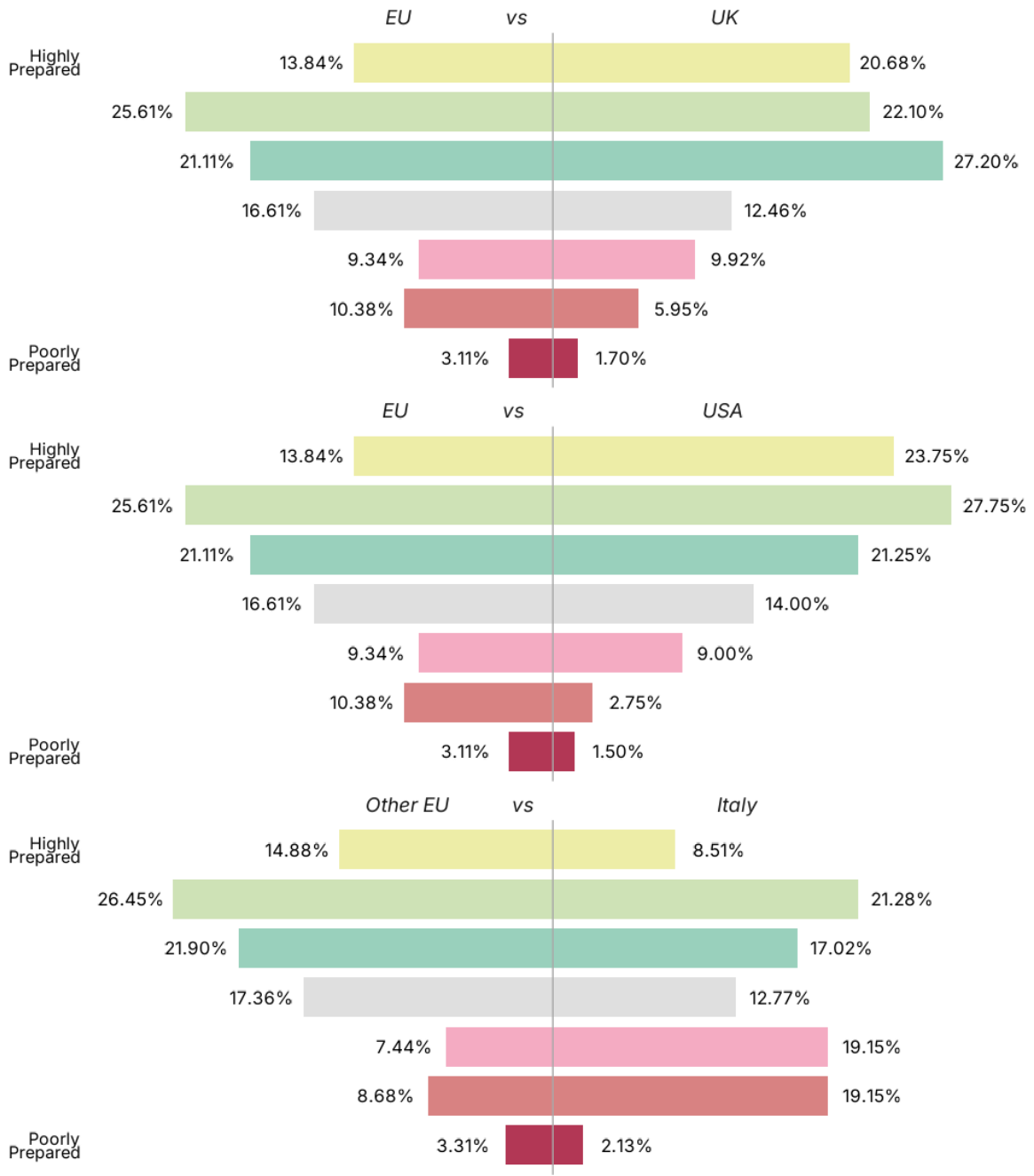

Figure 79 • Incident readiness perception by Country

The UK, USA, and EU appear to have a similar level of readiness when it comes to responding to cybersecurity incidents. Italy, on the other hand, shows lower preparedness, leaning more toward the "Poorly Prepared" end of the scale.

*How do different geographical regions perceive cyber insurance: as essential or non-essential for their organizations?*

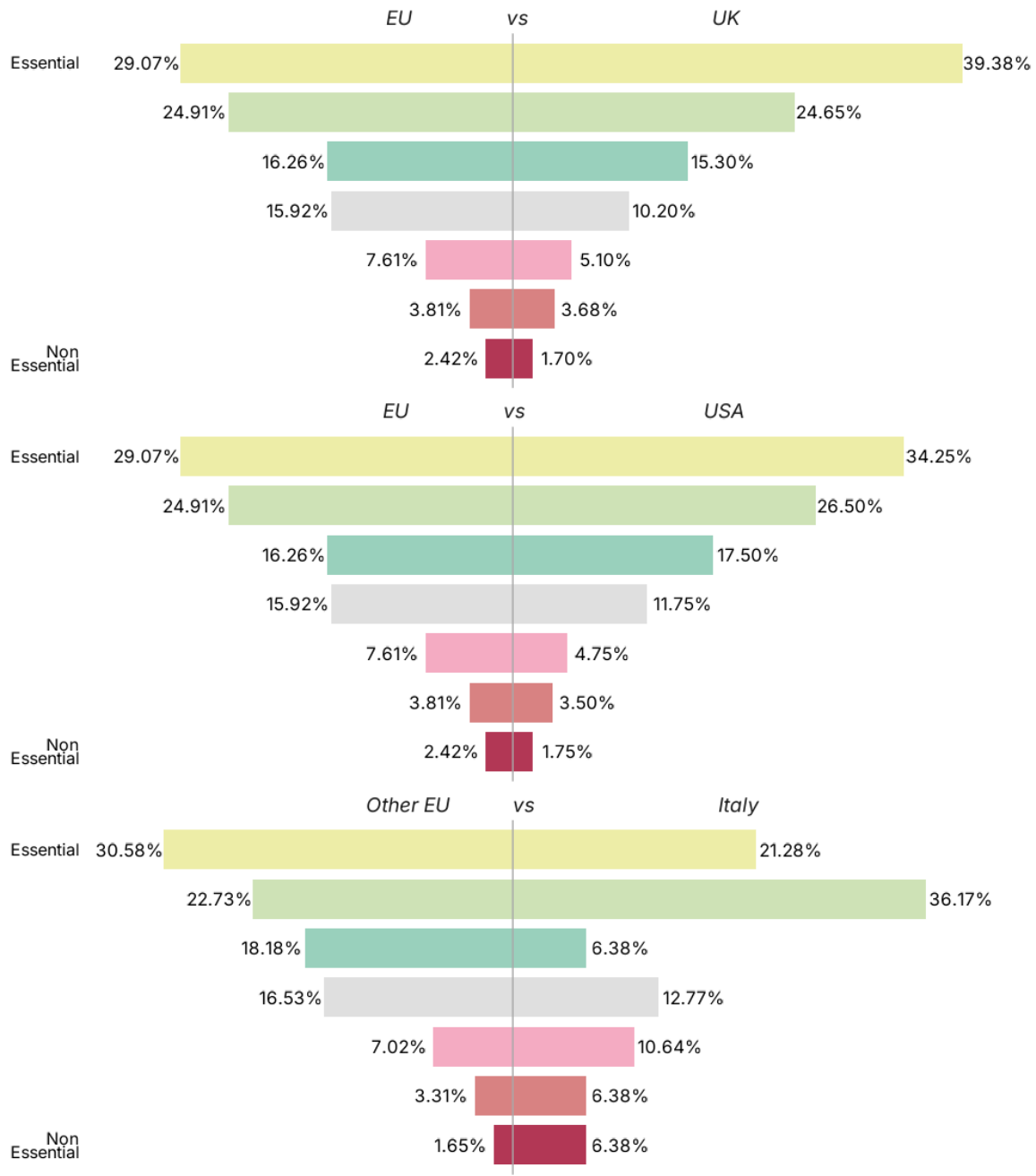

Figure 80 • Cyber Insurance perception by Country

On average, the UK, USA, and EU recognize cyber insurance as essential, progressively leaning toward the most positive end of the scale. Italy, meanwhile, aligns with the sixth level, still acknowledging the importance of cyber insurance, but without reaching the most extreme value.

*How effectively do different geographical regions find zero trust architecture in meeting their cybersecurity needs?*

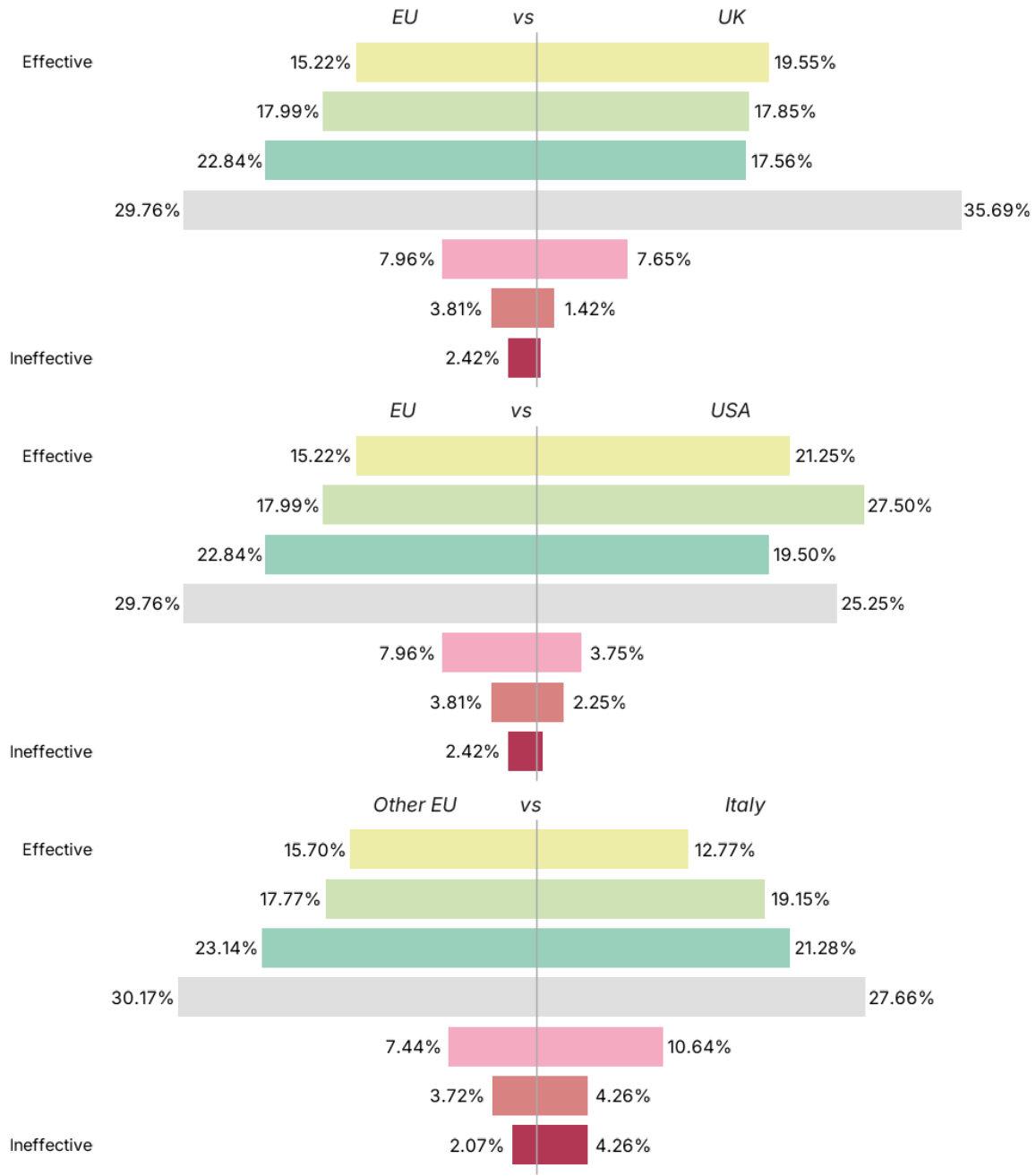

Figure 81 • Zero Trust Architecture perception by Country

The UK is the country that remains most aligned with the neutral point on the scale. Similarly, the USA, EU, and Italy also avoid leaning too far toward considering it essential. Among them, however, the USA stands out slightly for showing a stronger inclination in that direction.

*When allocating budgets, do geographical regions place greater emphasis on innovation or on security?*

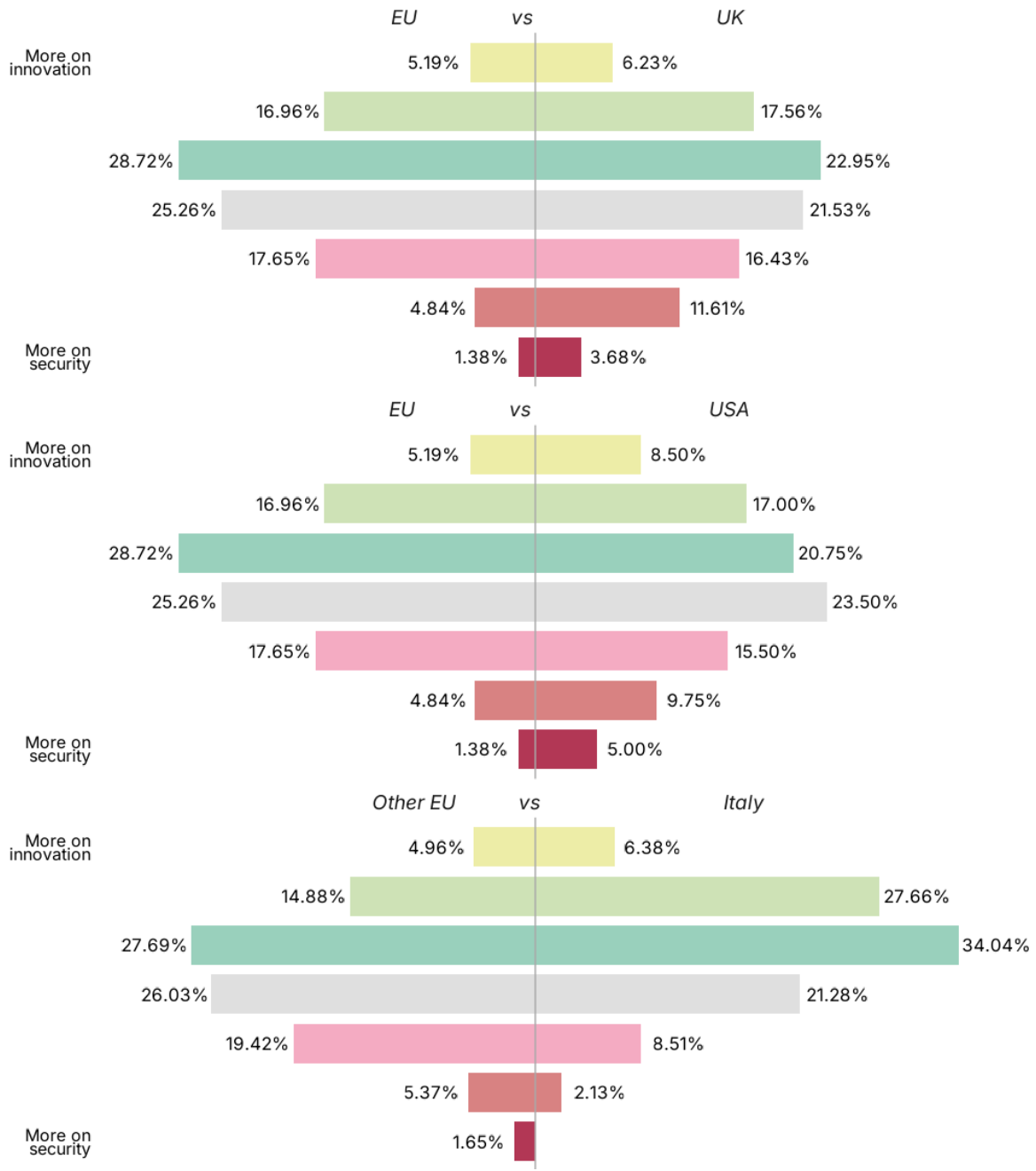

Figure 82 • Priorities in Budget Allocation by Country

The UK, USA, and EU maintain a largely similar distribution of results, while Italy aligns with them but places even greater emphasis on the neutral values of the scale.

*How do different geographical regions perceive employee training in cybersecurity: as a major priority or a minor one for their organizations?*

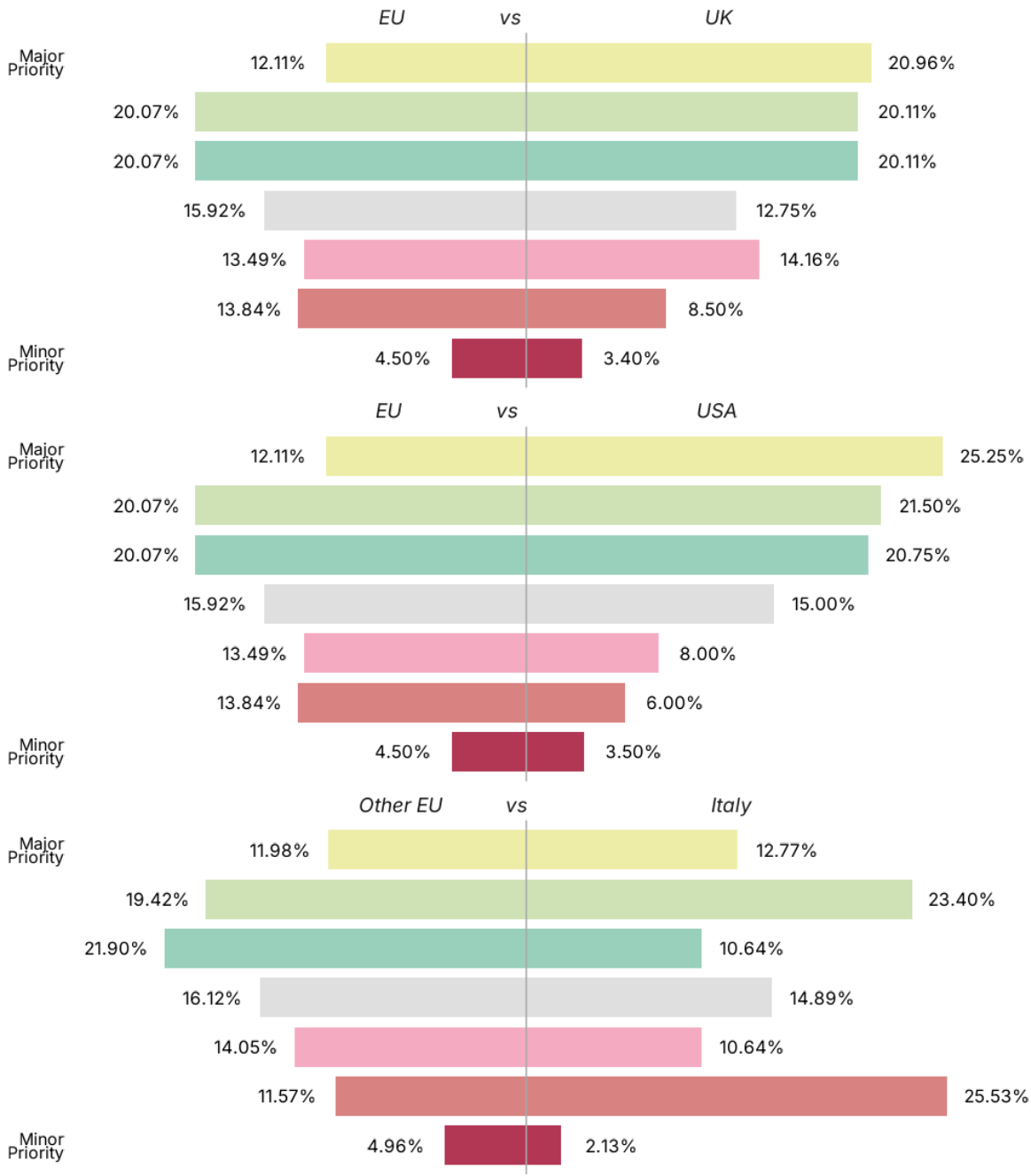

Figure 83 • Employee training in Cybersecurity by Country

The USA and UK clearly prioritize employee training in cybersecurity more than the EU and Italy. Concerningly, Italy stands out, with 25.53% of organizations recognizing employee training as a minor priority.

*How do different geographical regions view outsourcing cybersecurity?*

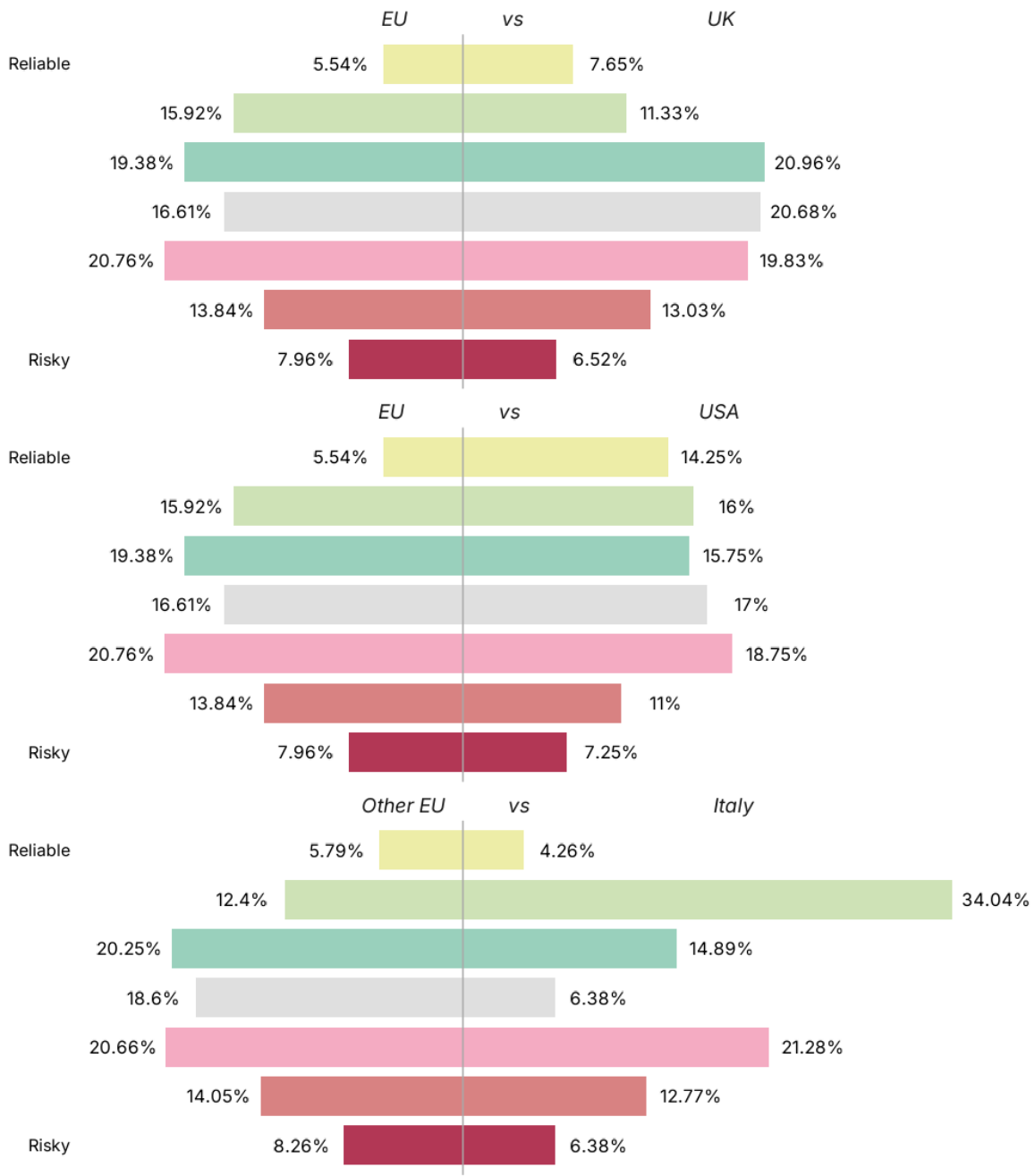

Figure 84 • Outsourcing Cybersecurity by Country

The UK, USA, and EU largely align in not explicitly stating whether outsourcing cybersecurity is reliable or risky, while Italy appears more inclined to embrace outsourcing.

*How do different geographical regions perceive automation in cybersecurity: beneficial or problematic?*

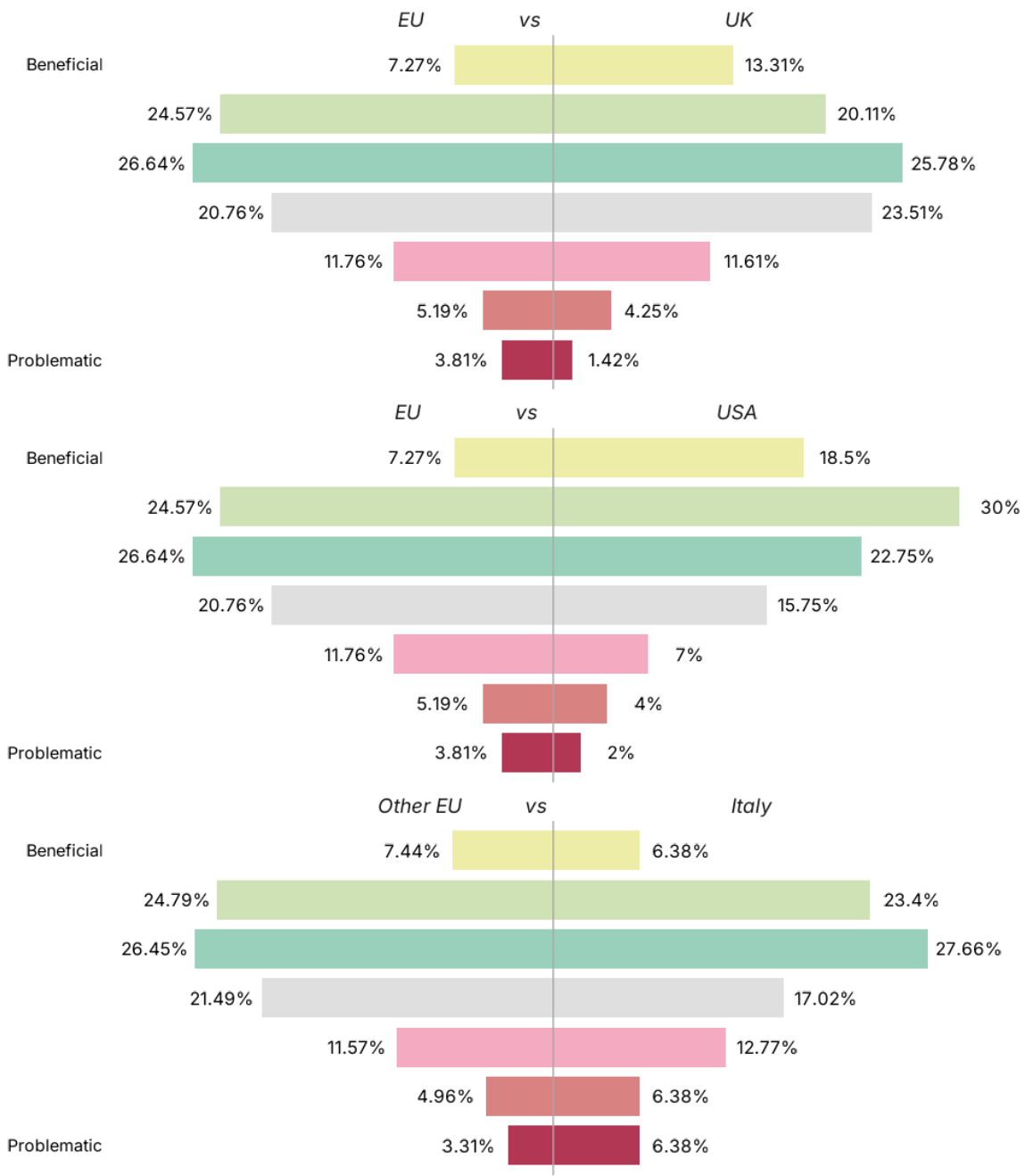

Figure 85 • Automation in Cybersecurity by Country

The USA is the most optimistic about the beneficial power of automation, followed by the UK. In contrast, the EU has a perception similar to Italy, leaning more toward the neutral and intermediate values on the scale.

*How much revenue do different geographical regions invest in cybersecurity?*

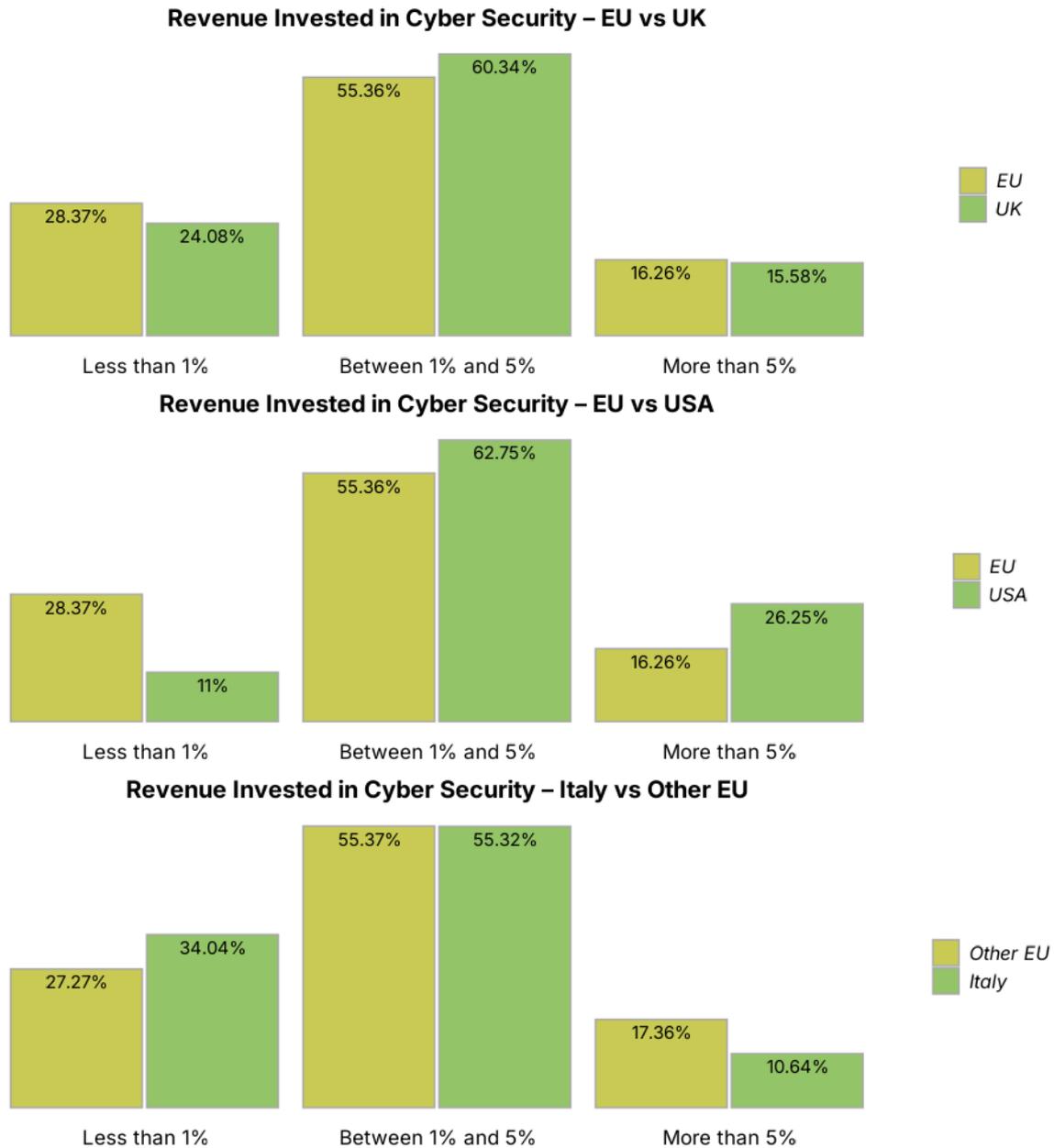

Figure 86 • Revenue Invested in Cybersecurity by Country

The United States clearly leads in cybersecurity investment, followed by the UK and, to a lesser extent, Europe. Italy aligns closely with the European average, although Italian respondents are more likely to report investing less than 1% of their budget in cybersecurity rather than more than 5%.

*How many cybersecurity incidents have occurred in the last 5 years across different geographical regions?*

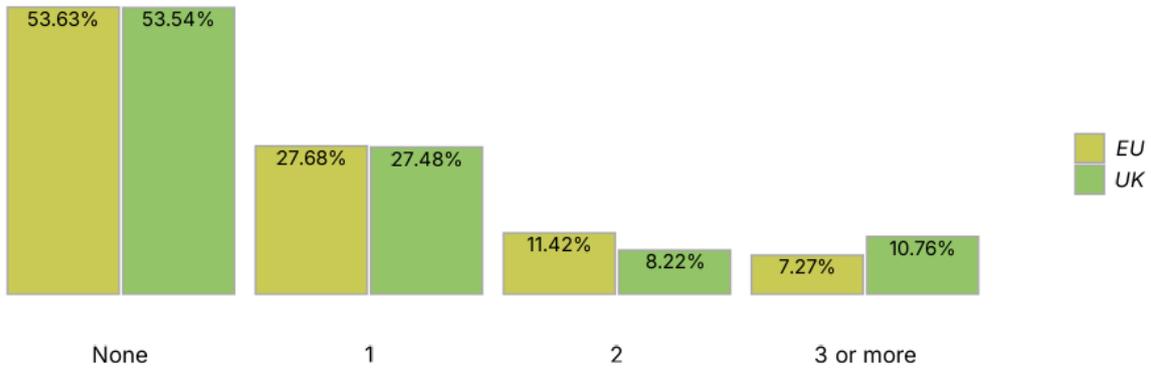

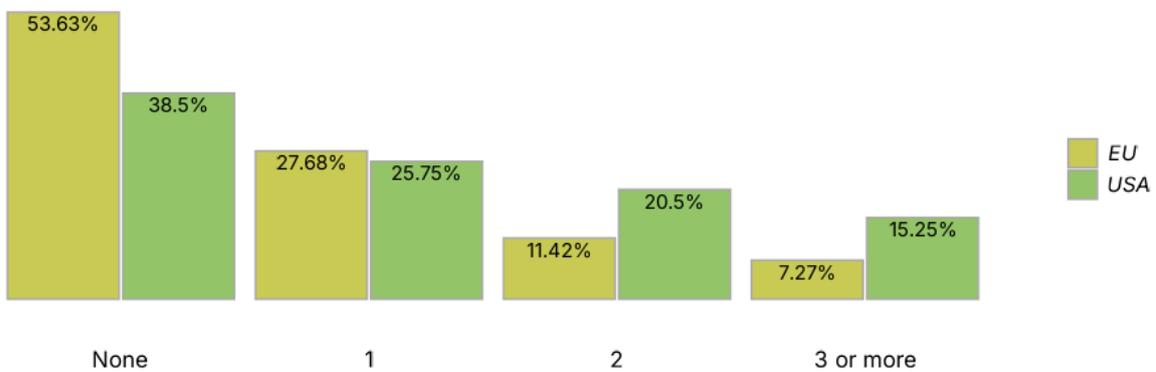

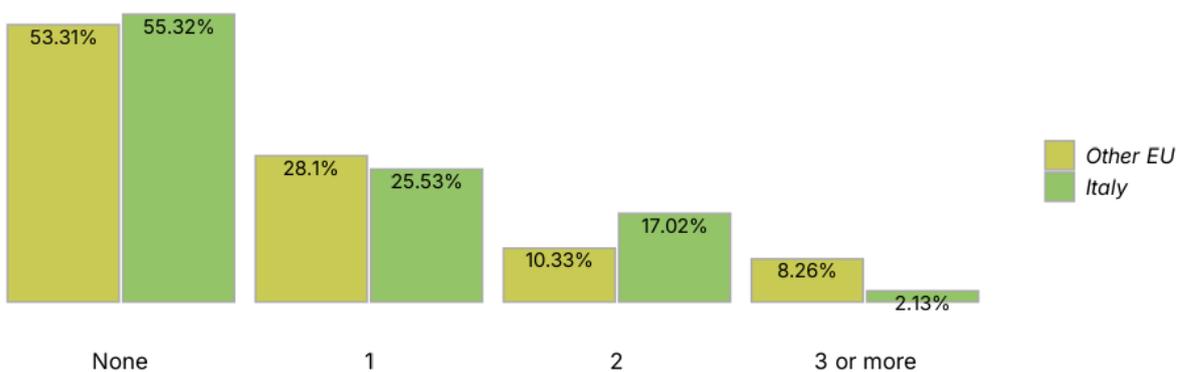

Figure 87 • Cybersecurity incidents in the last 5 years by Country

When it comes to incident history, the USA reports significantly higher exposure, having faced more cybersecurity incidents in the past five years, followed by the UK. In contrast, Italy and the broader European region report fewer. However, these numbers should be interpreted with caution. Fewer reported incidents may reflect limited detection capabilities, meaning attacks could have occurred without being noticed.

*How do different geographical regions respond to mitigate cyber attacks?*

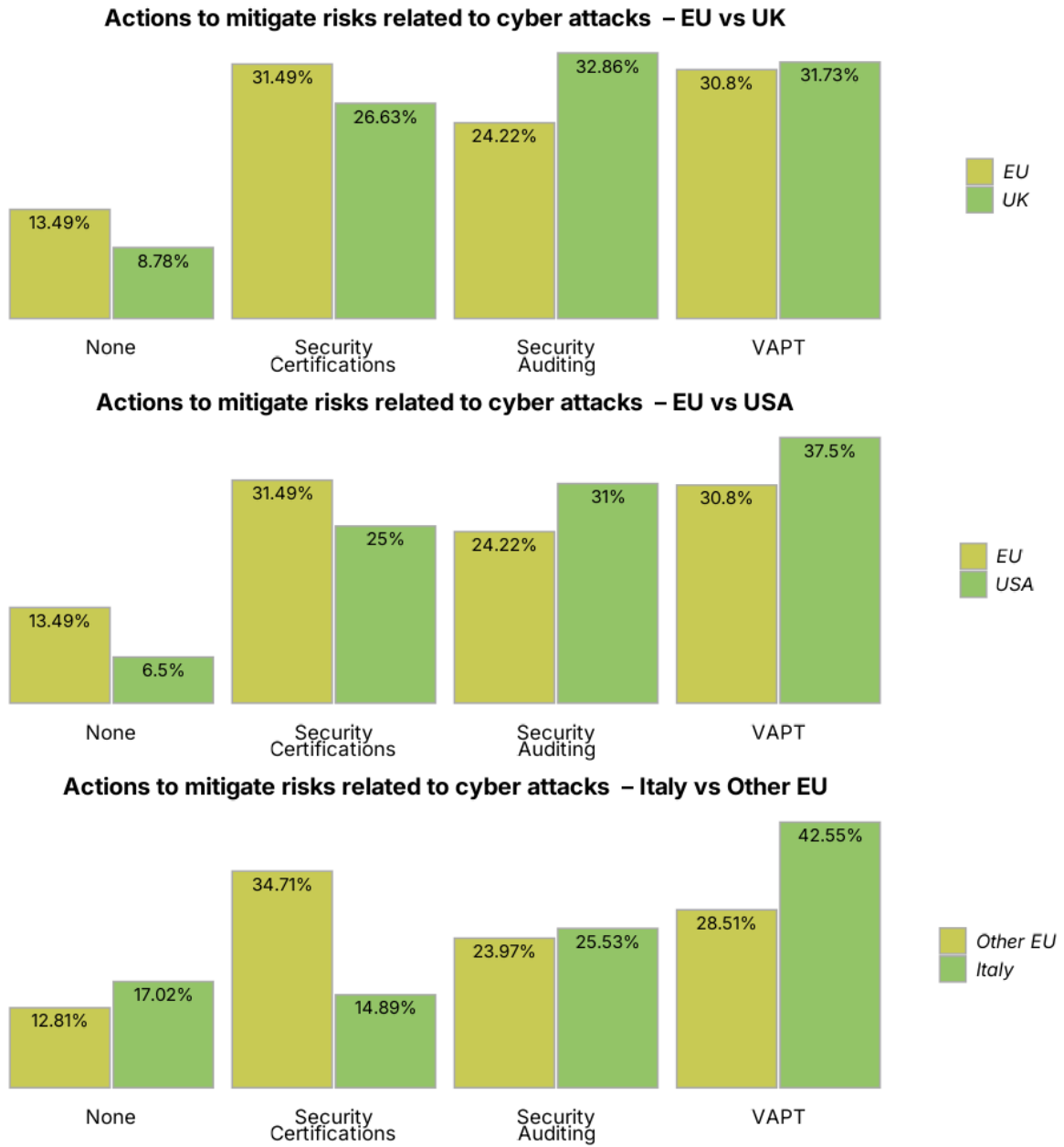

Figure 88 • Action to mitigate cyber-attacks by Country

Italy reports the highest VAPT adoption, even above the USA, UK, and Europe. Europe lags behind, while the USA shows the lowest share of organizations taking no action, indicating stronger baseline practices. Notably, there's a growing trend from no action to advanced measures like VAPT, which require greater investment. This makes the high adoption of VAPT particularly remarkable, especially in the case of Italy, suggesting that many organizations are willing to invest significantly in strengthening their cybersecurity posture.

*What are the planned responses across different geographical regions to mitigate cyberattacks?*

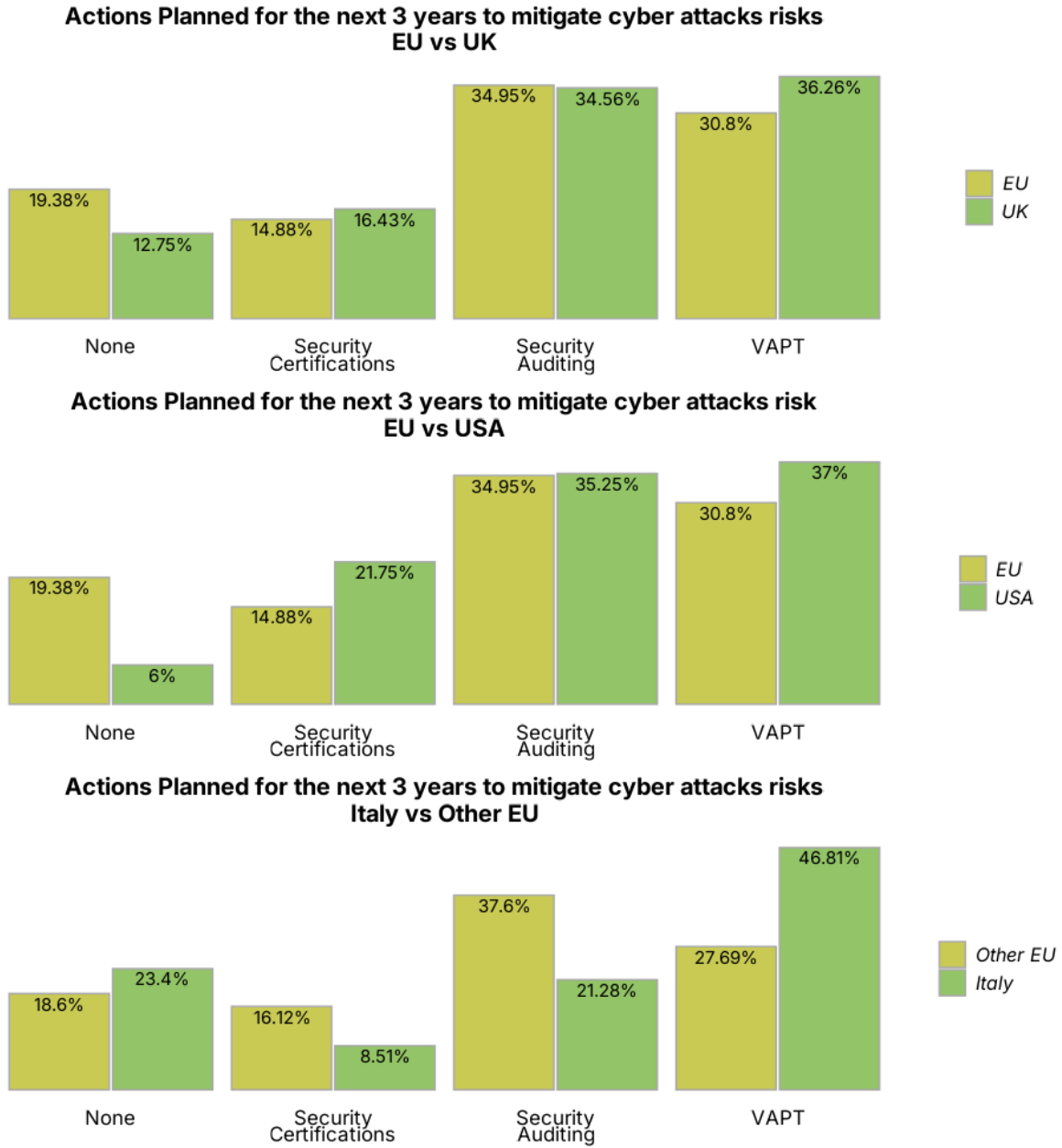

Figure 89 • Action planned to mitigate cyber-attacks by Country

This trend is further confirmed here, with even clearer evidence that companies recognize the importance of continued investment in cybersecurity through more advanced technical measures like VAPT. Surprisingly, Italy leads this trend.

*How much priority do different geographical regions place on including cybersecurity budgets during project planning?*

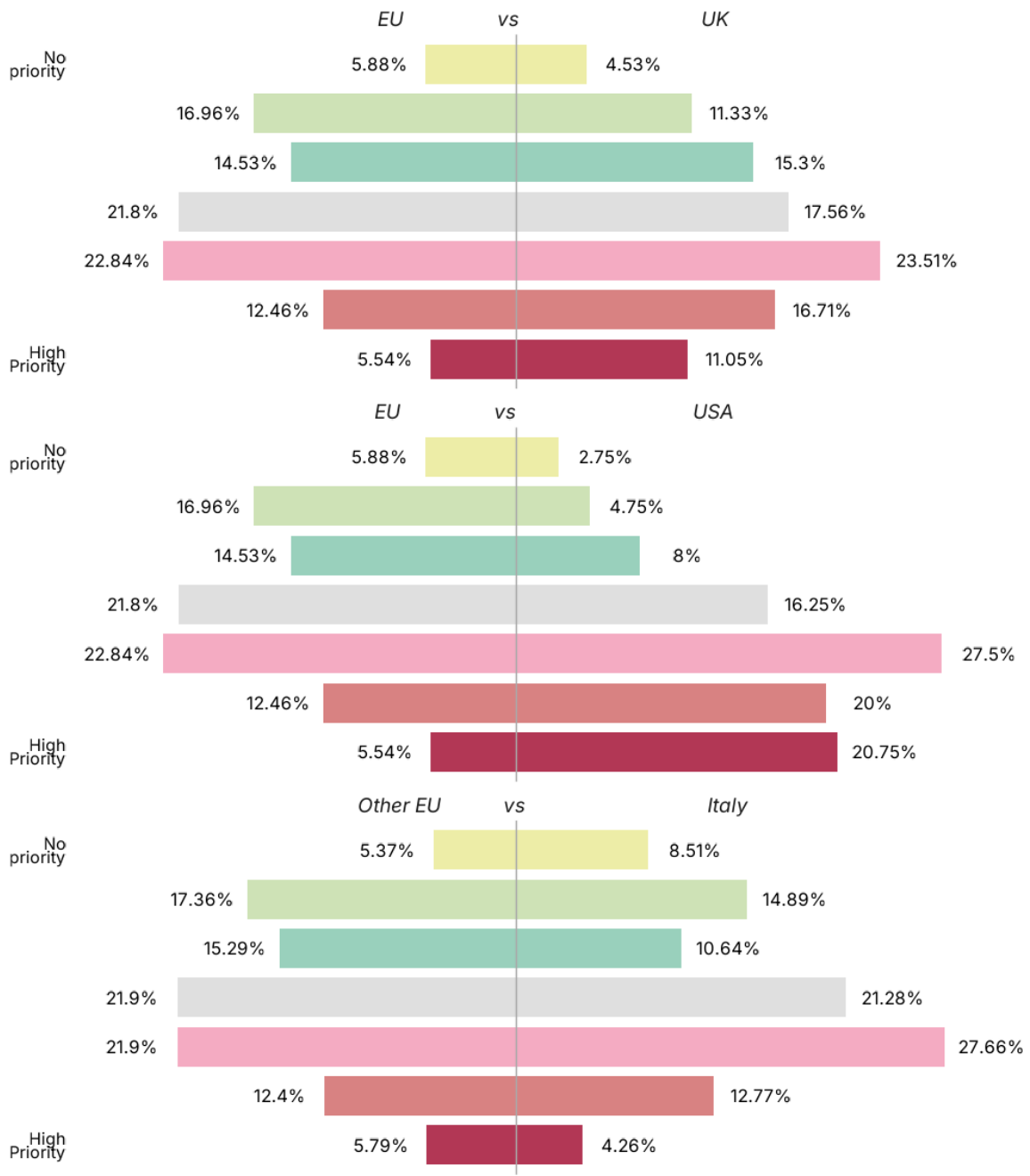

Figure 90 • Cybersecurity budgets allocated in project planning by Country

The USA places the highest priority on budget allocation for cybersecurity, as evidenced by the lowest percentage of responses indicating 'no priority' on the scale. In contrast, the EU, UK, and Italy show more balanced results.

*Across different geographical regions, are there clear tasks and defined timelines specifically allocated for implementing cybersecurity measures within projects?*

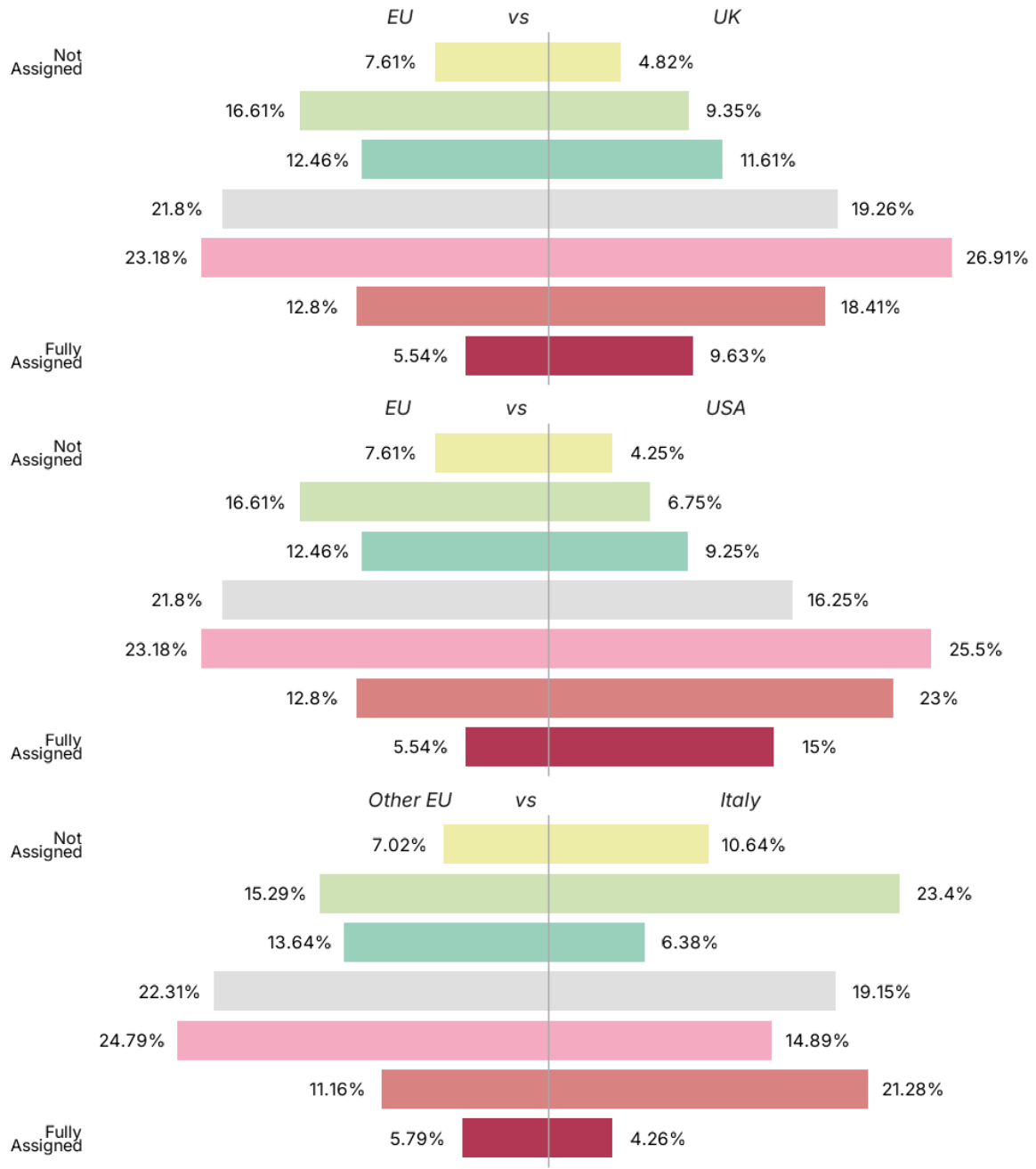

Figure 91 • Tasks and defined timelines specifically allocated for implementing Cybersecurity by Country

The EU is less inclined than the UK and USA to allocate clear tasks and timelines for cybersecurity, with Italy being even less proactive in this regard.

*How many resources do different geographical regions allocate to cybersecurity?*

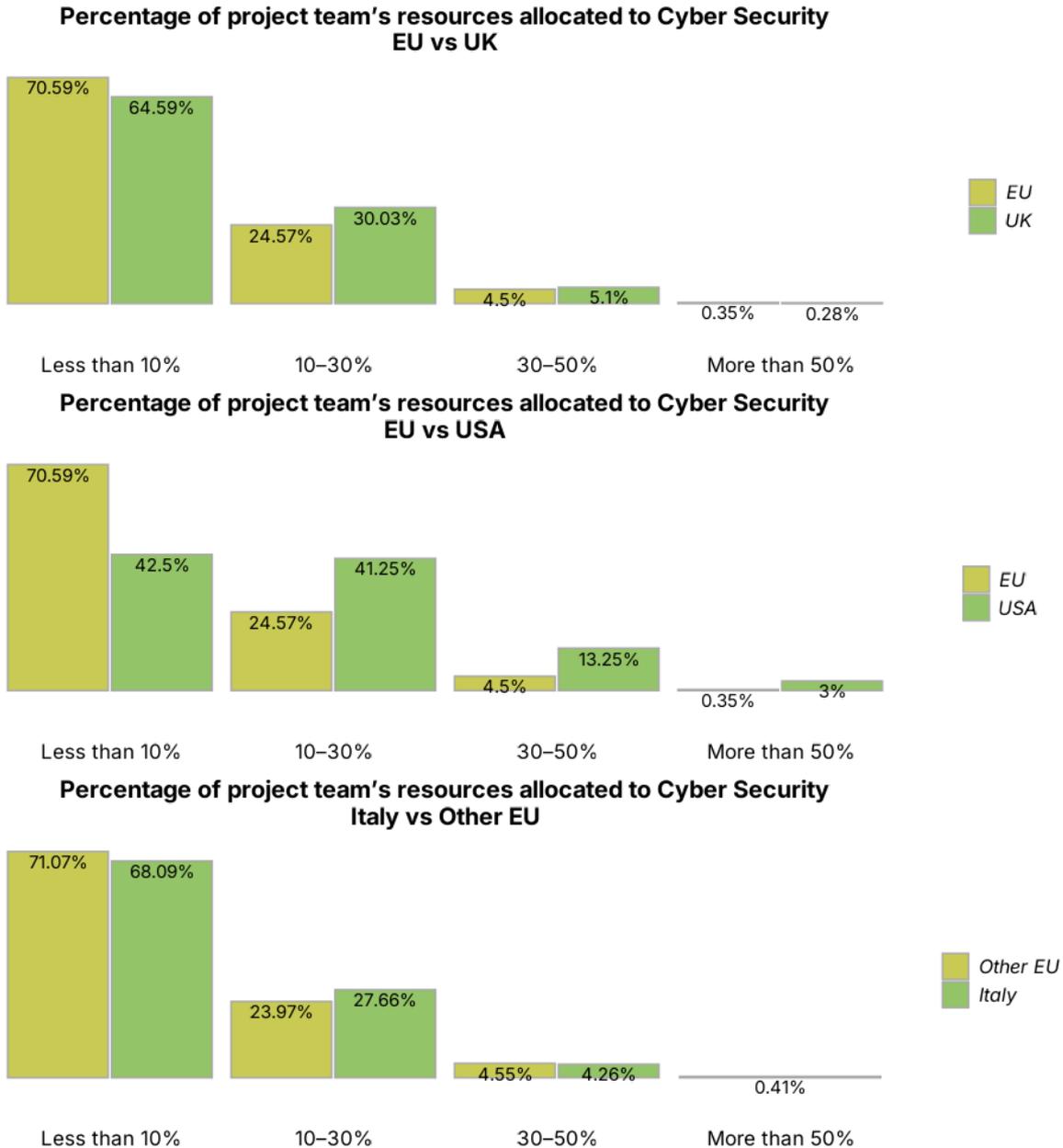

Figure 92 • Allocation of resources to cybersecurity by country

The USA stands out as the country most frequently reporting allocations in the 30–50% range, although this share remains relatively low overall. Notably, US respondents are also the least likely to allocate less than 10%, a threshold that dominates both the UK and broader Europe. Italy closely mirrors the European trend.

*How would different geographical regions rate the importance of planning for emergency scenarios, such as responses to potential cybersecurity incidents?*

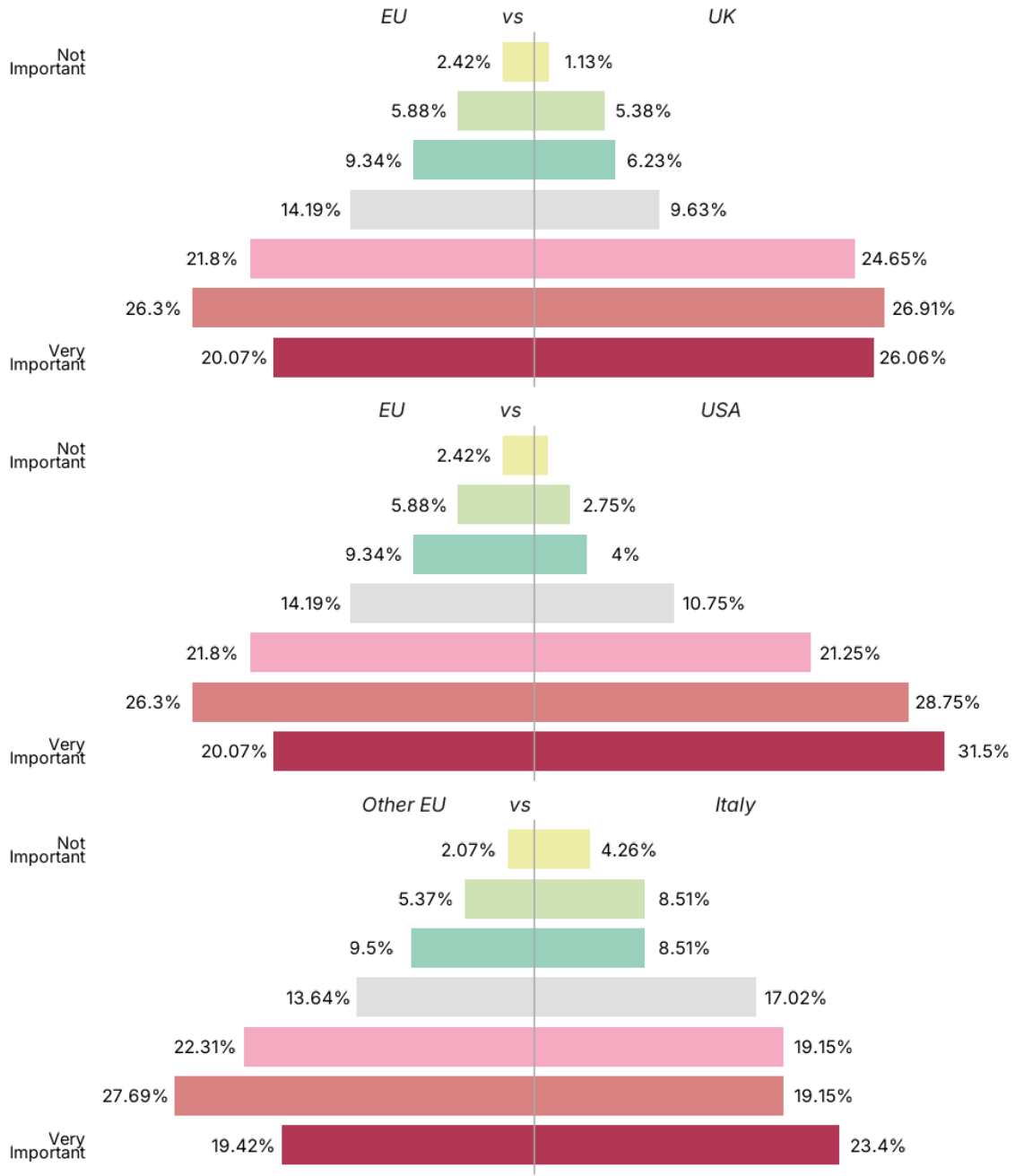

Figure 93 • Importance of planning for emergency scenarios by Country

The USA and UK are somewhat more inclined to assert the importance of planning for emerging scenarios, while the EU and Italy show a slightly broader distribution, leaning more toward the moderately positive values on the scale rather than the extremes.

*Do different geographical regions consider post-project vulnerability assessments and long-term cybersecurity protection as key components of their strategy?*

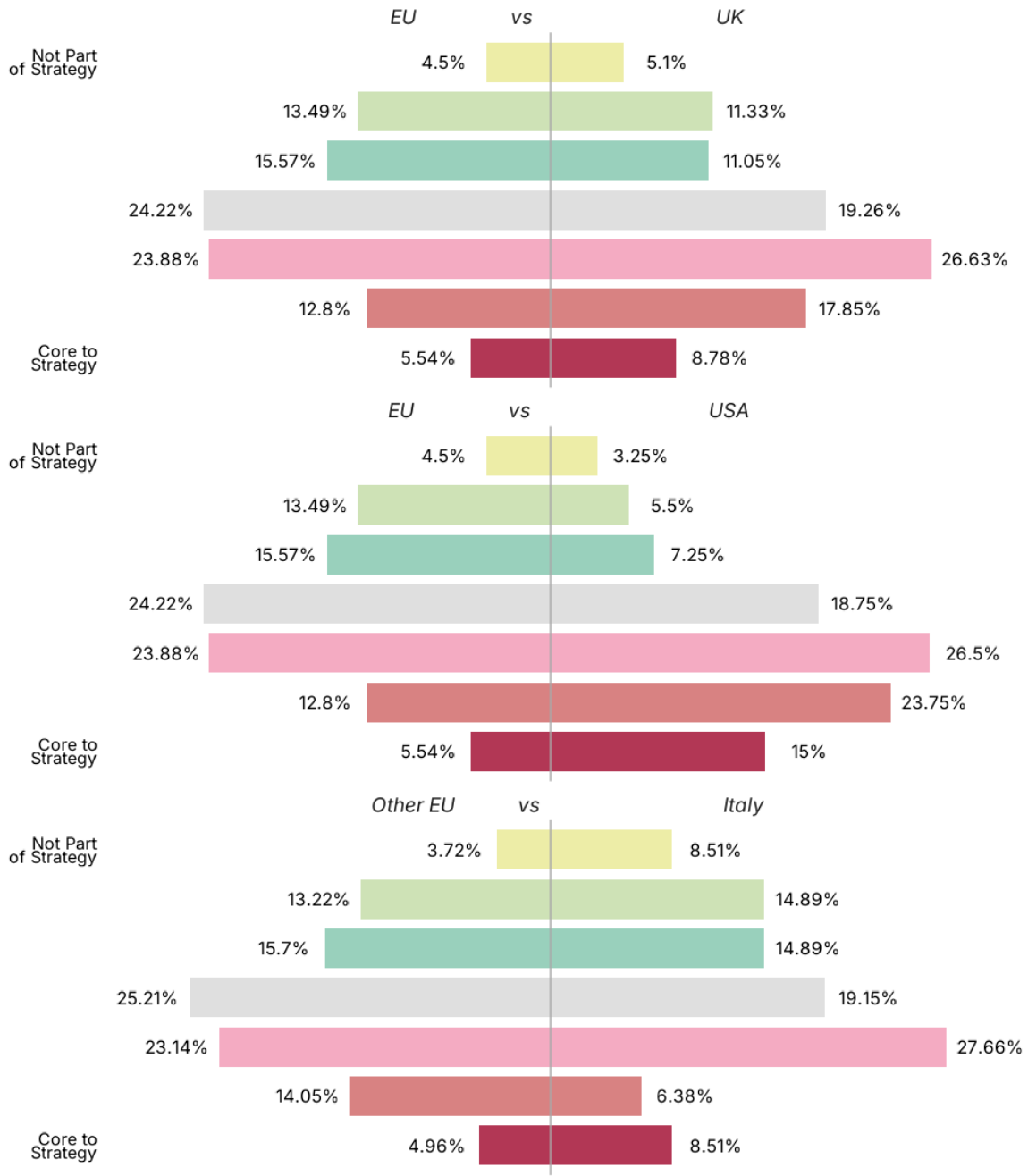

Figure 94 • Long-term Cybersecurity protection by Country

The UK and USA consider post-project vulnerability assessment to be core to their strategy, while the EU takes a more neutral stance. However, Italy stands out within the EU, prioritizing it more than other European countries.

# 6. APPENDIX: SURVEY QUESTIONS AND RESPONSES

## 6.1. RESPONDENT OVERVIEW: DEMOGRAPHIC ANALYSIS

The survey gathered responses from both SMEs and large enterprises, ensuring a balanced representation of both. As shown in the pie chart below, participation was evenly split between these two categories.

Among SMEs, the highest number of respondents came from companies with 21-50 employees, while in large enterprises, most responses came from companies with over 500 employees.

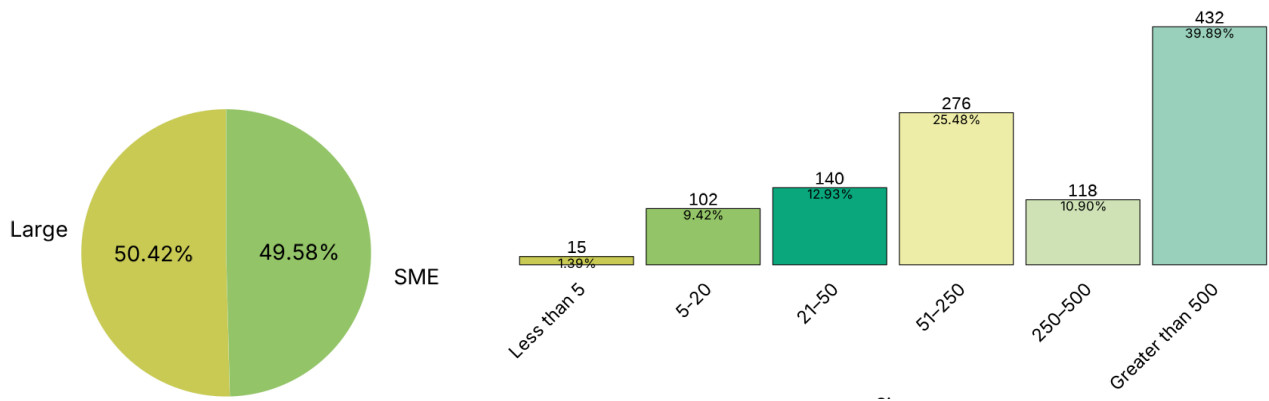

Figure 95 • Distribution of respondents by Company Size

The responses are well distributed across the USA, EU, and UK, with the majority coming from the EU, followed by the UK and then the USA.

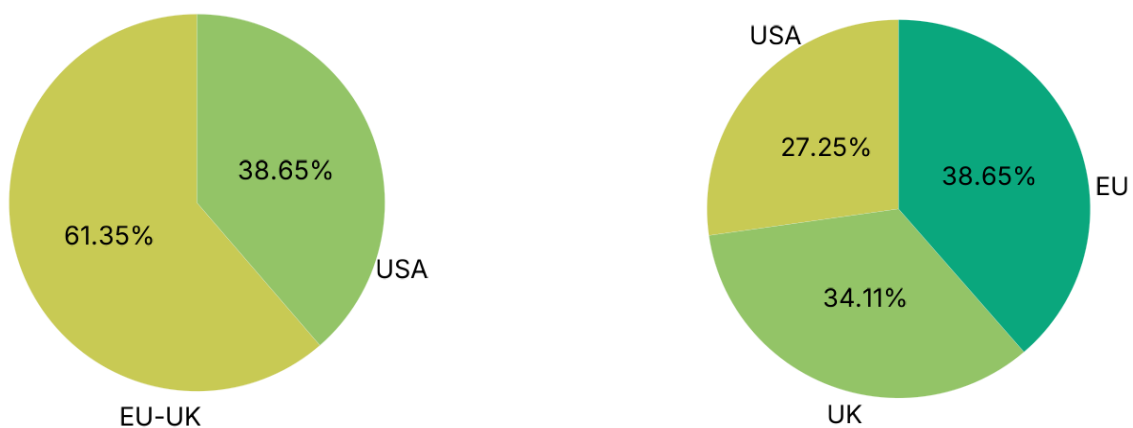

Figure 96 • Distribution of respondents by Country

The largest age group falls between 30-45 years old, making up 44.90% of respondents. The remaining responses are spread across other age ranges.

About the gender distribution in the dataset, it is skewed towards male respondents, with 70.40% male and 29.40% female participants.

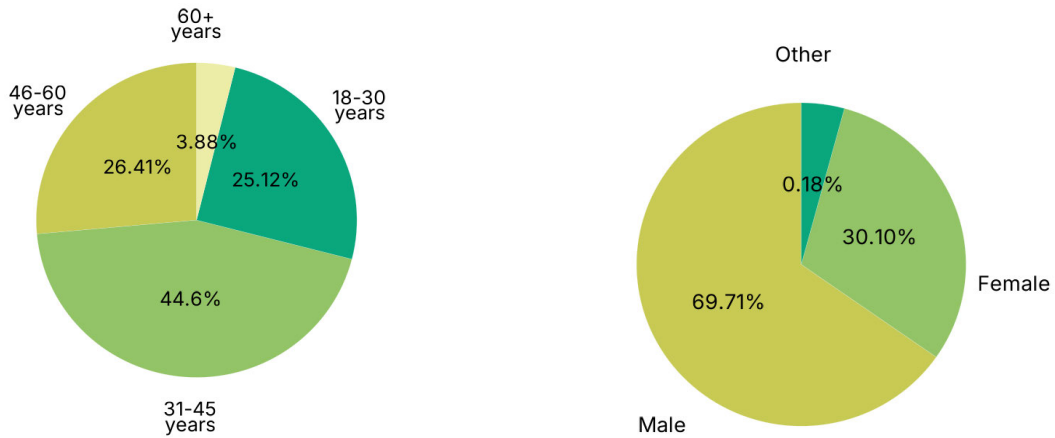

Figure 97 • Distribution of respondents by Age Category and by Gender

The majority of respondents come from the high-tech and B2C sectors.

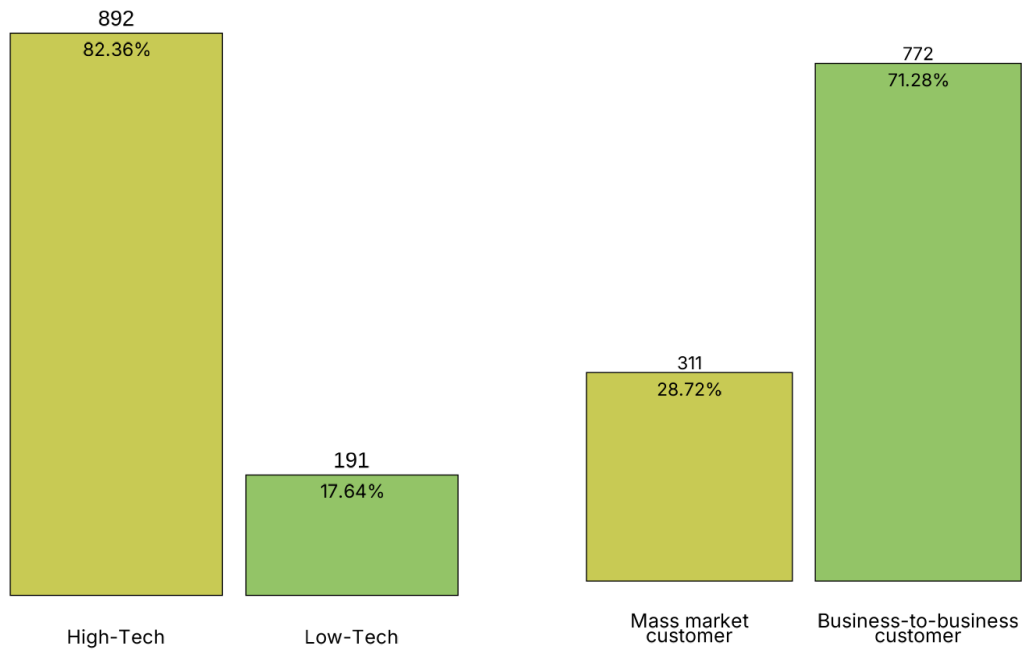

Figure 98 • Distribution of respondents by Sector

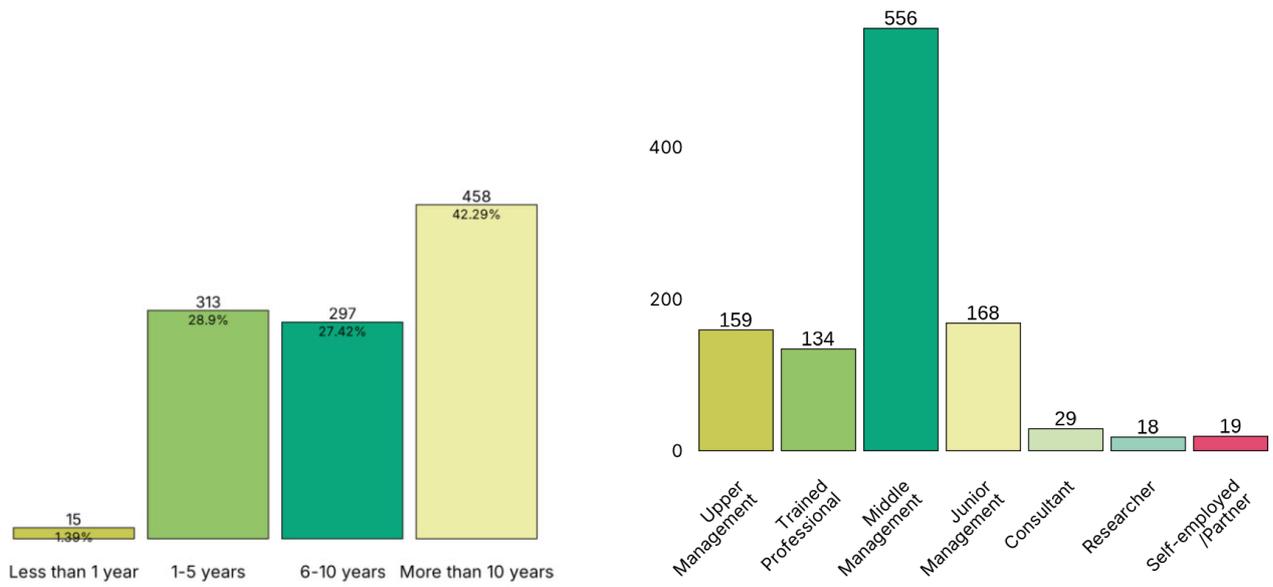

Figure 99 • Distribution of respondents by Experience Level and Role

# ACKNOWLEDGMENTS

The authors thank the participating managers for their valuable support and engagement throughout the study.

**Authors' Affiliations:** Silvia Tedeschi – IMT School for Advanced Studies Lucca (silvia.tedeschi@imtlucca.it), Giacomo Marzi – IMT School for Advanced Studies Lucca (giacomo.marzi@imtlucca.it), Marco Balzano – University of Trieste (marco.balzano@units.it), Gabriele Costa – IMT School for Advanced Studies Lucca (gabriele.costa@imtlucca.it).

**License:** This work is licensed under CC BY 4.0 - Free to share and adapt, requires attribution.

**Funding:** This research was supported by the "Resilienza Economica e Digitale" project (CUP D67G23000060001) funded by the Italian Ministry of University and Research (MUR) as "Department of Excellence" (Dipartimenti di Eccellenza 2023-2027, Ministerial Decree no. 230/2022).

**Data availability**: The authors do not have permission to share data.

**DOI:** 10.6084/m9.figshare.28881818

**To Cite this Industry Research Report:**

- **APA 7th:** Tedeschi, S., Marzi, G., Balzano, M., & Costa, G. (2025). *Managerial Insights on Investment Strategy in Cybersecurity: Findings from Multi-Country Research* [Industry research report]. https://doi.org/10.6084/m9.figshare.28881818

- **IEEE:** S. Tedeschi, G. Marzi, M. Balzano, and G. Costa, *Managerial Insights on Investment Strategy in Cybersecurity: Findings from Multi-Country Research*, Industry Research Report, 2025. [Online]. Available: https://doi.org/10.6084/m9.figshare.28881818

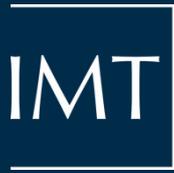

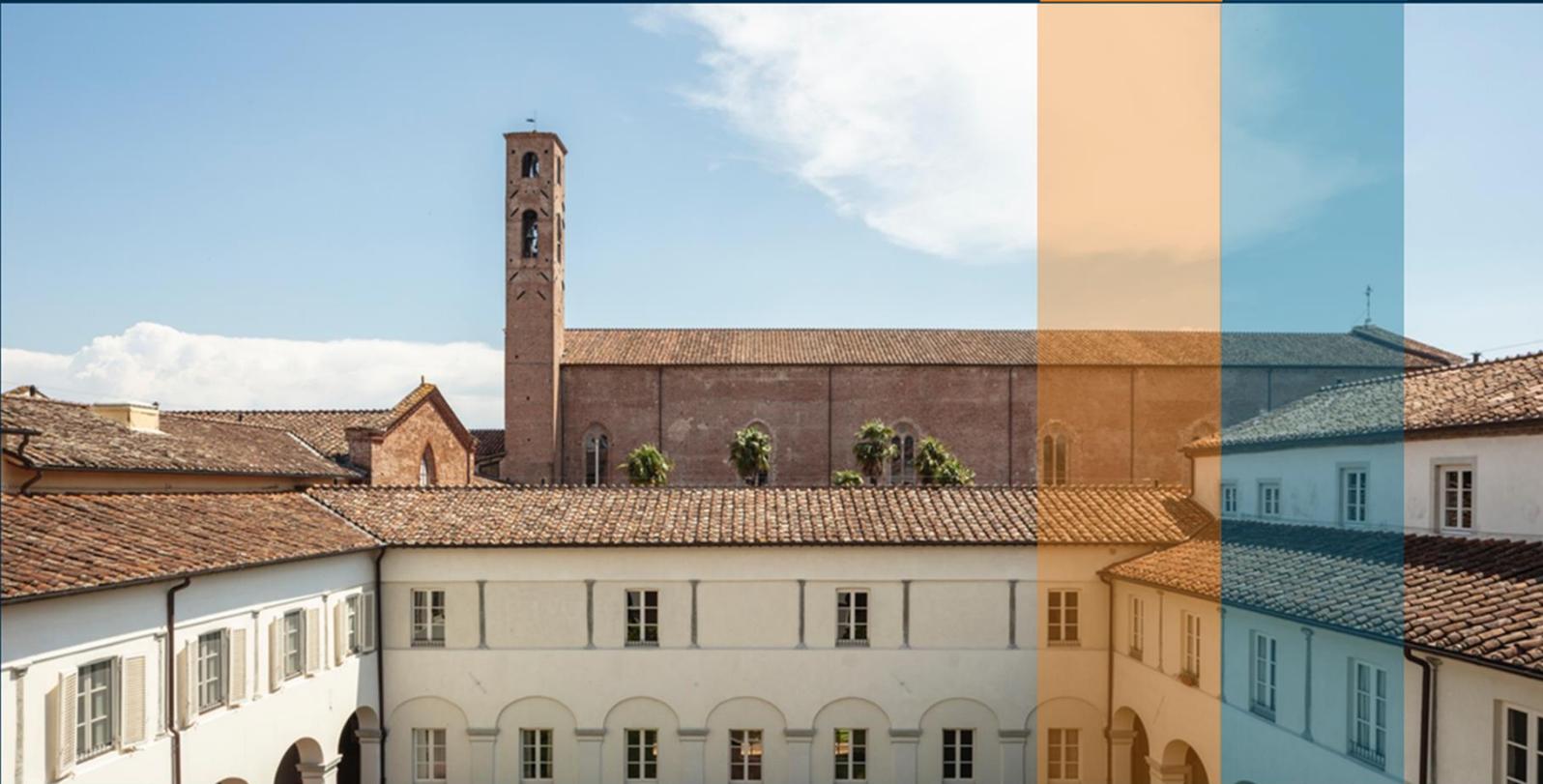

IMT School San Francesco Residential Campus